\documentclass{aa}  

\usepackage{natbib}
\usepackage{graphicx}
\usepackage{txfonts}
\usepackage{hyperref}
\usepackage{mathrsfs}
\usepackage{color}
\usepackage{mathabx}

\hypersetup{
   colorlinks=true,
   citecolor=blue,
   linkcolor=blue,
   urlcolor=black
}

\def\inumber{i}					
\def\vprod{\wedge}				
\def\sprod{\! \cdot \!}				
\def\Nset{\mathbb{N}}			
\def\Rset{\mathbb{R}}			
\def\define{\equiv}				
\def\sign{{\rm sign}}				
\def\isup{{\rm sup}}				
\def\iinf{{\rm inf}}				
\def\sphere{\mathcal{S}}			
\def\scale{\, \propto \,}			

\def\Xreg{X}		
\def\Yreg{Y}		
\def\alphareg{\alpha}	
\def\betareg{\beta}	
\def\Rdet{R}		

\newcommand{\notation}[4]{#1_{#2 ; #3}^{#4}}	
\newcommand{\ddroit}{{\rm d}}				
\newcommand{\vect}[1]{\boldsymbol{#1}}		
\newcommand{\normvar}[1]{\hat{#1}}			
\newcommand{\dd}[2]{\partial_{#2} #1}		
\newcommand{\ddd}[3]{\partial_{#2 #3} #1}	
\newcommand{\DD}[2]{\dfrac{\ddroit #1}{\ddroit #2}}			
\newcommand{\DDn}[3]{\dfrac{\ddroit^{#3} #1}{\ddroit #2^{#3}}} 
\newcommand{\Dpart}[2]{\dfrac{{\rm D} #1}{{\rm D} #2}} 
\newcommand{\real}[1]{\Re \left\{ #1 \right\}}	
\newcommand{\imag}[1]{\Im \left\{ #1 \right\}}	
\newcommand{\abs}[1]{\left| #1 \right|}		

\def\nab{\nabla}					
\def\grad{\nab}					
\def\div{\grad \sprod}				
\def\divh{\grad_{\rm h} \sprod}		

\def\llat{l}						
\def\mm{m}					
\def\Legendre{P}				
\def\SPH{Y}					

\newcommand{\SPHlm}[2]{\SPH_{#1}^{#2}} 

\def\Yquad{\SPHlm{2}{2}}			
\def\Ylm{\SPHlm{\llat}{\mm}}		
\def\xpoly{x}					

\def\nn{n}						
\def\Hough{\Theta}				
\def\spinpar{\nu}				
\def\mnu{{\mm,\spinpar}}			
\def\Laplace{\mathcal{L}^{\mnu}}		

\newcommand{\LegP}[1]{\Legendre_{#1}}			
\newcommand{\LegF}[2]{\Legendre_{#1}^{#2}}		
\newcommand{\expo}[1]{{\rm e}^{#1}}				
\newcommand{\integ}[4]{\int_{#3}^{#4} #1 {\rm d} #2 }	
\newcommand{\HoughF}[3]{\Hough_{#1}^{#2,#3}}		
\newcommand{\Houghval}[3]{\Lambda_{#1}^{#2,#3}}		

\def\Ggrav{\mathcal{G}}					
\def\SBconstant{\sigma_{\rm SB}}	

\def\rr{r}						
\def\col{\theta}					
\def\lon{\varphi}					
\def\time{t}					
\def\xx{x}						
\def\zz{z}						

\def\ivenus{{\rm Venus}}			
\def\ipla{{\rm p}}				
\def\istar{\star}					
\def\isolar{{\rm sol}}				
\def\smaxis{a}					
\def\norb{n_\istar}				
\def\spinrate{\Omega}			
\def\spinvect{\vect{\spinrate}}		
\def\smvenus{\smaxis_{\ivenus}}

\def\periode{P}					
\def\luminosity{L}				
\def\Mbody{M}					
\def\Rbody{R}					
\def\Mpla{\Mbody_\ipla}			
\def\Mstar{\Mbody_\istar}			
\def\Rpla{\Rbody_\ipla}			
\def\Psol{\periode_{\isolar}}		

\def\iearth{\Earth}				
\def\Mearth{\Mbody_{\iearth}}		
\def\Rearth{\Rbody_{\iearth}}		

\def\isun{\sun}					
\def\Msun{\Mbody_{\isun}}			
\def\Lsun{\luminosity_{\isun}}		

\def\iatm{{\rm atm}}				
\def\isol{{\rm gr}}				
\def\idrag{{\rm R}}				
\def\isurf{{\rm s}}				
\def\chartime{\tau}				
\def\freq{\sigma}				
\def\tauinf{\chartime_{\iinf}}		
\def\tausup{\chartime_{\isup}}		
\def\ggravi{g}					
\def\gvect{\vect{\ggravi}}			
\def\Hatm{H}					
\def\fdrag{\freq_\idrag}			

\def\nitrogen{{\rm N_2}}		
\def\carbondiox{{\rm CO_2}}	
\def\water{{\rm H_2 O}}		
\def\sulfuricacid{{\rm H_2 SO_4}}	
\def\psatwater{\pressure_{\water}}	

\def\iconv{{\rm conv}}	
\def\Pconv{\periode_{\iconv}}	
\def\irad{{\rm rad}}		
\def\taurad{\chartime_{\irad}}	
\def\gad{\Gamma_1}		
\def\kad{\kappa}		
\def\Rspec{\mathcal{R}_{\rm s}}	
\def\Rgp{\mathcal{R}_{\rm GP}}		
\def\Mgaz{M_{\iatm}}				
\def\pressure{p}				
\def\density{\rho}				
\def\Gpress{G}					
\def\temp{T}					
\def\flux{F}					
\def\Albedo{A}					
\def\yvar{y}					
\def\thermalinertia{I}				
\def\emissiv{\epsilon}			
\def\psurf{\pressure_{\isurf}}		
\def\Tsurf{\temp_{\isurf}}			
\def\heatcapa{C}				
\def\Cpgaz{\heatcapa_{\pressure}}			
\def\effheat{\varepsilon}			
\def\fluxstar{\flux_{\istar}}			
\def\Lstar{\luminosity_{\istar}}		
\def\Asurf{\Albedo_{\isurf}}		
\def\Csurf{\heatcapa_{\isurf}}		
\def\Ceff{\heatcapa}				
\def\Teff{\temp_{\rm e}}			
\def\Iatm{\thermalinertia_{\iatm}}	
\def\Isol{\thermalinertia_{\isol}}		
\def\emisurf{\emissiv_{\isurf}}		
\def\conductivity{k}				
\def\diffusivity{K}				
\def\katm{\conductivity_{\iatm}}		
\def\ksol{\conductivity_{\isol}}		
\def\Katm{\diffusivity_{\iatm}}		
\def\Ksol{\diffusivity_{\isol}}		
\def\Csol{\heatcapa_{\isol}}		
\def\Bsol{\mathfrak{B}_{\isurf}}		
\def\Bsolstat{\Bsol^{0}}			
\def\Bsolf{\Bsol^{\ftide}}			
\def\tausurf{\chartime_{\isurf}}		

\def\tauM{\chartime_{\rm M}}		

\def\ftide{\freq}					
\def\msigma{{\mm,\ftide}}			
\def\perturb{\delta}				
\def\gravpot{U}					
\def\vel{V}						
\def\Qheat{Q}					
\def\iforcing{{\rm T}}				
\def\ithermal{{\rm J}}				
\def\ibgd{0}					
\def\iLamb{{\rm L}}				
\def\imaxwell{{\rm M}}			
\def\ipeak{{\rm max}}				
\def\maxpress{q}				
\def\tauM{\chartime_{\imaxwell}}	
\def\maxM{\maxpress_{\imaxwell}}			
\def\sigmaM{\ftide_{\imaxwell}}		
\def\sigmaMsup{\sigmaM^{\isup}}	
\def\fpeak{\ftide_{\ipeak}}			
\def\omegapeak{\omeganorm_{\ipeak}}	
\def\maxpeak{\maxpress_{\ipeak}}	
\def\taupeak{\chartime_{\ipeak}}	

\def\iinc{{\rm inc}}
\def\deltaflux{\perturb \! \! \: \flux}	
\def\fluxsurf{\flux_{\isurf}}			
\def\deltaFatm{\deltaflux_{\iatm}}	
\def\deltaFinc{\deltaflux_{\iinc}}		
\def\deltaFrad{\deltaflux_{\irad}}		
\def\deltaQsol{\perturb \Qheat_{\isol}}	
\def\deltaQatm{\perturb \Qheat_{\iatm}}	
\def\coolingcoeff{K}				
\def\skinthickness{h}				
\def\skinhsol{\skinthickness_{\isol}^{\ftide}}	
\def\skinhatm{\skinthickness_{\iatm}^{\ftide}}	

\def\pbgd{\pressure_{\ibgd}}		
\def\rhobgd{\density_{\ibgd}}		
\def\Tbgd{\temp_{\ibgd}}			

\def\iturning{{\rm TP}}
\def\fturning{\ftide_{\iturning}}		
\def\fthermal{\ftide_{\ithermal}}		
\def\maxthermal{\maxpress_{\ithermal}} 		
\def\tauthermal{\chartime_{\ithermal}}

\def\heq{h}					
\def\tauJ{b_{\ithermal}}			
\def\omeganorm{\omega}			
\def\fturningnorm{\omeganorm_{\iturning}}
\def\omegaLamb{\omeganorm_{\iLamb}} 
\def\fLamb{\ftide_{\iLamb}}		
\def\hLamb{h_{\iLamb}}			
\def\vLamb{\vel_{\iLamb}}			
\def\Jtide{J}					
\def\Jsurf{\Jtide_{\isurf}}			
\def\Utide{\gravpot_\iforcing}		
\def\Vvect{\vect{\vel}}			
\def\Vr{\vel_\rr}					
\def\Vtheta{\vel_\col}				
\def\Vphi{\vel_\lon}				
\def\deltap{\perturb \pressure}		
\def\deltarho{\perturb \density}		
\def\deltaT{\perturb \temp}			
\def\deltaTsurf{\perturb \Tsurf}		
\def\deltapsurf{\perturb \psurf}		
\def\deltapsquad{\notation{\deltap}{\isurf}{2}{2,\ftide}} 
\def\deltaTsquad{\notation{\deltaT}{\isurf}{2}{2,\ftide}} 
\def\deltaFincquad{\notation{\deltaflux}{\iinc}{2}{2,\ftide}}

\def\kwave{\hat{k}}				
\def\kvert{\kwave_{\xx}}			
\def\Ulmsig{\notation{\gravpot}{\iforcing}{\llat}{\mm,\ftide}}	
\def\quanti{q}					
\def\quantisig{\quanti^{\ftide}}		

\def\Fms{\mathcal{F}^{\msigma}}	

\def\Vthetams{\Vtheta^{\msigma}}	
\def\Vphims{\Vphi^{\msigma}}	

\def\Gms{\Gpress^{\msigma}}		
\def\Jms{\Jtide^{\msigma}}		

\def\deltapn{\deltap_{\nn}^{\msigma}}			
\def\Gmsn{\Gpress_{\nn}^{\msigma}}				
\def\heqmsn{\heq_{\nn}^{\msigma}}

\def\deltapsurfl{\notation{\deltap}{\isurf}{\llat}{\msigma}} 

\def\torque{\mathcal{T}}			
\def\etatorque{\eta}				

\def\Aint{A}		
\def\Bint{B}		
\def\yspec{\yvar^{\left( 0 \right)}}	

\def\Pl{\LegP{\llat}}						
\def\Plm{\LegF{\llat}{\mm}}				
\def\Thetan{\HoughF{\nn}{\mm}{\spinpar}}		
\def\Lambdan{\Houghval{\nn}{\mm}{\spinpar}}	
\def\Lambdao{\Lambda_0}				

\def\ffunc{f}					
\def\iodd{{\rm odd}}				
\def\ieven{{\rm even}}			
\def\igcm{{\rm GCM}}			
\def\fodd{\ffunc_{\iodd}}			
\def\feven{\ffunc_{\ieven}}			
\def\fgcm{\ffunc_{\igcm}}			

\def\xmod{\chi}					
\def\iactBF{1}					
\def\iactHF{2}					
\def\aact{a}					
\def\bact{b}					
\def\xact{\xmod}				
\def\dact{d}					
\def\alf{\aact_\iactBF}			
\def\blf{\bact_\iactBF}			
\def\xlf{\xact_\iactBF}			
\def\dlf{\dact_\iactBF}			
\def\ahf{\aact_\iactHF}			
\def\bhf{\bact_\iactHF}			
\def\xhf{\xact_\iactHF}			
\def\dhf{\dact_\iactHF}			
\def\btrans{b_{\rm trans}}			

\def\Ffunction{\mathcal{F}}		
\def\ipara{{\rm par}}				
\def\Fpara{\Ffunction_{\ipara}}		
\def\torquepara{\torque_{\ipara}}	
\def\Flf{\Ffunction_{\iactBF}}		
\def\Fhf{\Ffunction_{\iactHF}}		

\def\xv{\ftide}
\def\yv{\torque}
\def\xvnorm{\normvar{\xv}}
\def\yvnorm{\normvar{\yv}}
\def\parx{\alpha_1}
\def\pary{\alpha_2}
\def\jcas{j}
\def\kcas{k}
\def\qpar{\mathfrak{p}}
\def\Ncas{N_{\qpar}}
\def\Nxv{N_{\xv}}

\def\xvj{\xv_{\jcas}}
\def\yvj{\yv_{\jcas}}

\def\xvjfirst{\xv_{\jcas,1}}
\def\xvjlast{\xv_{\jcas,\Nxv}}
\def\xvnormj{\xvnorm_{\jcas}}
\def\yvnormj{\yvnorm_{\jcas}}
\def\aj{\qpar_{\jcas}}
\def\ak{\qpar_{\kcas}}
\def\finterp{f}
\def\finterpj{f_{\jcas}}
\def\finterpk{f_{\kcas}}
\def\xvnorminf{\xvnorm_{\iinf}}
\def\xvnormsup{\xvnorm_{\isup}}
\def\Ffit{F}
\def\iref{{\rm ref}}

\def\taupeakref{\chartime_{\iref}}
\def\maxpeakref{\maxpress_{\iref}}
\def\qparref{\qpar_{\iref}}

\def\icar{0}

\def\taucar{\chartime_{\icar}}
\def\maxcar{\maxpress_{\icar}}

\newcommand{\eq}[1]{Eq.~(\ref{#1})}
\newcommand{\eqs}[2]{Eqs.~(\ref{#1}) and~(\ref{#2})}
\newcommand{\eqsto}[2]{Eqs.~(\ref{#1} - \ref{#2})}
\newcommand{\append}[1]{Appendix~\ref{#1}}
\newcommand{\units}[1]{~${\rm #1}$}
\newcommand{\sect}[1]{Sect.~\ref{#1}}
\newcommand{\fig}[1]{Fig.~\ref{#1}}
\newcommand{\figs}[2]{Figs.~\ref{#1} and~\ref{#2}}
\newcommand{\tab}[1]{Table~\ref{#1}}

\newcommand{\beqtwo}[2]{
\begin{equation}
\begin{array}{l l l}
\displaystyle #1 , & \mbox{and} & \displaystyle #2
\end{array}
\end{equation}
}

\newcommand{\comments}[1]{}

\definecolor{emerald}{rgb}{0.3,0.85,0.2}
\definecolor{smcolor}{rgb}{0.7,0.3,0.0}

\newcommand{\rec}[1]{#1}

\begin{document} 
  \title{A generic frequency dependence for the atmospheric tidal torque \\ of terrestrial planets}




  
  \author{P. Auclair-Desrotour
     \and
           J. Leconte
      \and
      	   C. Mergny
          }

  \institute{Laboratoire d'Astrophysique de Bordeaux, Univ. Bordeaux, CNRS, B18N, allée Geoffroy Saint-Hilaire, 33615 Pessac, France\\
              \email{pierre.auclair-desrotour@u-bordeaux.fr, jeremy.leconte@u-bordeaux.fr}
             }

  \date{Received ...; accepted ...}

  \abstract
   {Thermal atmospheric tides have a strong impact on the rotation of terrestrial planets. They can lock these planets into an asynchronous rotation state of equilibrium.}
   {We aim at characterizing the dependence of the tidal torque resulting from the semidiurnal thermal tide on the tidal frequency, the planet orbital radius, and the atmospheric surface pressure.}
   {The tidal torque is computed from full 3D simulations of the atmospheric climate and mean flows using a generic version of the LMDZ general circulation model (GCM) in the case of a nitrogen-dominated atmosphere. Numerical results are discussed with the help of an updated linear analytical framework. Power scaling laws governing the evolution of the torque with the planet orbital radius and surface pressure are derived.}
   {The tidal torque exhibits i) a thermal peak in the vicinity of synchronization, ii) a resonant peak associated with the excitation of the Lamb mode in the high frequency range, and iii) well defined frequency slopes outside these resonances. These features are well explained by our linear theory. Whatever the star-planet distance and surface pressure, the torque frequency spectrum -- when rescaled with the relevant power laws -- always presents the same behaviour. This allows us to provide a single and easily usable empirical formula describing the atmospheric tidal torque over the whole parameter space. With such a formula, the effect of the atmospheric tidal torque can be implemented in evolutionary models of the rotational dynamics of a planet in a computationally efficient, and yet relatively accurate way.}
   {}



\maketitle


\section{Introduction} 

Understanding the evolution of planetary systems has become a crucial question with the rapidly growing number of exoplanets discovered up to now. Terrestrial planets particularly retain our attention as they offer a fascinating diversity of orbital configurations, and possible climates and surface conditions. This diversity is well illustrated by Proxima-b, an exo-Earth with a minimum mass of 1.3~$M_\Earth$ orbiting Proxima Centauri \citep[][]{AE2016,Ribas2016}, and the TRAPPIST-1 system, which is a tightly-packed system of seven Earth-sized planets orbiting an ultracool dwarf star \citep[][]{Gillon2017,Grimm2018}. 

Characterizing the atmospheric dynamics and climate of these planets is a topic that motivated numerous theoretical works, both analytical and numerical \citep[e.g.][]{Pierrehumbert2011,HK2012,Leconte2013,HW2014,Wolf2017a,Wolf2017b,Turbet2018}. This tendency will be reinforced in the future by the rise of forthcoming space observatories such as the James Webb Space Telescope (JWST), which will unravel features of the planetary atmospheric structure by performing high resolution spectroscopy over the infrared frequency range \citep[][]{Lagage2015}. 

Constraining the climate and surface conditions of the observed terrestrial planets requires to constrain their rotation rate first because of the key role played by this parameter in the equilibrium atmospheric dynamics \citep[][]{Vallis2006,Pierrehumbert2010}. Particularly, it is important to know whether a planet is locked into the configuration of spin-orbit synchronization with its host star and the extent to which asynchronous rotation states of equilibrium might exist. Over long timescales, the planet rotation is driven by tidal effects, that is the distortion of the planet by its neighbors (star, planets and satellites) resulting from mutual distance interactions. Tides are a source of internal dissipation inducing a variation of mass distribution delayed with respect to the direction of the perturber. As a consequence, the planet undergoes a tidal torque, which modifies its rotation by establishing a transfer of angular momentum between the orbital and spin motions. 

Tides can be generated by forcings of different natures. First, the whole planet is distorted by the gravitational tidal potential generated by the perturber, and is driven by the resulting tidal torque toward spin-orbit synchronous rotation and a circular orbital configuration. Second, if the perturber is the host star, the atmosphere of the planet undergoes a heating generated by the day-night cycle of the incoming stellar flux. The variations of the atmospheric mass distribution generated by this forcing are the so-called thermal atmospheric tides \citep[][]{CL70}. 

As demonstrated by the pioneering study by \cite{GS1969} in the case of Venus, thermal tides are able to drive a terrestrial planet away from spin-orbit synchronization since they induce a tidal torque in opposition with that resulting from solid tides in the low frequency range. Hence, the competition between the two effects locks the planet into an asynchronous rotation state of equilibrium, which explains the departure of the rotation rate of Venus to spin-orbit synchronization.

The understanding of this mechanism has been progressively consolidated by analytical works based upon the classical tidal theory \citep[e.g.][]{ID1978,DI1980,ADLM2017a,ADLM2017b} or using parametrized models \citep[][]{CL2001,CL2003,Correia2003}. Over the past decade, the growing performances of computers have made full numerical approaches affordable, and the atmospheric torque created by the thermal tide was computed using general circulation models \citep[or GCM;][]{Leconte2015}. This approach remains complementary with analytical models owing to its high computational cost. However, it is particularly interesting since it allows to characterize the atmospheric tidal response of a planet by taking into account the atmospheric structure, mean flows and other internal processes by solving the primitive equations of fluid dynamics in a self-consistent way. 

By using a generic version of the LMDZ GCM \citep[][]{Hourdin2006}, \cite{Leconte2015} retrieved the frequency-dependence of the tidal torque predicted by ab initio analytical models \citep[][]{ID1978,ADLM2017a,ADML2018}. The torque increases linearly with the tidal frequency in the vicinity of synchronization. It reaches a maximum associated with a thermal time of the atmosphere and then decays in the high-frequency range. This behaviour is approximated at the first order by the Maxwell model, which describes the forced response of a damped harmonic oscillator. It shows evidence of the important role played by dissipative processes such as radiative cooling in Venus-like configurations. In order to better understand the action of the thermal tide on the planet rotation, this frequency-dependent behaviour has to be characterized. 

Thus, our purpose in this study is to investigate the dependences of the tidal torque created by the semidiurnal tide on the tidal frequency and on key control parameters. We follow along the line by \cite{Leconte2015} for the method, and treat the case of an idealized dry terrestrial planet hosting a nitrogen-dominated atmosphere and orbiting a Sun-like star. Hence, we recall in \sect{sec:basic_principle} the mechanism of the thermal atmospheric tide. In \sect{sec:method}, we detail the method and the physical setup of the treated case.

In \sect{sec:frequency_behaviour}, we compute the tidal torque exerted on the atmosphere from simulations using the LMDZ GCM and examine its dependence on the tidal frequency. We introduce in this section two new models for the thermally generated atmospheric tidal torque: an ab initio analytical model based upon the linear theory of atmospheric tides \citep[e.g.][]{CL70}, and a parametrized semi-analytical model derived from results obtained using GCM simulations. This later model describes in a realistic way the behaviour of the torque in the low-frequency range, where a thermal peak is observed. In addition, we investigate in this section the role played by the ground-atmosphere thermal coupling in the lag of the tidal bulge. 

In \sect{sec:exploration_space}, we examine the dependence of the tidal response on the planet orbital radius and surface pressure. We thus establish empirical scaling laws describing the evolution of the characteristic amplitude and timescale of the thermal peak with these two parameters. Combining together the obtained results, we finally derive a new generic formula to quantify the atmospheric tidal torque created by the thermal semidiurnal tide in the case of a $\nitrogen$-dominated atmosphere. We give our conclusions in \sect{sec:conclusions}.

\section{Basic principle}
\label{sec:basic_principle}

\comments{Insister sur la vision globale.}


We briefly recall in this section the main aspects of the mechanism of atmospheric tides in the case of terrestrial planets, and we introduce analytical expressions that will be used in the following to compute the resulting tidal torque. For the sake of simplification, we consider in this study the case of a spherical planet of radius $ \Rpla $ and mass $ \Mpla $, orbiting its host star, of mass $\Mstar$, circularly. The star-planet distance is denoted $ a $, the mean motion of the system $\norb $, and the obliquity of the planet is set to zero. We assume that the planet rotates at the spin angular velocity $ \spinrate $, which is positive if the spin rotation is along the same direction as the orbital motion, and negative otherwise. 

The atmosphere of the planet undergoes both the tidal gravitational and thermal forcings of the host star. Below a certain orbital radius, the planet is sufficiently close to the star to make gravitational forces predominate. Thus, its rotation is driven towards spin-orbit synchronization ($\spinrate = \norb$), which is the unique possible final state of equilibrium for the planet rotation in the absence of obliquity and eccentricity. Conversely, the predominance of the thermal tide enables the existence of asynchronous final rotation states of equilibrium, as showed in the case of Venus \citep[e.g.][]{GS1969,ID1978,DI1980,CL2001,ADLM2017a}. As a consequence, we ignore here the action of gravitational forces on the atmosphere. Note however that the action of these forces on the atmospheric tidal bulge will be taken into account to compute the tidal torque, as seen in the following. 

 The thermal forcing results from the day-night periodic cycle. The atmosphere undergoes heating variations due to the time-varying component of the incoming stellar flux $F$, which scales as the equilibrium one, $\fluxstar = \Lstar / \left( 4 \pi \smaxis^2 \right)$, where $\Lstar$ is the luminosity of the star. Hence, the absorbed energy induces a delayed variation of the atmospheric mass distribution. Let us assume the hydrostatic approximation (i.e. that pressure and gravitational forces compensate each other exactly in the vertical direction) and consider that the surface of the planet is rigid enough to support the atmospheric pressure variations with negligible distortions. It follows that the variation of mass distribution is directly proportional to the surface pressure anomaly,
 
\begin{equation}
\deltapsurf \left( \time , \col , \lon \right) = \sum_{\llat = 1}^{+ \infty} \sum_{\mm =- \llat}^\llat \deltapsurfl  \Ylm \left( \col , \lon \right) \expo{\inumber \ftide \time }. 
\label{deltaps_SH}
\end{equation}

\noindent where $ \llat$ and $\mm$ designate the \rec{latitudinal and longitudinal degrees} of a mode,  $ \col$ and $ \lon $ the colatitude and longitude in the reference frame co-rotating with the planet, $\time$ the time, $\Ylm $ the normalized spherical harmonics (see \append{app:normSH}), $ \deltapsurfl $ the associated components, \rec{and $\ftide = \mm \left(\spinrate - \norb \right)$ the associated forcing frequencies \citep[see e.g.][]{Efroimsky2012,Ogilvie2014}}\footnote{\rec{The expression of $\deltapsurf$ given by \eq{deltaps_SH} is similar to that of the tidal gravitational potential, which is derived by applying the addition theorem to the components of the potential expanded in Legendre polynomials \citep[see e.g.][Eq.~(47)]{Efroimsky2012}. }}.

The tidal torque exerted on the atmosphere is obtained by integrating the gravitational force undergone by the tidal bulge over the sphere. Hence, denoting $\Utide$ the tidal gravitational potential at the planet surface, the atmospheric tidal torque is defined in the thin-layer approximation ($\Hatm \ll \Rpla$) as \citep[e.g.][]{Zahn1966a}

\begin{equation}
\torque = \frac{1}{\ggravi} \integ{\dd{\Utide}{\lon}  \, \deltapsurf}{\sphere}{\sphere}{} ,
\label{torqueZahn}
\end{equation}

\noindent where the notation $\ggravi$ refers to the surface gravity of the planet, $\dd{}{\lon} $ to the partial derivative in longitude, $ \sphere $ to the sphere of radius $\Rpla$, and $\ddroit \sphere = \Rpla^2 \sin \col \ddroit \col \ddroit \lon$ to the surface element. 

Similarly as the surface pressure anomaly, $\Utide$ can be expanded in Fourier series of time and spherical harmonics,

\begin{equation}
\Utide \left( \time , \col , \lon \right) = \sum_{\llat = 1}^{+ \infty} \sum_{\mm=-\llat}^\llat \Ulmsig  \Ylm \left( \col , \lon \right) \expo{\inumber \ftide \time},
\end{equation}

\noindent where the $ \Ulmsig $ are the amplitudes of the different modes. Terms associated with $ \llat=1 $ do not contribute to the tidal torque since they just correspond to a displacement of the planet gravity center. Thus, the main components of the expansion are those associated with the quadrupolar semidiurnal tide, that is with \rec{degrees} $ \llat = \abs{\mm} = 2 $. Besides, since the $ \Ulmsig $ scale as $ \Ulmsig \scale \left( \Rpla / \smaxis \right)^\llat $, terms of greater order in $ \llat$ can be neglected with respect to the quadrupolar components if the radius of the planet $\Rpla$ is assumed to be small compared to the star-planet distance, which is the case in the present study.  

Thus, by substituting $\Utide$ and $\deltapsurf$ by their expansions in spherical harmonics in \eq{torqueZahn}, we note that the quadrupolar terms $ \llat = \abs{\mm} = 2$ only remain, as for the tidal potential $\Utide$, and we end up with the well-known expression of the semidiurnal quadrupolar torque in the thin layer approximation \citep[e.g.][]{Leconte2015},

\begin{equation}
\torque = \sqrt{\frac{24 \pi}{5}} \frac{\Mstar}{\Mpla} \frac{\Rpla^6}{\smaxis^3} \imag{ \deltapsquad},
\label{torque}
\end{equation}

\noindent with $\ftide = 2 \left( \spinrate - \norb \right)$, and where the notation $\Im$ refers to the imaginary part of a complex number ($\Re$ referring to the real part). In this expression, $\deltapsquad$ designates the component of \rec{degrees} $\llat=2$ and $\mm=2$ in the expansion on spherical harmonics given by \eq{deltaps_SH}. This complex quantity is the most important one since it encompasses the whole physics of the atmospheric tidal response. In the following, it will be calculated using a GCM.

The action of the torque on the planet is fully determined by the sign of the product $\etatorque = \sign \left( \ftide \right) \imag{ \deltapsquad}$. When $ \etatorque < 0 $ ($\etatorque>0$), the atmospheric tidal torque pushes the planet toward (away from) spin-orbit synchronization, $\abs{\spinrate - \norb } $ decays (increases). Positions for which $\eta=0$ correspond to the stable ($ \left. \ddroit \etatorque / \ddroit \ftide \right|_{\rm eq} < 0 $) or unstable ($ \left. \ddroit \etatorque / \ddroit \ftide \right|_{\rm eq} > 0 $) equilibrium rotation rates that the planet would reach if it were subject to atmospheric tides only, that is if solid tides were ignored in the case of a dry terrestrial planet.


\section{Method}
\label{sec:method}

\comments{Ajouter une petite introduction.}

As mentioned in the previous section, the guideline of the method is to compute the quadrupolar component of the surface pressure anomaly from 3D GCM simulations. We detail here the basic physical setup of these simulations in a first time, and the way $\deltapsquad $ is extracted from pressure snapshots in a second time. In the whole study, we focus on a Venus-sized planet orbiting a Sun-like star. 

A ``reference case'' of fixed surface pressure and star-planet distance is defined. Specifically, the surface pressure is set in this case to $\psurf = 10$\units{bar} and we assume that the planet is located at the Venus-Sun distance, that is $\smvenus = 0.723$\units{AU}. This configuration, characterized in \sect{sec:frequency_behaviour}, corresponds to the case illustrated by Fig.~1 of \cite{Leconte2015}, and seems thereby a convenient choice for comparisons with this early work. 

In \sect{sec:exploration_space}, two families of configurations will be studied, both including the reference case. In the first family, the surface pressure is set to $\psurf = 10$\units{bar} and the semi-major axis varies. Conversely, in the second family, planets have the same orbital radius, $\smaxis = \smvenus$, and various surface pressures.

\subsection{Physical setup of the 3D simulations}
\label{sec:physical_setup}

Apart from the surface pressure and star-planet distance, all simulations are based on a common physical setup. For the stellar incoming flux, the emission spectrum of the Sun is used. The planet is assumed to be dry, with no surface liquid water or water vapour, which allows us to filter out effects associated with the formation of clouds in the study of its atmospheric tidal response. The atmosphere is arbitrarily assumed to be nitrogen-dominated. \rec{However, a pure $\nitrogen$-atmosphere would be an extreme case for radiative transfer owing to the absence of radiator. Hence, we have to set a non-zero volume mixing ratio for carbon dioxyde to avoid numerical issues in the treatment of the radiative transfer with the LMDZ, which was originally designed to study the Earth atmosphere. Although any value could be used, we choose to set the value of the $\carbondiox$ volume mixing ratio to that of the Earth atmosphere at the beginning of the twenty-first century,} that is $\sim370$\units{ppmv} \citep[e.g.][]{Etheridge1996}. \rec{The mass ratio corresponding to this volume mixing ratio being negligible, we} use the value of $\nitrogen$ for the mean molecular mass of the atmosphere, $\Mgaz = 28.0134 $\units{g.mol^{-1}} \citep{Meija2016}. 

For a perfect diatomic gas, the ratio of heat capacities (also called first adiabatic exponent) is $\gad = 1.4$, and it follows that $\kad = \left( \Gamma_1 - 1 \right) / \gad = 0.285$ (the parameter $\kad$ can also be written $\kad = \Rgp / \left( \Mgaz \Cpgaz \right)$, where $\Rgp$ and $\Cpgaz$ stand for the perfect gaz constant and the thermal capacity per unit mass of the atmosphere, respectively). The effects of topography are ignored and the surface of the planet is thus considered as an isotropic sphere of albedo $\Asurf = 0.2$ and thermal inertia $\Isol = 2000 $\units{J.m^{-2}.s^{-1/2}.K^{-1}}, which is a typical value for Venus-like soils \citep[see e.g.][]{Lebonnois2010}\footnote{This is the value prescribed for nonporous basalts by \cite{Zimbelman1986}.}. All of these parameters remain unchanged for the whole study and are summarized in Table~\ref{parameters}. 


\begin{table}[h]
\centering
 \textsf{\caption{\label{parameters} List of the main GCM parameters. The subscripts $\iearth$ and $\isun$ refer to the Earth and the Sun, respectively.  }}
\begin{small}
    \begin{tabular}{ l l l l}
      \hline
      \hline
      \textsc{Parameter} & \textsc{Symbol} & \textsc{Value} & \textsc{Unit} \\ 
      \hline 
      \multicolumn{4}{c}{\emph{Planet characteristics}$^{\rm a}$} \\
      Planet mass & $ \Mpla $ &  0.815  & $ \Mearth$  \\
      Planet radius & $ \Rpla $ &  0.950  & $ \Rearth$  \\
      Surface gravity & $\ggravi$ & 8.87 & \units{m.s^{-2}} \\[0.3cm]
      \multicolumn{4}{c}{\emph{Star characteristics}$^{\rm a}$} \\
      Mass of the star & $\Mstar$ & 1.0 & $\Msun$ \\
      Luminosity of the star & $\Lstar$ & 1.0 & $\Lsun$ \\[0.3cm]
      \multicolumn{4}{c}{\emph{Atmospheric properties}$^{\rm b}$} \\
      Mean molecular mass & $\Mgaz$ & 28.0134 & \units{g.mol^{-1}} \\
      Adiabatic exponent & $\kad $ & 0.285 & -- \\[0.3cm]
      \multicolumn{4}{c}{\emph{Surface parameters}} \\
      Surface albedo & $\Asurf$ & 0.2 & -- \\
      Surface thermal inertia & $\Isol$ & 2000 & \units{J.m^{-2}.s^{-1/2}.K^{-1}}  \\
      \hline 
    \end{tabular}
    \begin{flushleft}
    $^{\rm a}$Source: \cite{CT2011}. \\
    $^{\rm b}$From, e.g. \cite{Meija2016}.
    \end{flushleft}
\end{small}
 \end{table}

Our simulations are performed with an upgraded version of the LMD general circulation model (GCM) specifically developed for the study of extrasolar planets and paleoclimates \citep[see e.g.][]{Wordsworth2010,Wordsworth2011,Wordsworth2013,Forget2013,Leconte2013}, and used previously by \cite{Leconte2015} for the study of atmospheric tides. The model is based of the dynamical core of the LMDZ 4 GCM \citep[][]{Hourdin2006}, which uses a finite-difference formulation of the primitive equations of geophysical fluid dynamics. Particularly, the following hypotheses are assumed:
\begin{itemize}
\item \emph{Hydrostatic approximation}. The pressure and gravitational forces compensate each other exactly along the vertical direction. 
\item \emph{Traditional approximation}. Components of the Coriolis approximation associated with a vertical motion of fluid particules or generating a force along the vertical direction are ignored.
\item \emph{Thin layer approximation}. The thickness of the atmosphere is considered as small with respect to the radius of the planet.
\end{itemize}

\noindent A spatial resolution of $32 \times 32 \times 26$ in longitude, latitude, and altitude is used for the simulations. 

The radiative transfer is computed in the model using a method similar to \cite{Wordsworth2011} and \cite{Leconte2013}. High-resolution spectra characterizing optical properties were preliminary produced for the chosen gas mixture over a wide range of temperatures and pressures using the HITRAN 2008 database \citep[][]{Rothman2009}. These spectra are interpolated every radiative timestep during simulations to determine local radiative transfers. The method is commonly used and has been thoroughly discussed in past studies \citep[e.g.][]{Leconte2013}. We thus refer the readers to these works for a detailed description. 

\begin{figure}[htb]
   \centering
   \includegraphics[width=0.48\textwidth,trim = 11cm 1.5cm 0cm 0cm,clip]{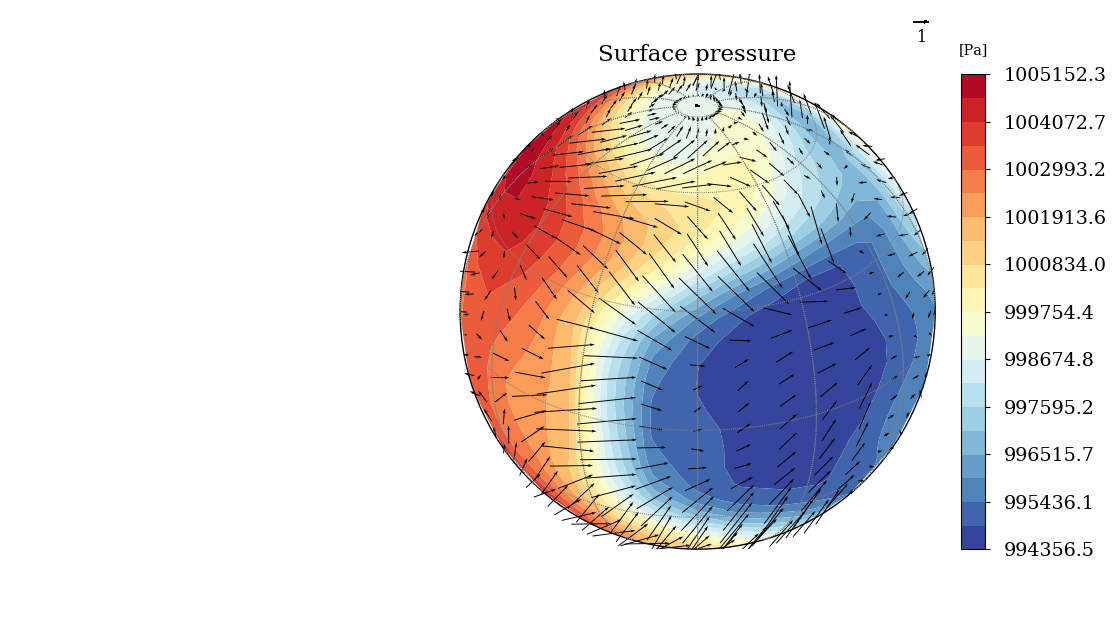}
   \caption{Surface pressure and horizontal winds computed with the LMDZ GCM for a Venus-sized terrestrial planet hosting a 10~bar atmosphere (reference case). In this study, the surface pressure anomaly is folded over one Solar day and expanded in spherical harmonics to calculate the atmospheric tidal torque using the formula given by \eq{torque}.}
       \label{fig:globe_GCM}%
\end{figure}

\begin{figure*}[htb]
   \centering
    \raisebox{1.3cm}{\includegraphics[height=0.02\textwidth]{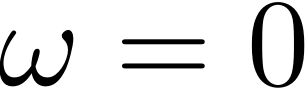}} \hspace{0.5cm} \hspace{0.2cm}
    \includegraphics[width=0.02\textwidth,trim = 1.3cm 7.3cm 32.2cm 2.1cm,clip]{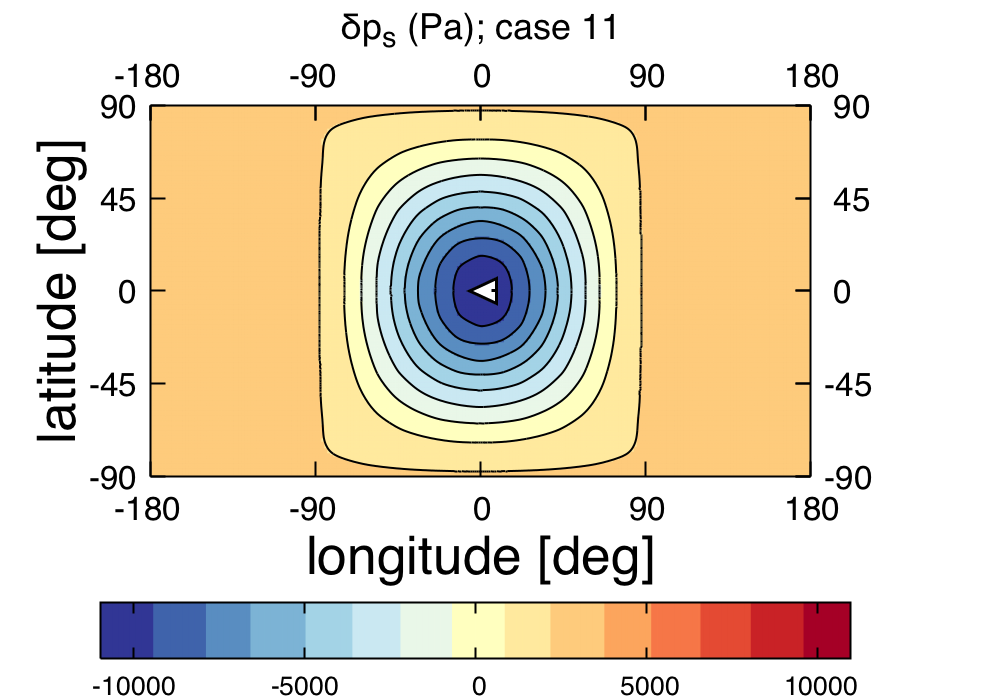} 
   \includegraphics[width=0.30\textwidth,trim = 3.2cm 7.3cm 4.5cm 2.1cm,clip]{auclair-desrotour_fig2b.png} \hspace{0.5cm}
   \includegraphics[width=0.30\textwidth,trim = 3.2cm 7.3cm 4.5cm 2.1cm,clip]{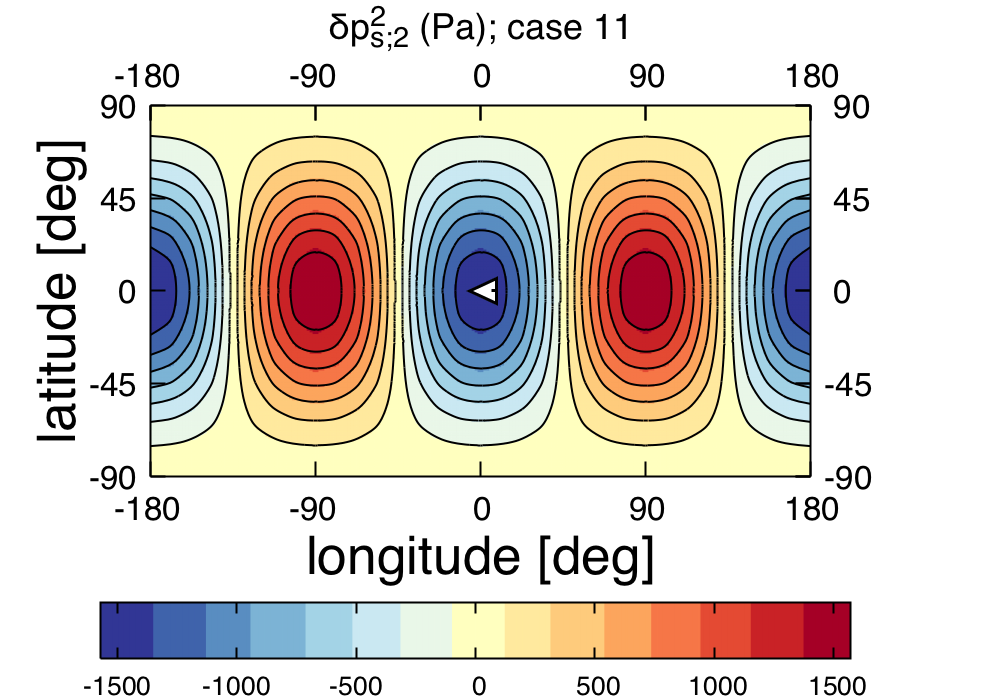} \\
   \raisebox{1.3cm}{\includegraphics[height=0.02\textwidth]{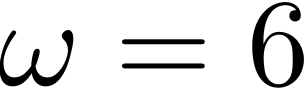}} \hspace{0.5cm} \hspace{0.2cm}
   \includegraphics[width=0.02\textwidth,trim = 1.3cm 7.3cm 32.2cm 3.0cm,clip]{auclair-desrotour_fig2b.png} 
    \includegraphics[width=0.30\textwidth,trim = 3.2cm 7.3cm 4.5cm 3.2cm,clip]{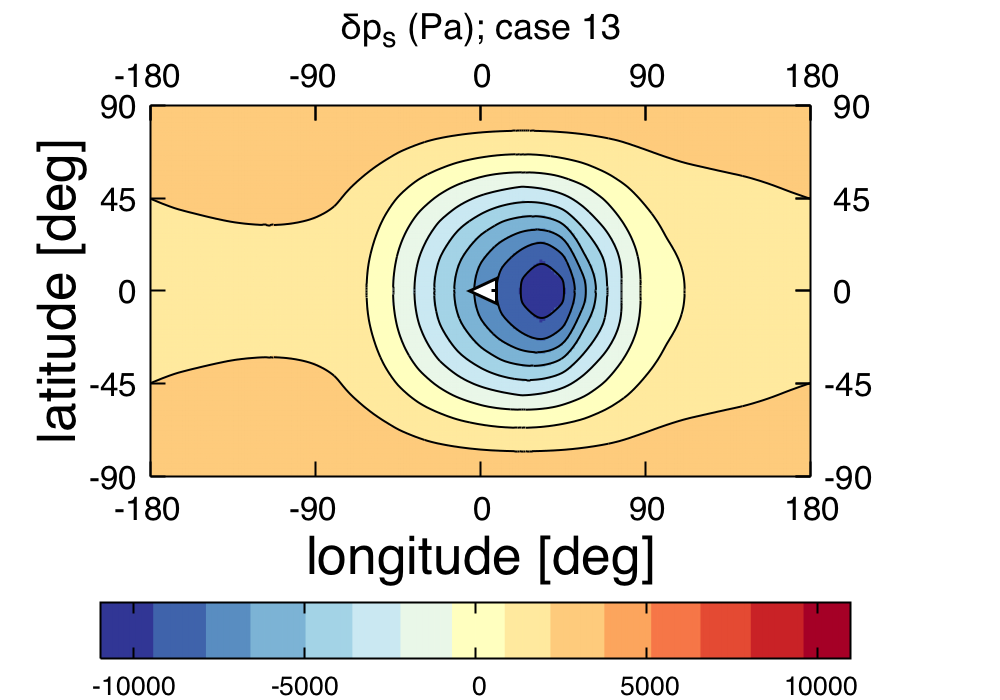} \hspace{0.5cm}
   \includegraphics[width=0.30\textwidth,trim = 3.2cm 7.3cm 4.5cm 3.2cm,clip]{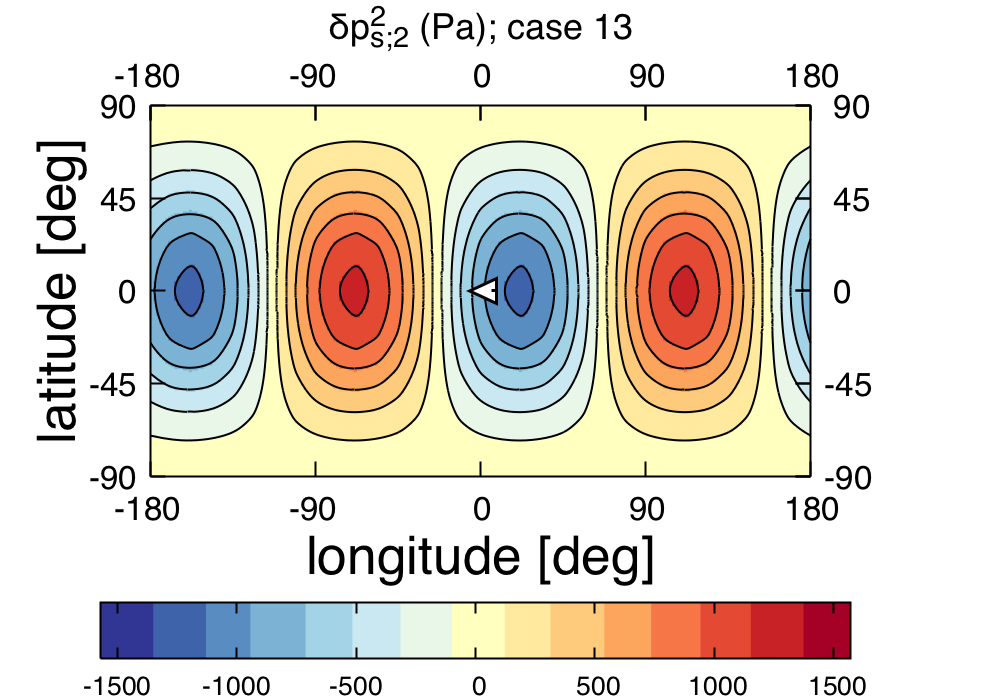} \\
   \raisebox{1.3cm}{\includegraphics[height=0.02\textwidth]{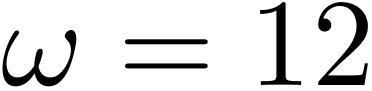}} \hspace{0.5cm}
   \includegraphics[width=0.02\textwidth,trim = 1.3cm 7.3cm 32.2cm 3.0cm,clip]{auclair-desrotour_fig2b.png} 
    \includegraphics[width=0.30\textwidth,trim = 3.2cm 7.3cm 4.5cm 3.2cm,clip]{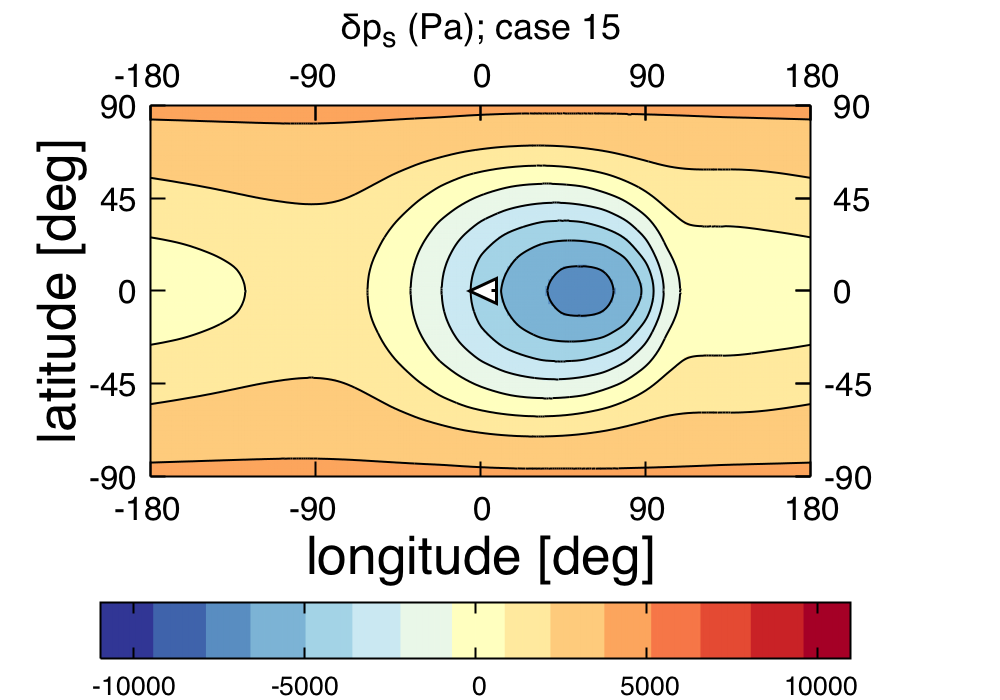} \hspace{0.5cm}
   \includegraphics[width=0.30\textwidth,trim = 3.2cm 7.3cm 4.5cm 3.2cm,clip]{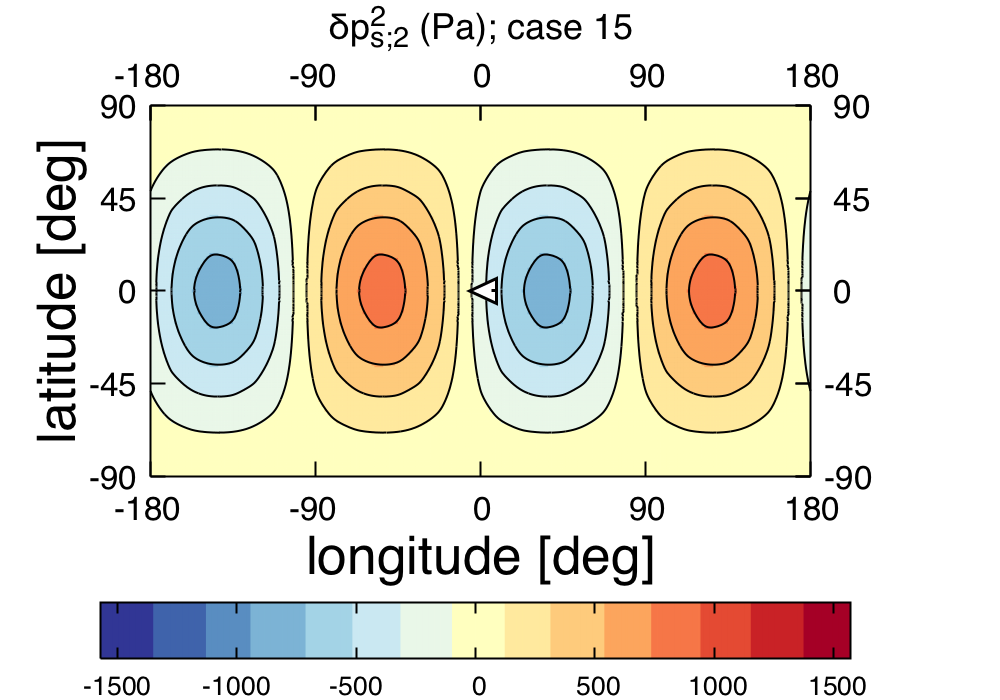} \\
   \raisebox{1.3cm}{\includegraphics[height=0.02\textwidth]{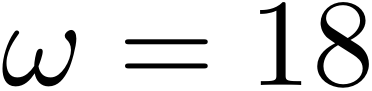}} \hspace{0.5cm}
   \includegraphics[width=0.02\textwidth,trim = 1.3cm 7.3cm 32.2cm 3.0cm,clip]{auclair-desrotour_fig2b.png} 
    \includegraphics[width=0.30\textwidth,trim = 3.2cm 7.3cm 4.5cm 3.2cm,clip]{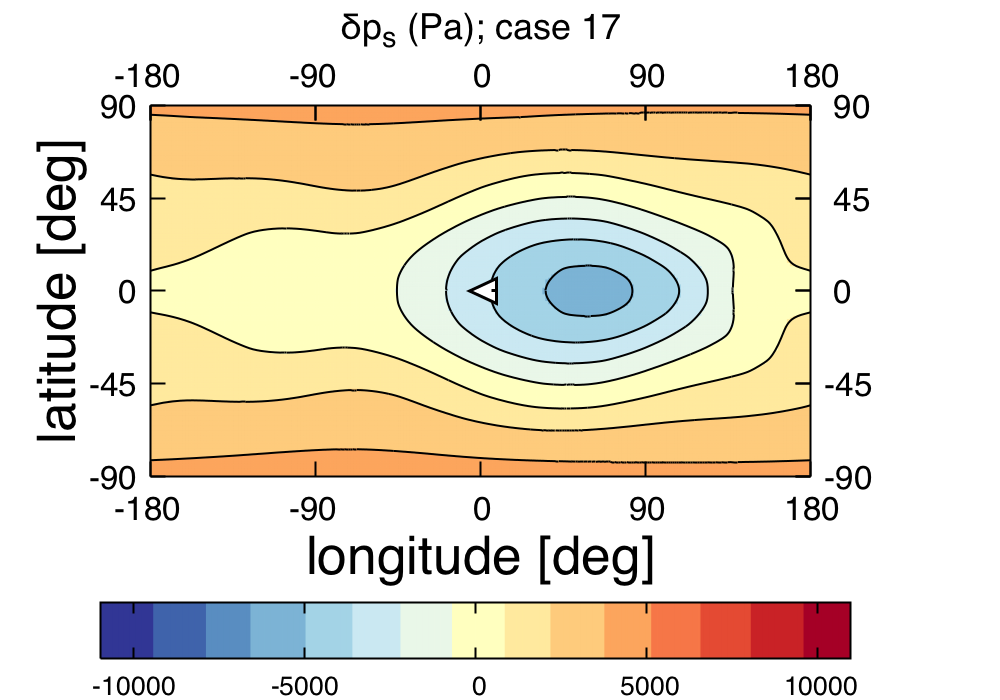} \hspace{0.5cm}
   \includegraphics[width=0.30\textwidth,trim = 3.2cm 7.3cm 4.5cm 3.2cm,clip]{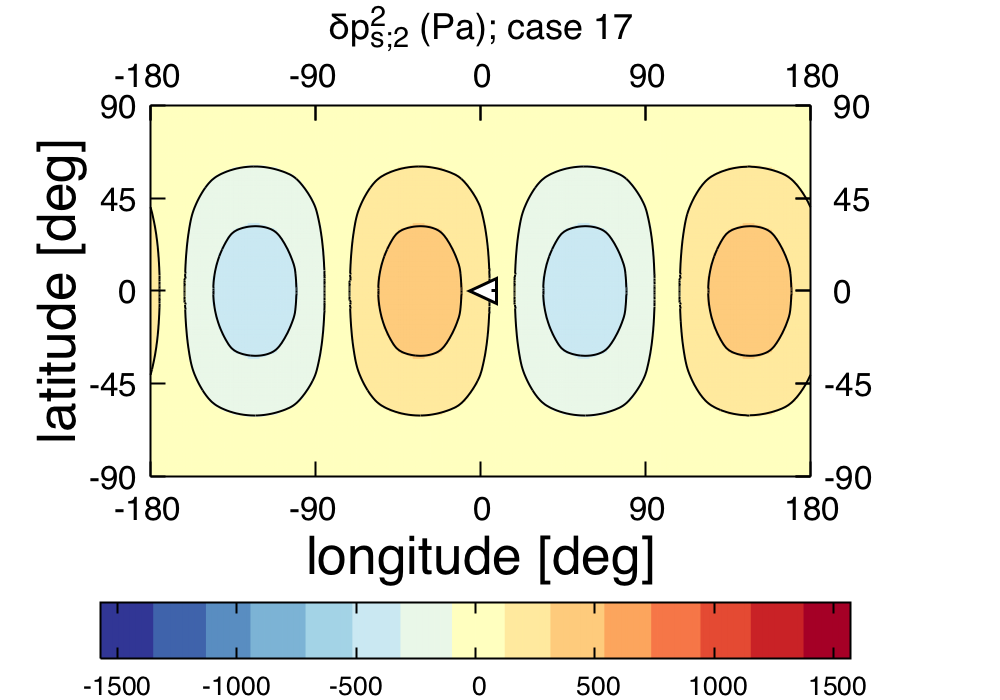} \\
   \raisebox{2.0cm}{\includegraphics[height=0.02\textwidth]{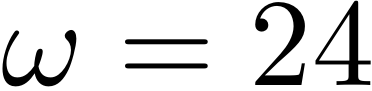}} \hspace{0.5cm}
   \includegraphics[width=0.02\textwidth,trim = 1.3cm 4.0cm 32.2cm 3.0cm,clip]{auclair-desrotour_fig2b.png} 
    \includegraphics[width=0.30\textwidth,trim = 3.2cm 4.0cm 4.5cm 3.2cm,clip]{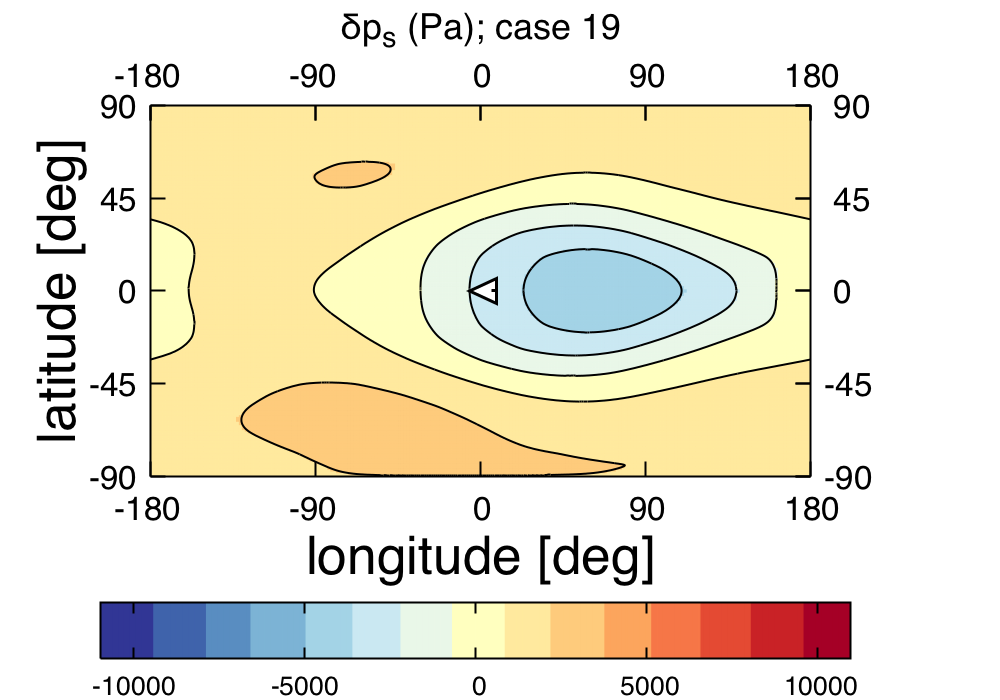} \hspace{0.5cm}
   \includegraphics[width=0.30\textwidth,trim = 3.2cm 4.0cm 4.5cm 3.2cm,clip]{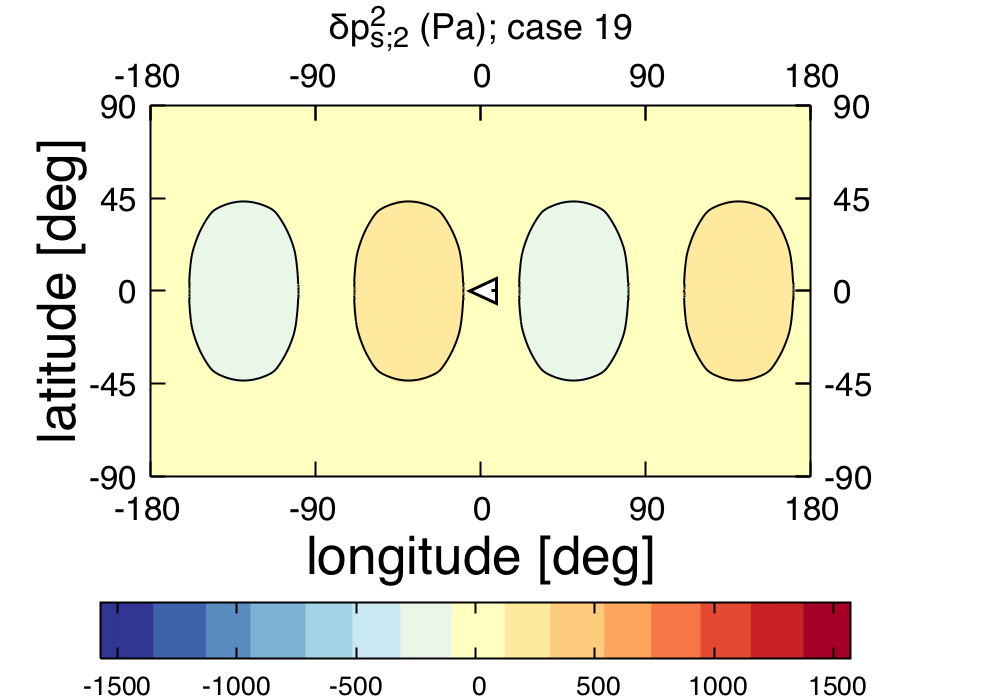} \\
   \hspace{2.7cm}
   \includegraphics[width=0.38\textwidth,trim = 3.0cm 0.0cm 4.5cm 21.0cm,clip]{auclair-desrotour_fig2n.png} \hspace{0.5cm}
   \includegraphics[width=0.38\textwidth,trim = 3.0cm 0.0cm 4.5cm 21.0cm,clip]{auclair-desrotour_fig2o.png} \\
    \hspace{2.7cm}
   \includegraphics[width=0.30\textwidth]{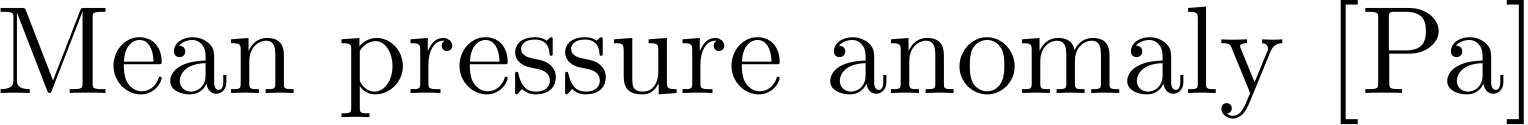} \hspace{2.0cm}
   \includegraphics[width=0.30\textwidth]{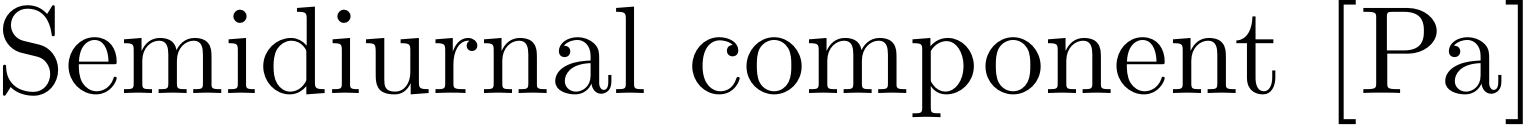}
   \caption{Surface pressure anomaly created by the thermal tide. {\it Left panels:} Daily averaged spatial distribution of the departure of the surface pressure from its mean value created by the thermal tide. {\it Right panels:} Spatial distribution of the semidiurnal component only. The surface pressure anomaly is computed for 300~Solar days and folded over one Solar day centered on the substellar point, whose location and direction of motion are shown with a white arrow. From top to bottom, the normalized forcing frequency $\omeganorm = \left( \spinrate - \norb \right) / \norb$ is increased from 0 (spin-orbit synchronization) to 24 (this corresponds to the length of the Solar day $\Psol = 9.36$\units{days}) for the reference case of the study ($\smaxis = \smvenus$ and $\psurf = 10$\units{bar}).  }
       \label{fig:surface_pressure}%
\end{figure*}

\begin{figure*}[htb]
   \centering
    \includegraphics[width=0.35\textwidth,trim = 1.5cm 2.1cm 7.5cm 2.5cm,clip]{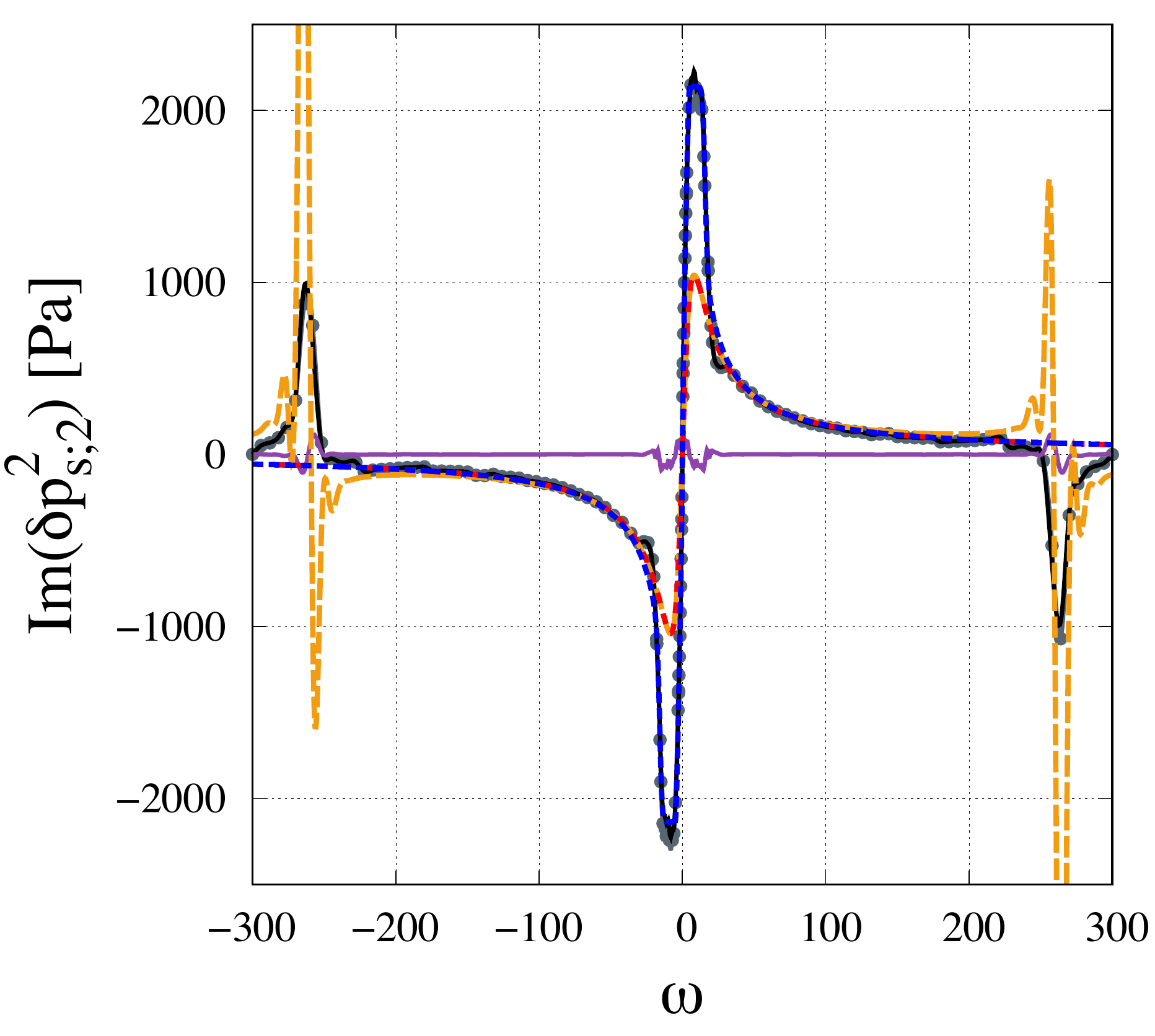} 
     \hspace{0.02\textwidth}
   \includegraphics[width=0.35\textwidth,trim = 1.5cm 2.1cm 7.5cm 2.5cm,clip]{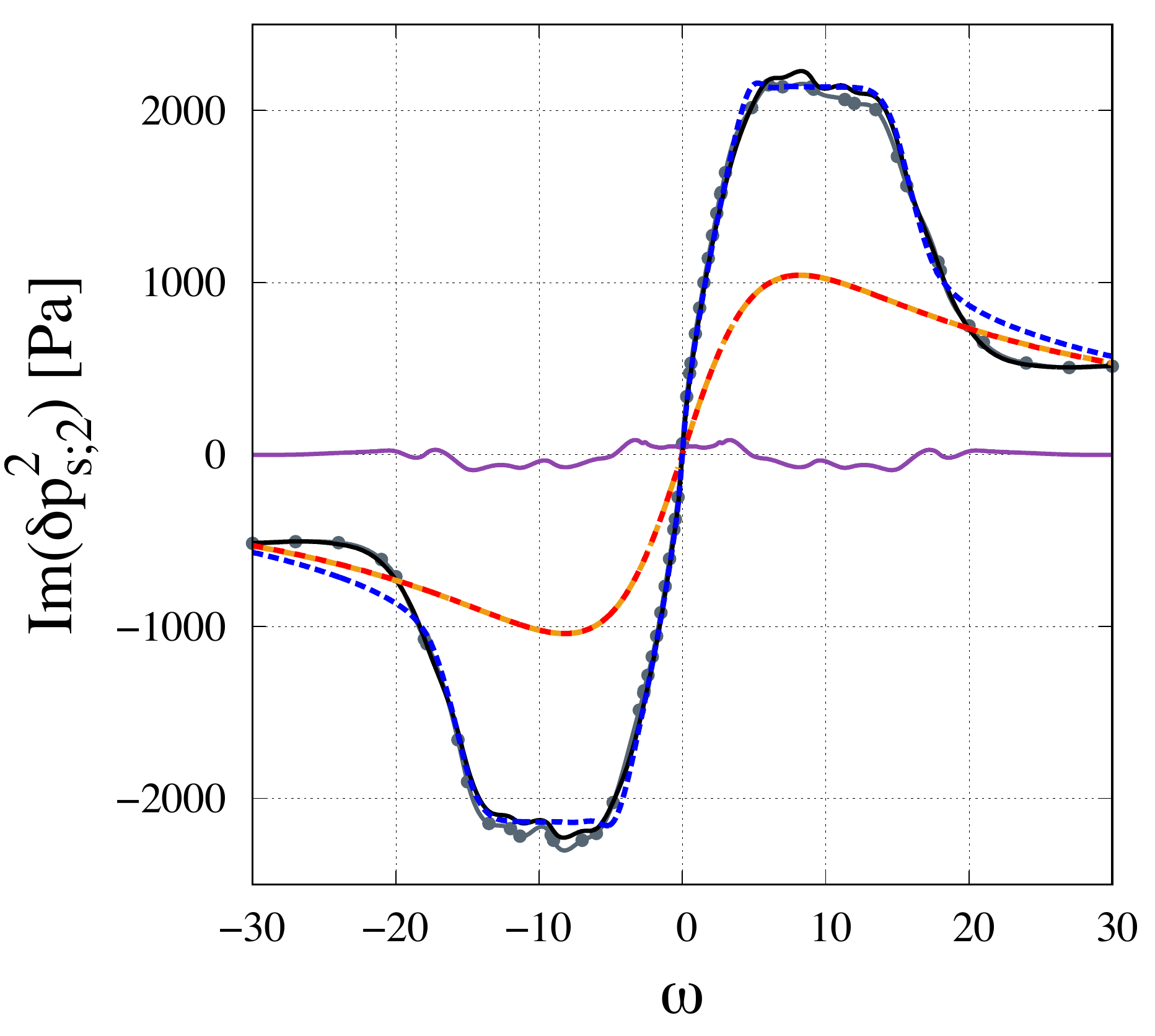} \\
    \includegraphics[width=0.35\textwidth,trim = 1.5cm 2.5cm 7.5cm 2.2cm,clip]{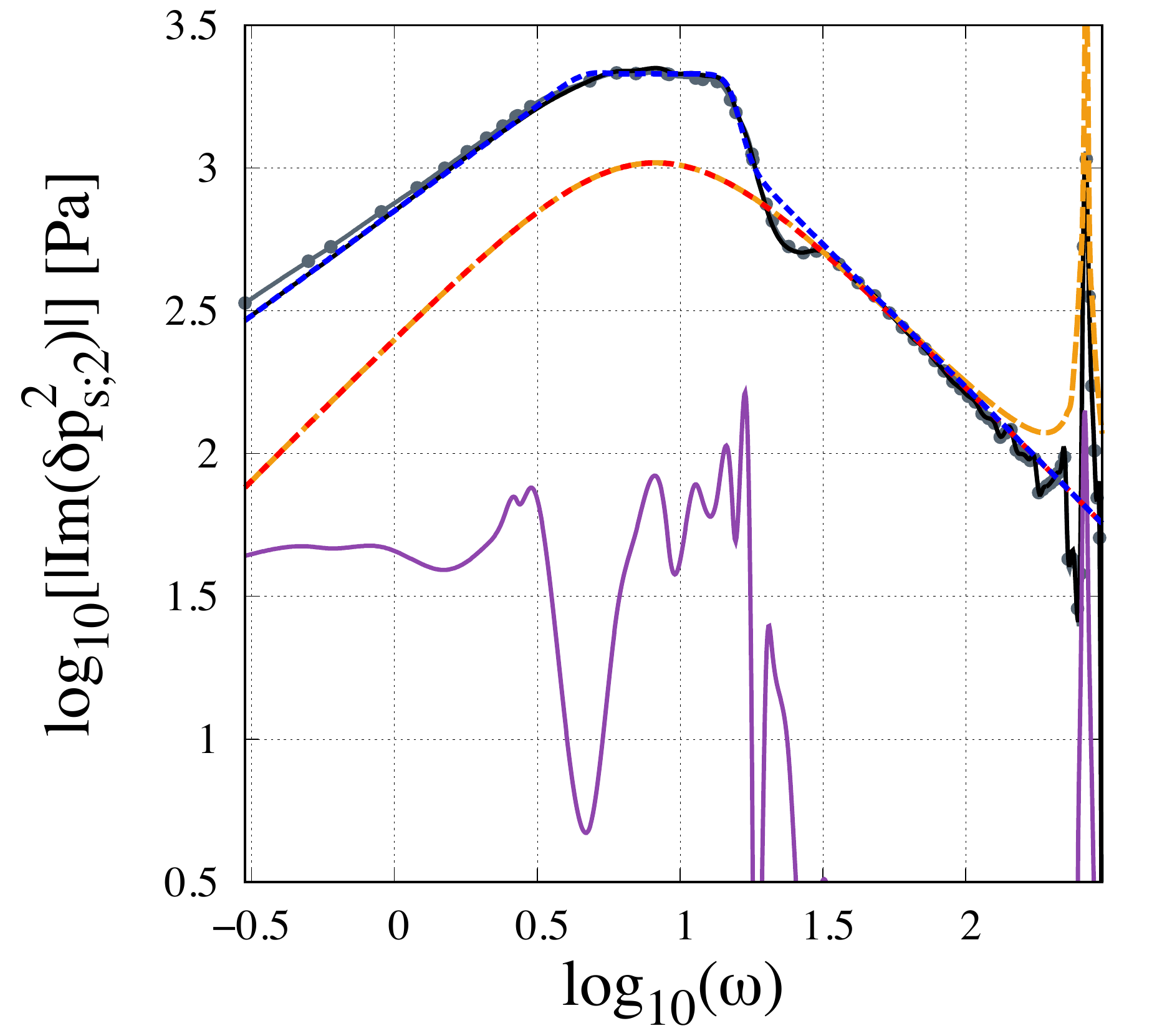} 
    \hspace{0.02\textwidth}
     \hspace{0.05\textwidth}
    \includegraphics[width=0.13\textwidth,trim = 20cm 7cm 1.5cm 2.5cm,clip]{auclair-desrotour_fig3c} 
    \hspace{0.16\textwidth}
   \caption{Imaginary part of the $\Yquad$ component of surface pressure variations as a function of the normalized forcing frequency $\omeganorm = \left( \spinrate - \norb \right)/\norb $ in the reference case ($\smaxis=\smvenus$ and $\psurf = 10$\units{bar}). {\it Top left:} Spectrum over the high-frequency range ($-300 < \omeganorm < 300)$ in linear scales. {\it Top right:} Spectrum over the low-frequency range ($-30<\omeganorm<30$) in linear scales. {\it Bottom left:} Spectrum in logarithmic scales (for $\omeganorm>0$). Numerical results obtained with GCM simulations are plotted (grey points), as well as the interpolating function $\fgcm$ (grey solid line), the odd and even functions of $\ftide$ defined by \eqs{fodd}{feven} (black and purple solid line, respectively), the function derived from the ab initio analytical model and given by \eqs{dps1}{dps2} (orange dashed line), its Maxwell-like approximation given by \eq{maxwell_ana} (red dotted line), and the parametrized function given by \eq{torque_para} (blue dashed line). }
       \label{fig:spectres_model}%
\end{figure*}

\subsection{Extraction of the quadrupolar surface pressure anomaly}
\label{ssec:extraction}



For a given planet, of fixed rotation, semi-major axis and surface pressure, the calculation of the quadrupolar surface pressure anomaly follows several steps.

First, the GCM is run for a period $\Pconv$ corresponding to the convergence timescale necessary to reach a steady cycle. Note that this period has to be specified for each doublet $\left( \smaxis , \psurf \right)$. As a first approximation, it depends on the radiative timescale of the deepest layers of the atmosphere $\taurad$, which scales as

\begin{equation}
\taurad  \scale  \frac{\psurf}{\ggravi} \frac{\Cpgaz}{4 \SBconstant \Teff^3},
\label{taurad}
\end{equation}

\noindent where $\Teff$ stands for the mean effective, or black body, temperature of the atmosphere \citep[see e.g.][Eq.~10]{SG2002}. In the reference case, we observe that the atmospheric state has converged toward a steady cycle after $\sim 5800$ Earth Solar days, and we thus use this value for calculations in this section. 

After this first step, a simulation is run for 300 Solar days of the planet, defined by $\Psol = 2 \pi / \abs{ \spinrate - \norb }$, except in the case of the spin-orbit synchronization ($\spinrate = \norb$), where there is no day-night cycle (in this case, the simulation is simply run for 3000 Earth Solar days). At the end of the simulation, we have at our disposal a time series of snapshots of the surface pressure given as a function of the longitude and latitude. 

The third step consists in post-processing these data. We first remove the constant component, that is the mean surface pressure. Then, we proceed to a change of variable: the time coordinate is replaced by the Solar zenith angle, so that snapshots are all centered on the substellar point. Since meteorological fluctuations can be considered as a perturbation varying randomly over short timescales, we filter them by folding the surface pressure anomaly over one Solar day. 

We finally apply a spherical harmonics transform to the resulting averaged surface pressure snapshot in order to get the complex coefficient $\deltapsquad$ associated with the semidiurnal tidal mode of \rec{degrees} $\llat=2$ and $\mm=2$ (see \eq{deltaps_SH}). The method is illustrated by \fig{fig:surface_pressure} in the reference case ($\smaxis=\smvenus$ and $\psurf = 10$\units{bar}). 

This procedure provides the value of the tidal torque for a given forcing frequency. In practice, the torque is computed over an interval of the normalized frequency $\omeganorm = \left( \spinrate - \norb \right) /\norb$ centered on synchronization ($\omeganorm= 0$) with $\norb$ fixed, the planet rotation rate being deduced from $\omeganorm$ \rec{\citep[the normalized frequency $\omeganorm $ is employed here instead of $\ftide$ to follow along the line by][]{Leconte2015}}. Typically, we use $-30 \leq \omeganorm \leq 30$ to study the low-frequency regime of the atmospheric tidal response and $-300 \leq \omeganorm \leq 300$ to study the high-frequency regime. 

The frequency range is thus divided into $N$ intervals, meaning that the whole above procedure has to be repeated $N+1$ times to construct a frequency-spectrum of the tidal torque. \rec{The size of an interval is defined as $\Delta \omeganorm \define \left( \omeganorm_{\rm sup} - \omeganorm_{\rm inf} \right) / N$. For instance, for the exploration of the parameters space detailed in  \sect{sec:exploration_space}, $N = 20$, $\omeganorm_{\rm inf} = -30$, $\omeganorm_{\rm sup} = 30$, and thus $\Delta \omeganorm = 3$.}

\section{Frequency behaviour of the atmospheric tidal torque}
\label{sec:frequency_behaviour}

The apparent complexity of the physics involved in thermal atmospheric tides requires that we opt for a graduated approach of the problem. Hence, before investigating the dependence of the tidal torque on the planet orbital radius and atmospheric surface pressure as mentioned above, we have to preliminary characterize how it varies with the tidal frequency. To address this question, we consider the reference case ($\psurf= 10$\units{bar} and $\smaxis = \smvenus $).


\subsection{Characterization of the reference case}

In order to characterize the reference case, frequency-spectra of the atmospheric torque created by the semidiurnal thermal tide are computed in low-frequency and high-frequency ranges. For convenience, we introduce the function $\fgcm \left( \ftide \right) $, which is the interpolating function of GCM results with cubic splines. Noting that the tidal torque should be an odd function of the tidal frequency in the absence of rotation (or if the effect of rotation on the tidal response were negligible), we also introduce the function $\fodd$, defined by

\begin{equation}
\fodd \left( \ftide \right) = \frac{1}{2} \left[ \fgcm\left(  \ftide \right) - \fgcm \left( -\ftide \right) \right],
\label{fodd}
\end{equation}

\noindent which is the odd function $\ffunc$ minimizing for any $\sigma$ the distance defined by

\begin{equation}
d \left( \sigma \right) = \left| \left| f_{\rm GCM} \left( \sigma \right) - f \left( \sigma \right) \right|  - \left| f_{\rm GCM} \left( - \sigma \right) - f \left(- \sigma \right) \right|  \right|.
\end{equation}

The complementary function $\feven$, such that $\fgcm = \fodd + \feven$, is defined by

\begin{equation}
\feven = \frac{1}{2} \left[ \fgcm \left( \ftide \right) + \fgcm \left( - \ftide \right) \right],
\label{feven}
\end{equation}

\noindent and provides a measure of the impact of Coriolis effects on the tidal torque. The data, the interpolating function $f_{\rm GCM}$ and its components $\fodd$ and $\feven$ are plotted in \fig{fig:spectres_model} as functions of the normalized tidal frequency $\omeganorm = \ftide / \left( 2 \norb \right) $ in linear and logarithmic scales. Additional functions of the frequency are plotted in dashed lines. They correspond to the ab initio analytical (``Ana.''), Maxwell, and parametrized (``Param.'') models that will be introduced and discussed further. 


We first consider the low-frequency range ($-30 \leq \omeganorm \leq 30$). The reference case of our study exactly reproduces the results plotted in Fig.~1 of \cite{Leconte2015}, with a maximum slightly greater that 2000 Pa located around $\omeganorm \sim 5$. We introduce here the maximal value of the peak $\maxpeak \define \max \left\{ \fodd \left( \ftide \right) \right\} $ and the associated frequency $\fpeak$, such that $\fodd \left( \fpeak \right) = \maxpeak$, timescale $\taupeak \define \fpeak^{-1}$ and normalized frequency $\omegapeak = \fpeak / \left( 2 \norb \right) $.

The tidal torque is negative for $\ftide <0$ and positive otherwise, which corresponds to the typical behaviour of the thermally induced atmospheric tidal response in the the vicinity of synchronization, as discussed in \sect{sec:basic_principle}. As shown by early studies \citep[][]{GS1969,ID1978,DI1980,CL2001}, thermal atmospheric tides thus tends to drive the planet away from synchronous rotation and determines its non-synchronized rotation states of equilibrium. 

In the zero-frequency limit, the torque scales as $\torque \scale \ftide^{\alpha} $ , with $\alpha \approx 0.73$. In the high-frequency range ($20 \lesssim \abs{\omeganorm} \leq 300$), it scales as $ \torque \scale  \ftide^{-1}$ with a remarkable regularity (see \fig{fig:spectres_model}, bottom left panel) and exhibits a resonance at $\omeganorm \approx 260$. We will see in the next section that these features can be explained using the linear theory of atmospheric tides \citep[][]{Wilkes1949,Siebert1961,LC1969}.

We note that the spectrum of $\fgcm$ exhibits a slight systematic asymmetry with respect to the synchronization. This feature is obvious in the low-frequency range, where $\abs{ \fgcm\left( -\ftide \right)} > \abs{  \fgcm \left( \ftide \right) }$, and tends to vanish while $\abs{\ftide} $ increases. Particularly, a small departure between $\fgcm$ and $\fodd$ can be observed around the extrema of the tidal torque, and we note that the atmosphere undergoes a non-negligible tidal torque at synchronization ($\ftide = 0$), although the perturber does not move in the reference frame co-rotating with the planet. 

This asymmetry is an effect of the Coriolis acceleration, which comes from the fact that $\abs{\spinrate \left( -\ftide \right)} \neq \abs{\spinrate \left( \ftide \right)}$ \rec{(in the low-frequency range, the spin rotation rate is not proportional to the tidal frequency)}. The Coriolis acceleration affects the atmospheric general circulation by generating strong zonal jets through the mechanism of non-linear Rossby waves pumping angular momentum equatorward \citep[e.g.][]{SP2011}. These jets induce a Doppler-like angular lag of the tidal bulge with respect to the direction of the perturber.

\subsection{Ab initio analytical model}
\label{ssec:analytical_model}
\label{ssec:resonance_lamb}

 The behaviour of the torque in the high-frequency range can be explained with the help of the linear theory of thermal atmospheric tides \citep[][]{Wilkes1949,Siebert1961,LC1969}. In \append{app:analytical_model}, by using an ab initio approach, we compute analytically the atmospheric tidal torque created by the semidiurnal thermal tide in the idealized case of an isothermal atmosphere undergoing the tidal heating of the planet surface. The atmospheric structure is here characterized by the constant pressure height 

\begin{equation}
\Hatm = \frac{\Rspec \Tsurf}{\ggravi},
\end{equation}

\noindent \rec{where $\Rspec$ and $\Tsurf$ designate the specific gas constant and the surface temperature, respectively.} It allows to renormalizes the altitude $\zz$ with the introduction of the pressure height scale $\xx = \zz / \Hatm$. In the analytic model, we choose for the heat per unit mass inducing the tidal response  the vertical profile $\Jtide = \Jsurf \expo{- \tauJ \xx} $, where $\Jsurf$ is the heat per unit mass at the planet surface, and $\tauJ$ a dimensionless optical depth corresponding to the inverse of the characteristic thickness of the heated layer. Note that the limit $\tauJ \rightarrow + \infty$ corresponds to the case studied by \cite{DI1980}, where the vertical profile of heat is approximated by a Dirac distribution. 

The surface pressure anomaly is obtained by solving the vertical structure equation of the dominating mode with the above profile of the forcing. We refer the reader to the appendix for the detail of approximations and calculations made to get this result. Particularly, note that dissipative processes are ignored since they are associated to timescales that are supposed to far exceed typical tidal periods in the high-frequency range. 

The solution takes two different forms depending on the way $\ftide$ compares to the frequency characterizing the turning point, where the vertical wavenumber annihilates (see \append{app:analytical_model}),

\begin{equation}
\fturning = \sqrt{\frac{4 \kad \Hatm \Lambdao \ggravi}{\Rpla^2}}.
\label{fturning}
\end{equation}

\noindent The notation $\Lambdao$ designates here the eigenvalue of the predominating mode in the expansion of perturbed quantities on the basis of Hough functions (see \eq{hough_expansion}). This mode is the gravity mode of latitudinal wavenumber $\nn = 0$ in the indexing notation used by \cite{LS1997}. Its eigenvalue $\Lambdao$ can be approximated as a constant provided that $\norb \ll \abs{\spinrate}$. 

Hence, introducing the equivalent depth of the mode,

\begin{equation}
\heq \define \frac{\Rpla^2 \ftide^2}{\Lambdao \ggravi},
\label{heq}
\end{equation}

\noindent we obtain, for $\abs{\ftide} \leq \fturning$,

\begin{equation}
\imag{\deltapsurf} = \ftide^{-1} \psurf \frac{\kad \Jsurf}{g \Hatm} \frac{\frac{\Hatm}{\heq} \left( \tauJ + \frac{1}{2} + \kad \right) - \frac{1}{2} \left( \tauJ + 1 \right) }{\left[ \tauJ \left( \tauJ + 1 \right) + \frac{\kad \Hatm}{\heq}  \right] \left( \frac{\Hatm}{\heq} - \frac{1}{\Gamma_1} \right) } ,
\label{dps1}
\end{equation}

\noindent and, for $\abs{\ftide} > \fturning$, 

\begin{equation}
\imag{\deltapsurf} =  \frac{ \ftide^{-1} \psurf \frac{\kad \Jsurf}{\ggravi \heq} \left[  \tauJ + \frac{1}{2} \left( 1 - \sqrt{1 - \frac{4 \kad \Hatm}{\heq}} \right) \right]}{ \left[ \tauJ \left( \tauJ + 1 \right) + \frac{\kad \Hatm}{\heq}  \right] \left[ \frac{\Hatm}{\heq} - \frac{1}{2} \left( 1 + \sqrt{1 - \frac{4 \kad \Hatm}{\heq}} \right) \right] }. 
\label{dps2}
\end{equation}

\noindent We recall that $\kad = \Rspec / \Cpgaz$, \rec{where $\Cpgaz$ designates the heat capacity per unit mass,} and $\gad = 1 / \left( 1 - \kad \right)$ the adiabatic exponent at constant entropy \citep[][]{GZ2008}. The solution given by \eqs{dps1}{dps2} provides a useful diagnosis about the frequency-behaviour of the torque in the high-frequency range. 

The most striking feature of this behaviour is the peak that can be observed in \fig{fig:spectres_model} (top and bottom left panels) at the normalized frequency $\omeganorm  \approx 260$. This peak correspond to the fundamental resonance of the atmospheric vertical structure associated with the propagation of the Lamb mode \citep[e.g.][]{Lindzen1968,Bretherton1969,LB1972,Platzman1988,Unno1989}, which is an acoustic type wave of long horizontal wavelengh. In an inviscid, isothermal atmosphere, the Lamb mode is characterized by the equivalent depth $h_{\rm L} = \gad \Hatm $ \citep[][]{LB1972}. In the asymptotic regime, where $\norb \ll \abs{\spinrate}$, the characteristic Lamb frequency follows from \eq{heq}, 

\begin{equation}
\fLamb = \sqrt{\frac{\Lambdao \ggravi \hLamb}{\Rpla^2}} = \sqrt{\frac{\gad}{4 \kad}} \fturning. 
\label{fLamb}
\end{equation}

\noindent By noticing that $\fLamb > \fturning$ in the case of a diatomic gas ($\gad = 1.4$) and substituting $\heq$ by $\hLamb$ into the corresponding expression of the solution -- that is \eq{dps2} -- we can easily observe that the tidal torque is singular at $\abs{\ftide} = \fLamb$. The resonance hence occurs when the phase velocity of the forced mode equalizes the characteristic Lamb velocity $\vLamb = \sqrt{\ggravi \hLamb}$.

With the numerical values given by Table~\ref{parameters} and the mean surface temperature computed from GCM simulations ($\Tsurf \approx 316$\units{K}), the isothermal approximation leads to $\Hatm \approx 10.6$\units{km} and $\hLamb \approx 15$~km for the reference case. Besides, $\Lambda_0 \approx 11.1 $ in the adiabatic asymptotic regime of high rotation rates. It thus follows that $\omegaLamb \approx 308$, and we recover the order of magnitude of the frequency identified in Fig.~\ref{fig:spectres_model} (top left panel) using GCM simulations (i.e. $\omegaLamb \approx 260$).

The observed departure between values of $\omegaLamb$ obtained in analytical and numerical approaches can be explained by the dependence of the resonance on the atmospheric vertical structure \citep[see e.g.][]{Bretherton1969,LB1972}. The analytical value corresponds to the case of an isothermal atmosphere of temperature $\Tsurf$. In reality, the mean temperature vertical profile is characterized by a strong gradient in the troposphere, the temperature decaying linearly from $\sim 316$\units{K} at $\zz = 0$ to $\sim 160$\units{K} at $\zz \approx 25$\units{km} in GCM simulations. As a consequence, the mean pressure height scale of the tidally heated layer is less than the surface pressure heigh scale, which leads to smaller equivalent depth and resonance frequency for the Lamb mode.

The other interesting feature highlighted by \fig{fig:spectres_model} is the scaling law of the torque $\torque \scale \ftide^{-1}$ in the range of intermediate frequencies, that is between the thermal and Lamb resonances, typically. This behaviour is described by the analytical model. As discussed before (see \eq{fLamb}), $\fturning$ and $\fLamb$ are close to each other. The intermediate-frequency range thus corresponds to the case $\abs{\ftide} < \fturning$, which leads us to consider the solution given by \eq{dps1}. We place ourselves in the configuration where characteristic timescales are clearly separated, that is $\abs{\ftide} \ll \fturning$ and $\norb \ll \abs{\spinrate}$ in the meantime. As $\Hatm / \heq \scale \sigma^{-2}$, the preceding condition implies that $\Hatm / \heq \gg 1$. It follows that

\begin{equation}
\frac{\frac{\Hatm}{\heq} \left( \tauJ + \frac{1}{2} + \kad \right) - \frac{1}{2} \left( \tauJ + 1 \right) }{ \frac{\Hatm}{\heq} - \frac{1}{\Gamma_1} } \sim \tauJ + \frac{1}{2} + \kad. 
\end{equation}

By invoking the strong optical thickness of the atmosphere in the infrared ($\tauJ \gg 1$), we remark that we recover analytically the scaling law $\torque \scale \ftide^{-1}$ observed in \fig{fig:spectres_model} from the moment that the condition $1 \ll \Hatm / \heq \ll \kad^{-1} \tauJ^2$ is satisfied. This provides a definition for the intermediate frequency-range, which is now the range corresponding to $\fthermal \ll \abs{\ftide} \ll \fLamb $, where we have introduced the thermal frequency 

\begin{equation}
\fthermal = \frac{\fturning}{2 \sqrt{  \tauJ \left( \tauJ + 1 \right)}} \ll \fturning. 
\end{equation}

\noindent Basically, $\fthermal$ is the frequency for which the vertical wavelength of the mode and the characteristic depth of the heated layer are of the same order of magnitude. 

From the moment that $\abs{\ftide} \ll \fLamb$ (or $\Hatm / \heq \gg 1$), \eq{sol1} can be approximated by the function

\begin{equation}
\imag{\deltapsurf}  \approx \frac{2 \maxthermal \tauthermal \ftide }{1 + \left( \tauthermal \ftide \right)^2},
\label{maxwell_ana}
\end{equation}

\noindent where the associated characteristic timescale $\tauthermal$ and maximal amplitude of the pressure anomaly $\maxthermal$ are

\beqtwo{\tauthermal = \fthermal^{-1}}{\maxthermal =  \frac{\kad \Jsurf \left( \tauJ + \frac{1}{2} + \kad \right)}{ \ggravi \Hatm \fturning \sqrt{\tauJ \left( \tauJ + 1 \right)} }\psurf .}

\noindent We recognize in the form of the function given by \eq{maxwell_ana} the well-known Maxwell model, which is commonly used to describe the dependence of the tidally dissipated energy on the forcing frequency in the case of solid bodies \citep[e.g.][]{Efroimsky2012,Correia2014}. Its use in the case of thermal atmospheric tides is discussed in the next section.

\subsection{Discussion on the Maxwell model}

Analytic ab initio approaches based on a linear analysis of the atmospheric tidal response -- including this work (cf. previous section) -- predict that the imaginary part of surface pressure variations can be expressed as a function of the forcing frequency $\ftide = 2 \left( \spinrate - \norb \right) $ as \citep[e.g.][]{ID1978,ADLM2017a}

\begin{equation}
 \Im \left\{ \deltapsquad \right\}_{\imaxwell}= \frac{2 \maxM \tauM \ftide}{1 + \left( \tauM \ftide \right)^2}, 
\label{maxwell}
\end{equation}

\noindent the notations $ \tauM$ and $\maxM$ referring to an effective thermal time constant and the amplitude of the maximum (located at $ \ftide = \tauM^{-1}$), respectively (the factor 2 sets the maximal amplitude to $\maxM$). This functional form corresponds to the so-called Maxwell model mentioned above. It describes the behaviour of an idealized forced oscillator composed of a string and a damper arranged in series \citep[][]{Greenberg2009,Efroimsky2012,Correia2014}. 

Note that other works based upon different approaches converged toward the functional form of the Maxwell model. For instance, \cite{CL2001} used the parametrized function $ f \left( \ftide \right) = \ftide^{-1} \left(1 - \expo{-\gamma \ftide^2} \right)  $ ($\gamma$ being a real parameter, see Eq.~(26) of the article) to mimic the behaviour of the atmospheric tidal torque, while \cite{Leconte2015} retrieved Eq.~(\ref{maxwell}) empirically by analyzing results obtained from simulations run with the LMDZ GCM.

\comments{Remarque importante sur notre modèle et l'approche de Dobrovolskis et Ingersoll.}

An important remark should be made here concerning the behaviour of the tidal torque in the vicinity of the synchronization (i.e. for $\ftide \approx 0$). To our knowledge, most of early works using the classical tidal theory to study the spin rotation of Venus and ignoring dissipative processes obtained a torque scaling as $\torque \scale \ftide^{-1}$, and thus singular at the synchronization \citep[e.g.][]{DI1980,CL2001,CL2003}. This is precisely the reason that led \cite{CL2001} to introduce the regular ad hoc parametrized function mentioned above. Conversely, \cite{ID1978} and, later, \cite{ADLM2017a}, derived a Maxwell-like tidal torque analytically by introducing a characteristic thermal time associated with boundary layer processes and radiative cooling. These early results may let think that dissipative processes are a necessary ingredient for a regular tidal torque to exist at the synchronization. 

Although dissipative processes definitely regularize the atmospheric tidal torque at the synchronization \citep[e.g.][]{ADLM2017a}, we showed in \sect{ssec:analytical_model} that regularity also naturally emerges from approaches ignoring them when the vertical structure equation is solved in a self-consistent way. For a sufficiently small frequency, namely $\abs{\ftide} \ll \fthermal$, the torque derived from our analytic solution in the absence of dissipative mechanisms scales as $\torque \scale \ftide$. Therefore, it seems that the singularity at $\ftide = 0$ obtained by early works could result from oversimplifying hypotheses, such as neglecting the three-dimensional aspect of the tidal response or tidal winds. For instance, note that our analytical model asymptotically converges towards the function obtained by \cite{DI1980} when the vertical profile of tidal heating tends towards the Dirac distribution used by these authors (i.e. when $\tauJ \rightarrow + \infty$).  

The above statement means that the analytical solutions given by \eqs{dps1}{dps2} can be used in practice over the whole range of tidal frequencies without leading to unrealistic behaviors at the vicinity of synchronization, notwithstanding the fact that they were derived assuming that characteristic timescales associated with dissipative processes far exceed the tidal period.

In studies taking into account dissipative processes \citep[e.g.][]{ID1978,ADLM2017a}, the parameter $\tauM$ of \eq{maxwell} can be interpreted as an effective timescale associated with the radiative cooling of the atmosphere in the Newtonian cooling approximation, where radiative losses are assumed to be proportional to temperature variations \citep[][]{LM1967,ADLM2017a,ADL2018a}. These early analytical works established the following expression of the tidal torque \citep[see e.g.][Eq.~2]{ID1978},

\begin{equation}
\torque = \frac{3 \pi \norb^2 \Rpla^4 \varepsilon \fluxstar}{8 \Cpgaz \Tsurf} \frac{\ftide}{\ftide^2 + \tauM^{-2}},
\end{equation}

\noindent where $\effheat$ stands for the effective fraction of the incoming flux absorbed by the atmosphere. Substituting $\Im \left\{ \deltapsquad \right\}$ by \eq{maxwell} in \eq{torque} and comparing the obtained result with the preceding expression leads to a relationship between the Maxwell thermal time and maximum, which is

\begin{equation}
\frac{\maxM}{\tauM} = \frac{3}{128} \sqrt{\frac{5}{6 \pi}} \frac{\Ggrav \Mpla \effheat \Lstar}{\Cpgaz \Tsurf \Rpla^2 \smaxis^2},
\label{q0tau0}
\end{equation}

\noindent the notation $\Ggrav$ referring to the gravitational constant. 

Assuming that the atmosphere is optically thin in the visible frequency range and that the surface temperature corresponds to a black body equilibrium, we write the mean surface temperature as 

\begin{equation}
\Tsurf = \left[  \frac{\left( 1 - \Asurf \right) \Lstar }{16 \pi \SBconstant \smaxis^2} \right]^{\frac{1}{4}},
\label{Ts_BB}
\end{equation}

\noindent where we have introduced the Stefan-Boltzmann constant $\SBconstant$ and the surface albedo $\Asurf$. By substituting $\Tsurf$ by \eq{Ts_BB} in \eq{q0tau0}, we obtain that the ratio $\maxM / \tauM$ does not depend on the surface pressure and scales as

\begin{equation}
\frac{\maxM}{\tauM}  \approx  \frac{3}{128} \sqrt{\frac{5}{6 \pi}} \frac{\Ggrav \Mpla \effheat \Lstar}{\Cpgaz  \Rpla^2 } \left[  \frac{\left( 1 - \Asurf \right) \Lstar }{16 \pi \SBconstant } \right]^{-\frac{1}{4}}  a^{-3/2}. 
\label{SLana}
\end{equation} 

\noindent with $\effheat = 1 - \Asurf$ if the atmosphere is optically thick in the infrared. This relationship between $\tauM$ and $\maxM$ means that the two parameters of the Maxwell model (\eq{maxwell}) can theoretically be reduced to the effective thermal timescale only, which is determined by complex boundary layer and dissipative processes in the general case. The scaling law given by \eq{SLana} will be tested using GCM simulations in \sect{sec:exploration_space}.

We now compare the Maxwell model to numerical results by assimilating the Maxwell amplitude and timescales to the maximum value of $\fodd$ and its associated timescale, respectively. The ab initio analytical solution given by \eqs{dps1}{dps2} (``Ana.'') and its Maxwell-like form, derived for $\abs{\ftide} \ll \fLamb$ and given by \eq{maxwell_ana} (``Maxwell''), are both plotted in \fig{fig:spectres_model} as functions of the normalized forcing frequency ($\omeganorm$). The numerical values of $\fLamb$ and $\fturning$ used for the plot are determined by the eigenfrequency of the resonance associated with the Lamb mode in GCM simulations, that is $\omegaLamb \approx 260$. We arbitrarily choose to set $\fthermal = \fpeak$ (correspondence between the numerically-derived and the Maxwell maxima), which determines the value of $\tauJ$ (i.e. $\tauJ \approx 14$). Finally, the maximum $\maxthermal$ is obtained by fitting the slope in the intermediate frequency-range to numerical results ($\maxthermal \approx 1042$\units{Pa}), and provides the value of the parameter $\Jsurf$ (i.e. $\Jsurf \approx 0.05$\units{W.kg^{-1}}). 

Figure~\ref{fig:spectres_model} highlights the fact that the Maxwell model does not allow to recover the behaviour of the torque in the low-frequency regime. The functional form given by numerical results and the Maxwell function clearly differ in this regime. Particularly, the maximal amplitude obtained from GCM simulations is about twice larger than that given by the model. Note that a smaller departure between the Maxwell and numerical maxima would certainly be obtained by fitting the Maxwell function to the whole spectrum of numerical results, and not only to the peak. However, this would also lead to overestimate the Maxwell timescale, and the fit would not be satisfactory either. A a consequence, a novel parametrized model has to be introduced to better describe the behaviour of the tidal torque in the low-frequency range. This is the purpose of the next section.

\subsection{Introduction of a new parametrized model}

\comments{Sous-section dans laquelle on introduit le modèle à 7 paramètres avec les coefficients numériques obtenus pour le cas de référence. Il faut souligner le fait que ce modèle rend compte de la décroissance abrupte du couple de marée au niveau de la résonance. }

It has been shown that the ab initio analytic model described in \sect{ssec:analytical_model} and \append{app:analytical_model} reproduces the main features of the tidal torque in the high-frequency range, namely the resonance associated with the Lamb mode and the asymptotic scaling law $\torque \scale \ftide^{-1}$. However, in the low-frequency range, the behaviour of the torque appears to be a little bit more complex than that predicted by the model, which reduces to a simple Maxwell function. This is not surprising since the atmospheric tidal response at low tidal frequencies involves complex non-linear mechanisms, interactions with mean flows, and dissipative processes, which are clearly outside of the scope of the classical tidal theory used to establish the solution given by \eqs{dps1}{dps2}. 

Yet, the frequency dependence of the tidal torque has to be characterized in the vicinity of synchronization as this is where its action of the planetary rotation is the strongest. Our effort has thus to be concentrated on the low-frequency regime and the transition with the high-frequency regime. As they treat the full non-linear 3D dynamics of the atmosphere in a self-consistent way, GCM simulations are particularly useful in this prospect. 

To make oneself an intuition of the behaviour of the torque, it is instructive to look at the logarithmic plot of \fig{fig:spectres_model} (bottom left panel), which enables us to identify the different regimes at first glance. We basically observe two tendencies, highlighted in the plot by slopes taking the form of a straight line, in the zero-frequency limit ($\log \left( \omeganorm \right) \lesssim 0.5$) and the high-frequency asymptotic regime ($\log \left( \omeganorm \right) \gtrsim 1.5$). In the interval $0.5 \lesssim \log \left( \omeganorm \right) \lesssim 1.5$, the tidal torque reaches a maximum and undergoes an abrupt decay. 

Considering these observations, it seems relevant to approximate the logarithm of the torque by linear functions corresponding to the low and high-frequency regimes, and multiplied by sigmoid activation functions. By introducing the notation $\xmod \define \log \omega$, we thus define the parametrized function as

\begin{align}
\label{fonction_para}
\Fpara \left( \xmod \right) \define & \left( \alf \xmod + \blf \right) \Flf \left( \xmod \right) + \left( \ahf \xmod + \bhf \right) \Fhf \left( \xmod \right) \\
  & + \btrans \left[ 1 - \Flf \left( \xmod \right) - \Fhf \left( \xmod \right)  \right],  \nonumber
\end{align}



\noindent where $\btrans \approx \log \left( \maxpeak \right) $ is the level of the transition plateau,  $\alf$, $\blf$, $\ahf$ and $\bhf$ the dimensionless coefficients of linear functions describing asymptotic regimes, and $\Flf$ and $\Fhf$ two sigmoid activation functions expressed as 

\beqtwo{\Flf \left( \xmod \right) \define \frac{1}{1 + \expo{\left( \xmod - \xlf \right)/\dlf}} }{\Fhf \left( \xmod \right) \define \frac{1}{1 + \expo{- \left( \xmod - \xhf \right) / \dhf}}.}

\noindent In these expressions, the dimensionless parameters $\xlf$ and $\xhf$ designate the cutoff frequencies of $\Flf$ and $\Fhf$ in logarithmic scale, and $\dlf$ and $\dhf$ the widths of transition intervals. The corresponding tidal torque is given by 

\begin{equation}
\torquepara =  10^{\Fpara \left( \log \abs{\omeganorm} \right) } \sign \left( \omeganorm \right). 
\label{torque_para}
\end{equation}

As the scaling law $\torque \scale \ftide^{-1}$ was derived from the ab initio model of \sect{ssec:analytical_model} in the high-frequency range, we enforce it by setting $\ahf = -1$. The eight left parameters are then obtained by fitting the function given by \eq{fonction_para} to numerical results (as done previously, the odd function $\fodd$ is used). We thus end up with

\begin{equation}
\begin{array}{llll}
\alf = 0.734, & \blf = 2.85, & \dlf = 0.0100 , & \xlf = 0.637 ,   \\ 
\ahf = -1 , & \bhf = 4.23 , & \dhf = 0.0232, & \xhf = 1.20, \\
 \btrans = 3.33, & & &
\end{array}
\end{equation}



\noindent and plot the model function $\Fpara$ in \fig{fig:spectres_model} using these numerical values (``Param.''). 

As shown by \fig{fig:spectres_model}, the parametrized function defined by \eq{fonction_para} describes important features that escaped the Maxwell function, such as the fact that the tidal torque does not scale linearly with the forcing frequency in the zero-frequency limit, and the rapid decay characterizing the transition between low and high-frequency regimes.

\subsection{Dependence of the tidal torque on the atmospheric composition}

\begin{figure}[htb]
   \centering
   \includegraphics[width=0.48\textwidth,trim = 1.5cm 2.2cm 1.5cm 2.5cm,clip]{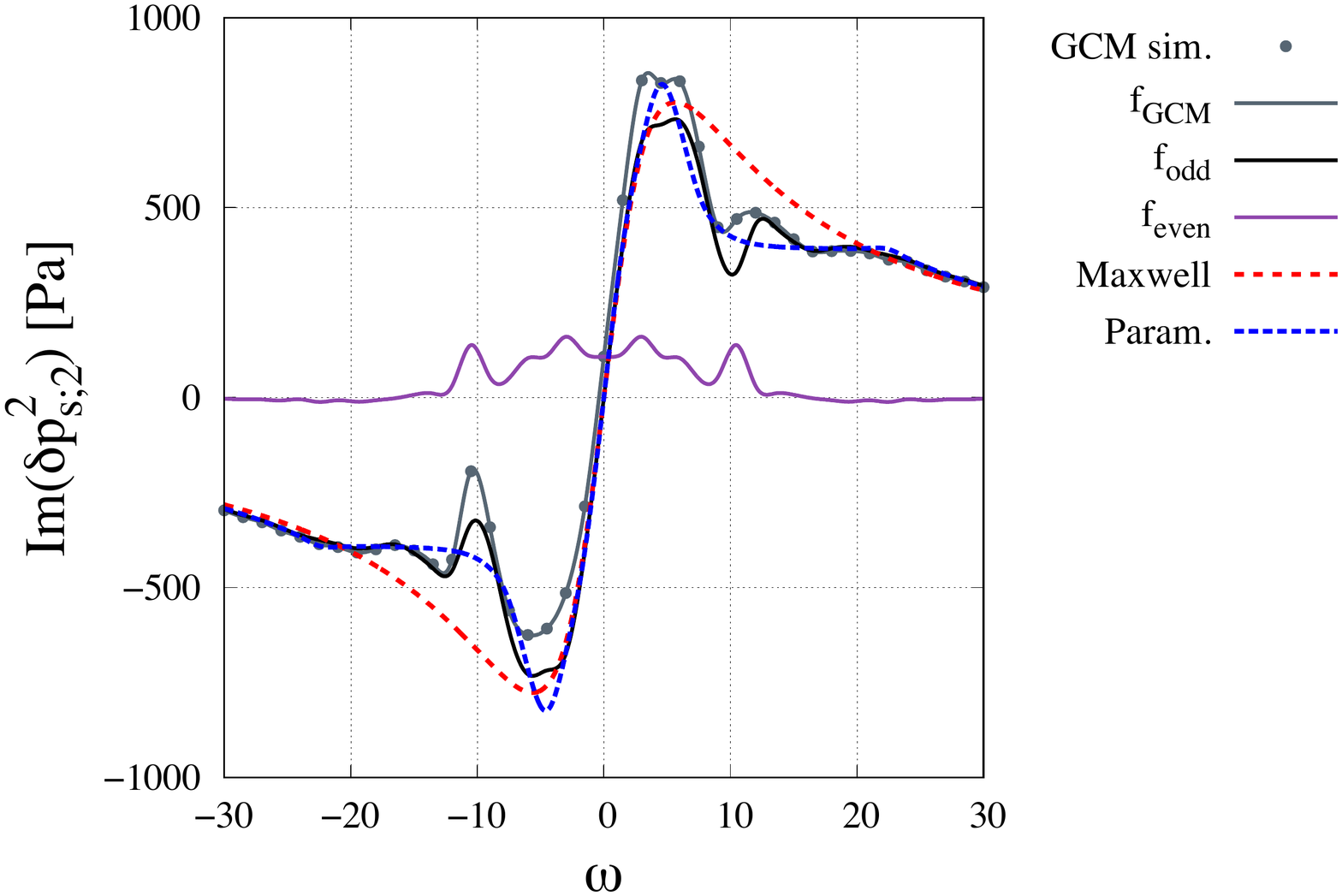}
   \caption{Imaginary part of the $\Yquad$ component of surface pressure variations as a function of the normalized forcing frequency $\omeganorm = \left( \spinrate - \norb \right) / \norb$ in the reference case ($\smaxis = \smvenus$ and $\psurf = 10$\units{bar}) with a $\carbondiox$-dominated atmosphere. Numerical results obtained with GCM simulations are plotted (grey points), as well as the interpolating function $\fgcm$ (dashed grey line), the odd and even functions of $\ftide$ defined by \eqs{fodd}{feven} (solid black and purple lines, respectively), the Maxwell function given by \eq{maxwell_ana} (red dotted line), and the parametrized function given by \eq{fonction_para} (blue dashed line) with the parameters: $\alf = 0.549 $, $\blf = 2.52$, $\ahf = - 1$, $\bhf = 3.94$, $\xlf = 0.798$, $\dlf = 0.072$, $\xhf = 1.35$, $\dhf = 0.0098$, and $\btrans = 2.59$.}
       \label{fig:spectres_CO2}%
\end{figure}

As it clearly has a strong impact, the dependence of the tidal torque on the atmospheric composition has to be discussed. In \append{app:dependence_composition}, we treat the case of a $\carbondiox$-dominated atmosphere with a mixture of water and sulfuric acid ($\sulfuricacid$) comparable to that hosted by the Venus planet. The obtained spectrum and the associated functions introduced above are plotted in \fig{fig:spectres_CO2}, and shall be compared to those computed for the $\nitrogen$-dominated atmosphere, plotted in \fig{fig:spectres_model} (right panel). Several interesting features may be noted. 

First, the tidal torque exerted on the $\carbondiox$-dominated atmosphere is more than twice weaker than that exerted on the $\nitrogen$-dominated atmosphere. Particularly, peaks are strongly attenuated. This results from the vertical distribution of tidal heating. Because of the optical thickness of carbon dioxide in the visible frequency range, an important part of the incoming stellar flux is absorbed above clouds. This is not the case of the $\nitrogen$-dominated atmosphere, where most of the flux reaches the planet surface and and is re-emited in the infrared frequency range, leading to the thermal forcing of dense atmospheric layers located at high pressure levels.

Second, we observe a greater asymmetry between the negative and positive frequency ranges, the function $\feven$ being not negligible with respect to $\fodd$. This is also an effect of the vertical distribution of tidal heating. In the case of the $\nitrogen$-dominated atmosphere, most of the tidal torque is generated by density variations occurring at low altitudes, where the fluid is well coupled to the solid part of the planet by frictional forces. Switching from $\nitrogen$ to $\carbondiox$ decreases the contribution of these layers, while it increases the contribution of layers located at pressure levels where the strong zonal jets mentioned above are generated. 

Despite the clear interest there is to study the tidal response of $\carbondiox$-dominated atmospheres for the similarity of configurations they offer with the Venus planet, we choose to focus in this work on $\nitrogen$-dominated atmospheres owing to their simpler frequency-behaviour. 


\subsection{The surface-atmosphere coupling}
\label{ssec:coupling}

The specific role played by the surface thermal response is not taken into account in linear models used to establish the Maxwell-like behaviour of the tidal torque described by \eq{maxwell} \citep[e.g.][]{DI1980,ADLM2017a}. In these early works, the thermal forcing is assumed to be in phase with the stellar incoming flux, which amounts to considering that thermal tides are caused by the direct absorption of the flux. This approximation seems realistic in the case of Venus-like planets given that their atmospheres are optically thick in the visible range, and sufficiently dense to neglect their interactions with the surface. 

However, it appears as a rough approximation in the case of optically thin atmospheres, where most of the stellar flux reaches the surface. In this case, thermal tides are mainly caused by the absorption of the flux emitted by the surface in the infrared range, which is delayed with respect to the stellar incoming flux owing to the surface inertia and dissipative processes such as thermal diffusion. Our ${\rm N_2}$-dominated atmosphere belongs to the second category. Thus, the role played by the thermal response of the ground should be considered in the present study to explain the observed difference between the obtained tidal torque and the Maxwell model. 

Concerning this point, we note that \cite{Leconte2015} included the heat capacity of the surface $\Csurf$ in the simplified model they used to establish the Maxwell-like behaviour of the tidal torque (see Sect.~4 in the Material and Methods of their article). Hence, by introducing the heat capacity of the atmosphere/surface system $\Ceff = \Cpgaz  \psurf / \ggravi + \Csurf$ and the emission temperature ($\Teff$), they expressed the relationship between surface temperature variations $\deltaT$ and the variations of the incoming stellar flux $\deltaFinc$ as 

\begin{equation}
\delta \Tsurf = \frac{\deltaFinc}{\sigmaM \Ceff} \frac{1}{1 + i \ftide/ \sigmaM},
\label{Tsurf_JL2015}
\end{equation}

\noindent with $\sigmaM = 4 \SBconstant \Teff^3 / \Ceff$ (the subscript $\imaxwell$ refers to the Maxwell-like form of the function given by \eq{Tsurf_JL2015}\rec{)}. As we generally observe that $\Teff \approx \Tsurf$ in our GCM simulations of a 10~bar atmosphere (the mean surface temperature of the planet is well approximated by the black body equilibrium temperature, given by \eq{Ts_BB}, in this case), this model implies that $\sigmaM$ should be always less than $\sigmaMsup = 4 \SBconstant \Tsurf^3 \ggravi / \left( \Cpgaz \psurf \right)$. However, in light of typical values of $\tauM$ obtained with the GCM (see \tab{scaling_laws}), it appears that the above formula for $\sigmaMsup$ leads to underestimate $\sigmaM$ by a factor 10 to 100 for the case treated in the present study. 


To understand the role played by the ground in the atmospheric tidal response, we adopt an ab initio approach describing thermal exchanges at the surface-atmosphere interface. Following along the line by \cite{Bernard1962} \citep[see also][]{ADLM2017a}, we write the local budget of perturbative power inputs and losses,

\begin{equation}
 \deltaFinc - 4 \SBconstant \Tsurf^3 \deltaTsurf + \deltaFatm - \deltaQsol - \deltaQatm = 0,
\end{equation}

\noindent where we have introduced the small variations of the incoming stellar flux $\delta F$, surface temperature $\deltaTsurf$, radiative heating by the atmosphere $\deltaFatm$ and diffusive losses in the ground $\deltaQsol$ and in the atmosphere $\deltaQatm$. Owing to the absence of water, latent heats associated with changes of states are ignored. 

In the general case, $\deltaTsurf$ and $\deltaQatm$ are coupled with the atmospheric tidal response. Particularly, in the Newtonian cooling approximation (i.e. variations of the emitted flux are proportional to temperature variations), $\deltaFatm$ can be expressed as 

\begin{equation}
\deltaFatm = \integ{\coolingcoeff \left( \xx,\col,\lon \right) \deltaT \left( \xx,\col,\lon \right)}{\xx}{0}{+ \infty},
\label{deltaFatm}
\end{equation}

\noindent where $K$ designates an effective coefficient of Newtonian cooling. In order to avoid mathematical complications, we ignore this coupling by assuming either that $\left| \deltaFatm \right| \ll 4 \SBconstant \Tsurf^3 \abs{\deltaTsurf} $, or, following \cite{Bernard1962}, that the variation of the atmospheric flux scales as $\deltaFatm \scale \deltaTsurf$ in a similar way as the variation of the flux emitted by the ground. This allows us to simplify radiative terms by writing 

\begin{equation}
4 \SBconstant \Tsurf^3 \deltaTsurf - \deltaFatm =  4 \SBconstant \Tsurf^3 \emisurf \deltaTsurf,
\end{equation}

\noindent where $\emisurf \approx 1$ stands for the effective emissivity of the surface. 

With the above approximations, surface temperature variations can be written for a given mode as $\deltaTsurf^{\ftide} = \Bsolf \deltaFinc^\ftide$. We thus end up with (see detailed calculations in \append{app:surface_response})

\begin{equation}
\Bsolf = \frac{\Bsolstat}{1 + \left[ 1 + \sign \left( \ftide \right) \inumber \right] \sqrt{\tausurf \abs{\ftide}}},
\label{Bs_model}
\end{equation}

\noindent where $\Bsolstat = \left( 4 \SBconstant \Tsurf^3 \emisurf \right)^{-1}$, and $\tausurf$ designates the characteristic timescale of the surface thermal response, which depends on the thermal inertia of the ground $\Isol$ and of the atmosphere $\Iatm$ at the interface, and is expressed as 

\begin{equation}
\tausurf = \frac{1}{2} \left( \frac{\Isol + \Iatm}{4 \SBconstant \Tsurf^3 \emisurf} \right)^2.
\label{taus}
\end{equation}

\begin{figure}[htb]
   \centering
   \includegraphics[width=0.48\textwidth,trim = 1.5cm 3.5cm 1.5cm 3.5cm,clip]{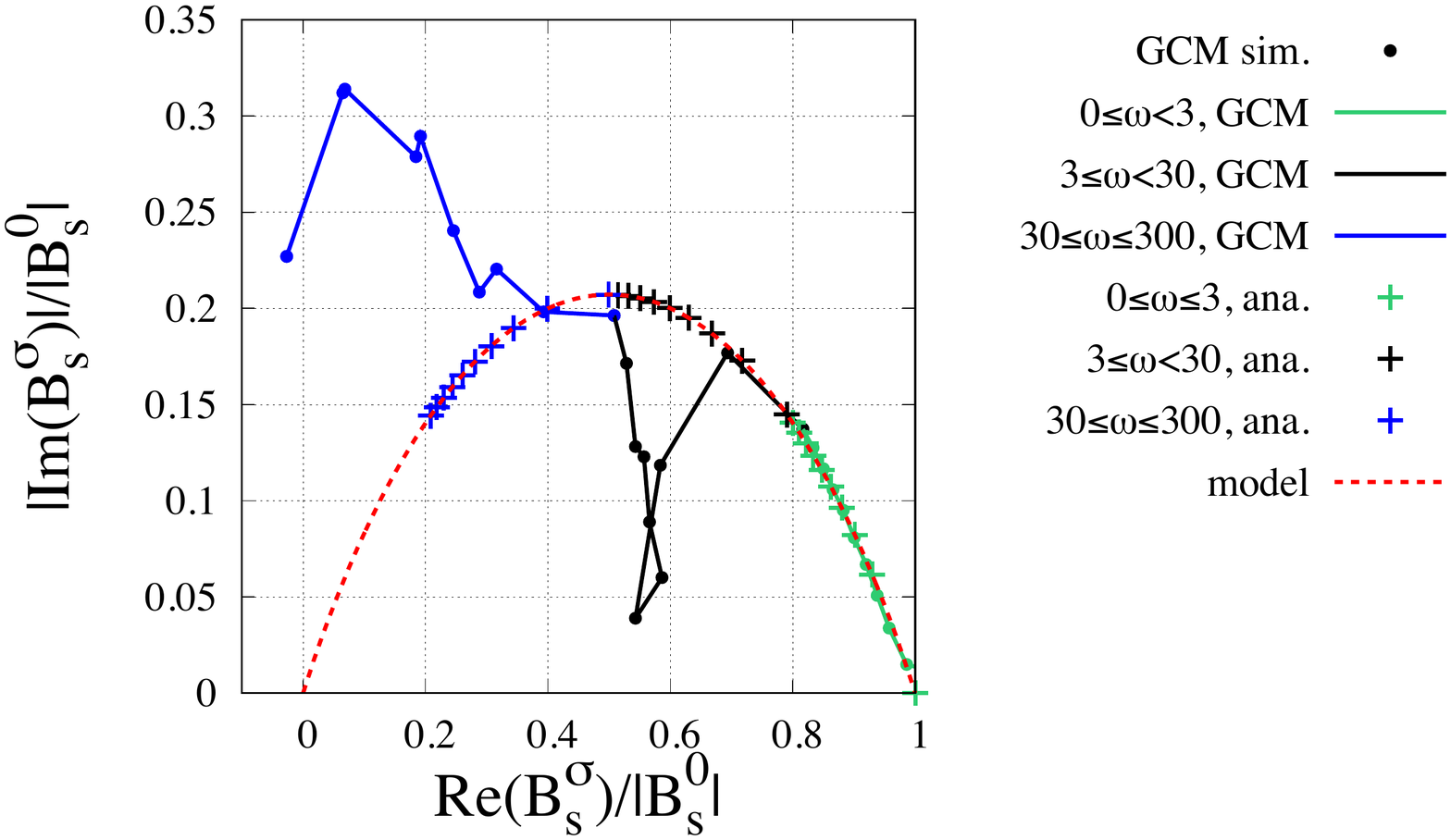} 
   \caption{Nyquist plot of the transfer function $\Bsolf$ associated with the $\Yquad$ component of the semidiurnal tide, i.e. such that $\deltaTsquad = \Bsolf \deltaFincquad $. The imaginary part of the normalized function $\Bsolf/ \abs{\Bsolstat}$ is plotted in absolute value as a function of the real part in the reference case of the study ($\smaxis = \smvenus$ and $\psurf = 10$\units{bar}). Values obtained using GCM simulations are indicated by points. They are interpolated by a green line in the range $0 \leq \omeganorm < 3$ (step $\Delta \omeganorm = 0.3$), a black line in the range $3 \leq \omeganorm < 30$ (step $\Delta \omeganorm = 3$), and a blue line in the range $30 \leq \omeganorm \leq 300 $ (step $\Delta \omega = 30$), $\omeganorm = \left( \spinrate - \norb \right) / \norb$ being the normalized tidal frequency. Similarly, values obtained in these ranges with the model given by \eq{Bs_model} for the surface thermal time $\tausurf = 0.3$\units{days} are designated by crosses. The red line corresponds to the function itself.}
       \label{fig:Bgr_Nyquist}%
\end{figure}

We compare this model to numerical results by extracting the $\Yquad$ component of the surface temperature distribution $\deltaTsquad$ provided by GCM simulations, as previously done for the surface pressure distribution. The obtained values are plotted in the complex plane in \fig{fig:Bgr_Nyquist}. In this plot, the horizontal and vertical axes correspond to the real and imaginary parts of the normalized transfer function $\Bsolf / \Bsolstat$ (such that $\deltaTsquad = \Bsolf \deltaFincquad $), respectively. Normalization is obtained by fitting numerical results with the function given by \eq{Bs_model} in the low-frequency range ($0\leq \omeganorm < 3$). 


Figure~\ref{fig:Bgr_Nyquist} shows a good agreement between the functional form of the model and numerical results in the zero-frequency limit. However, we observe that the value of the thermal time $\tausurf \sim 0.3$\units{days} obtained by fitting \eq{Bs_model} to numerical results in the low-frequency range is a decade smaller than the theoretical value given by \eq{taus}, $\tausurf \approx 4.6$\units{days} (we use values given by \tab{parameters}, set $\emisurf = 1$ and neglect $\Iatm$), which shows the limitations of the approach detailed above.

While the forcing frequency increases, the behaviour of the function interpolated using numerical results starts to change radically. In the vicinity of the resonance ($\ftide \sim \fpeak$), the imaginary part of $\Bsolf$ decays abruptly whereas its theoretical analogous keeps growing. This divergence suggests a strong radiative coupling between the surface and the atmosphere, which comes from the fact that the emission of the atmosphere to the surface $\deltaFatm$ (see \eq{deltaFatm}) can no longer be neglected, as done in the model. The abrupt variation of the surface thermal lag around the resonance partially explain the behavior of the tidal torque in this range. Nevertheless, to better understand it, one should study the whole dynamics of the atmospheric tidal response, which is beyond the scope of this work. 


In the high-frequency range, that is for $ \ftide  \gg \tausurf^{-1}$, the model predicts that the amplitude of temperature variations should tend to zero. Yet, we observe that $\abs{\deltaTsquad}$ increases until reaching a maximum before decaying. This maximum corresponds here to a resonance whose frequency coincides with that of the main Lamb mode identified previously, in \sect{ssec:resonance_lamb} \citep[see][]{Lamb1917,Vallis2006}.

\begin{figure*}[htb]
   \centering
   \includegraphics[height=0.25\textheight,trim = 1.5cm 2.1cm 8.5cm 3.0cm,clip]{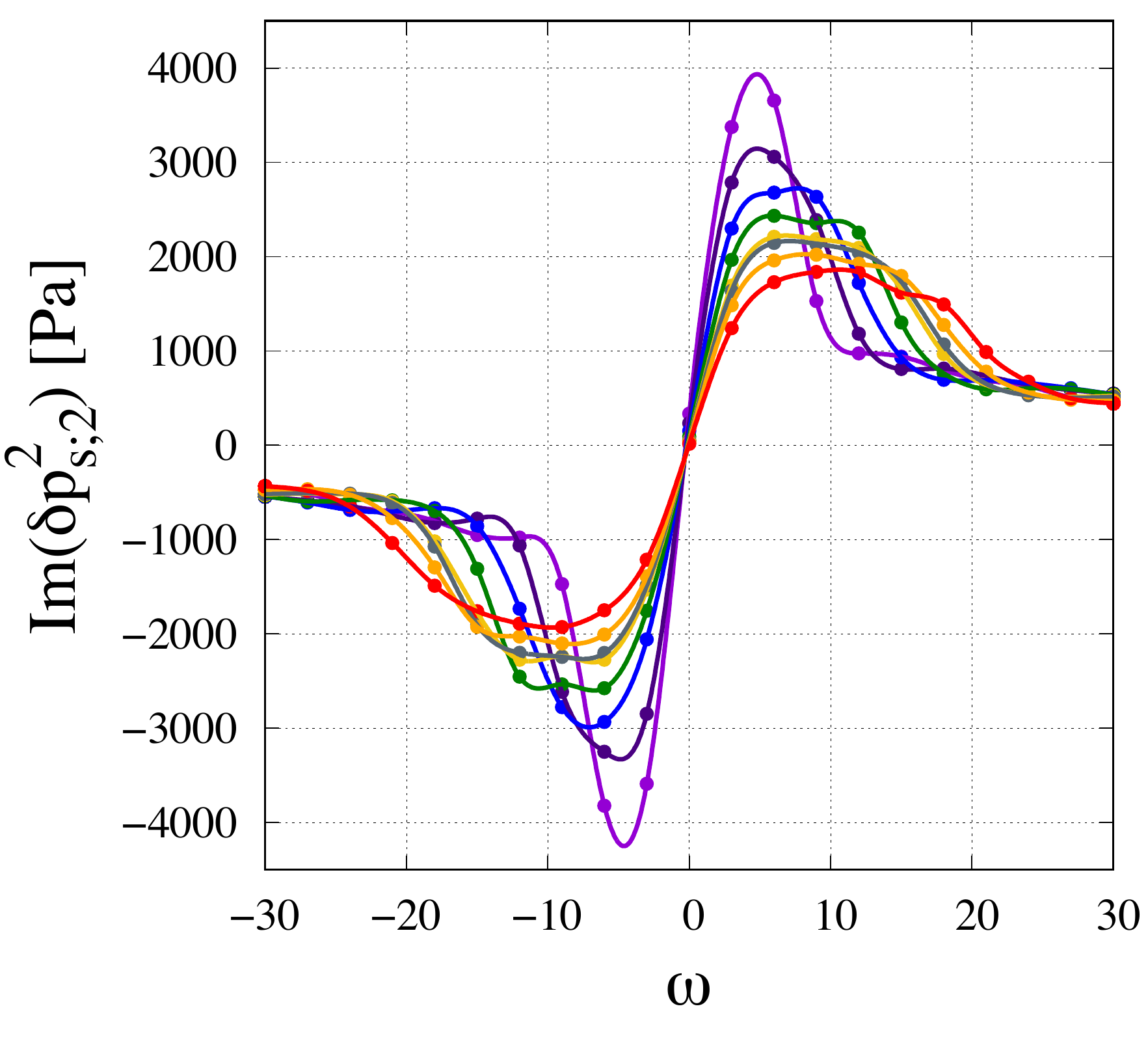} \hspace{0.5cm}
   \includegraphics[height=0.25\textheight,trim = 1.5cm 2.1cm 1.5cm 2.5cm,clip]{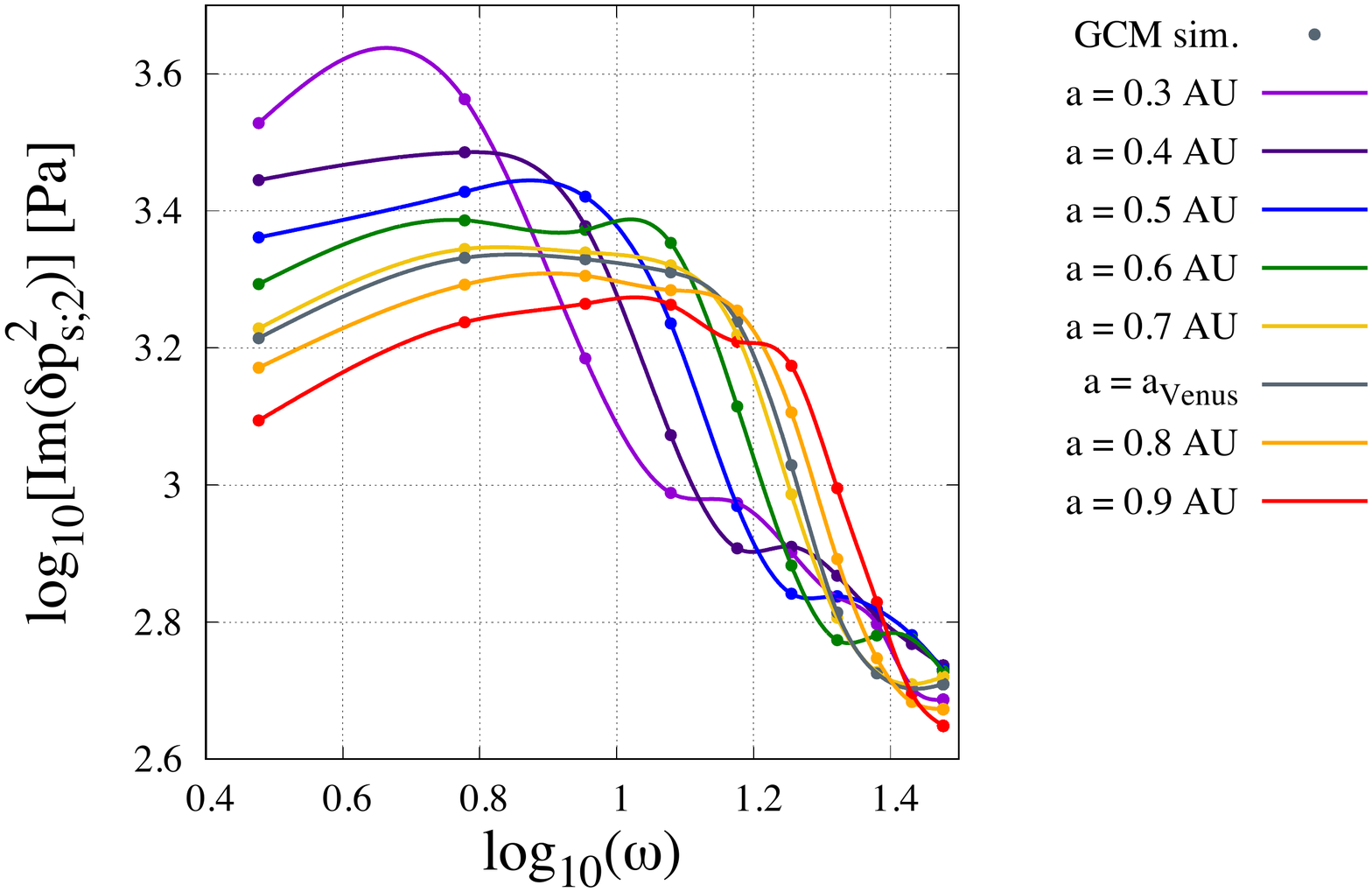} \\
   \includegraphics[height=0.25\textheight,trim = 1.5cm 2.1cm 8.5cm 3.0cm,clip]{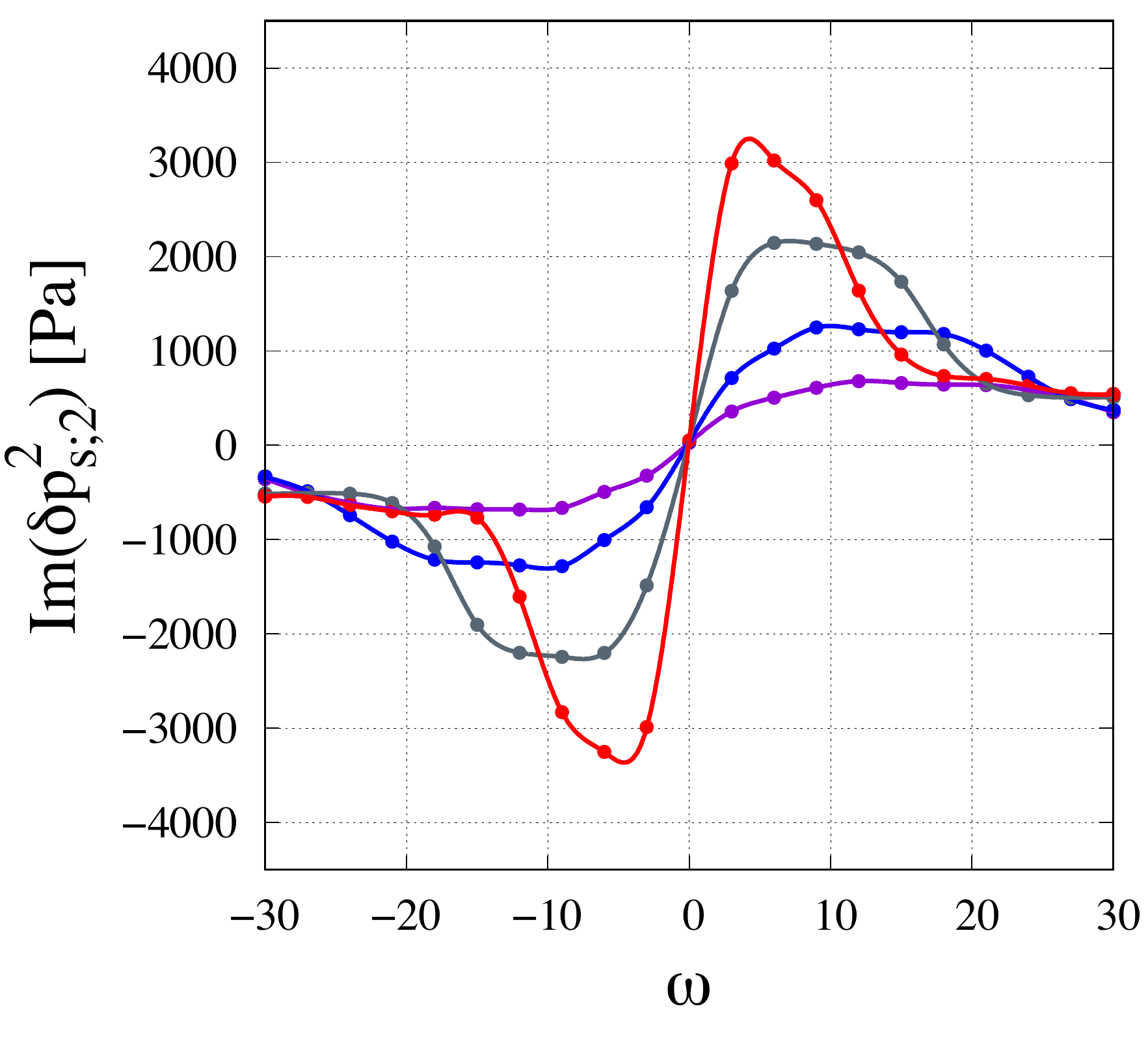} \hspace{0.5cm}
   \includegraphics[height=0.25\textheight,trim = 1.5cm 2.1cm 1.5cm 2.5cm,clip]{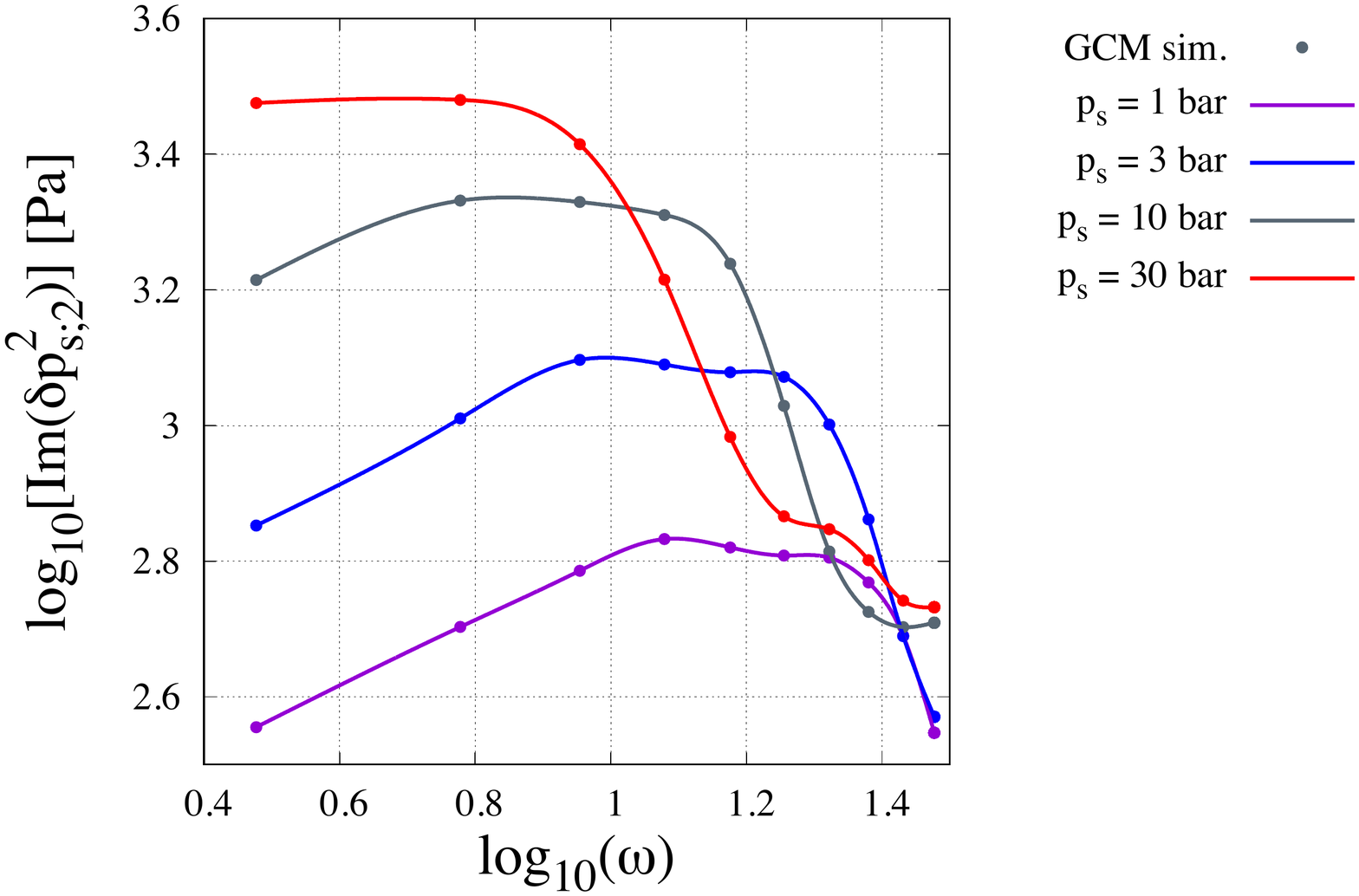}
   \caption{Imaginary part of the $\Yquad$ harmonic of surface pressure oscillations (Pa) as a function of the normalized tidal frequency $ \omeganorm = \left( \spinrate - \norb \right)/ \norb $ in linear (left panels) and logarithmic scales (for $\omeganorm >0$). {\it Top:} Spectra obtained with GCM numerical simulations for a fixed surface pressure, $ \psurf = 10 $~bar, and various values of the star-planet distance: $\smaxis = 0.3-0.9$\units{AU} with a step $\Delta \smaxis = 0.1$\units{AU}. {\it Bottom:} Spectra obtained for a fixed star planet-planet distance, $\smaxis = \smvenus$ and various values of the surface pressure : $\psurf= 1,3,10,30$\units{bar}. Numerical results are designated by points and interpolated with cubic splines. The reference case of the study ($\smaxis = \smvenus$ and $ \psurf = 10$\units{bar}) corresponds to the grey line in all plots. }
       \label{fig:spectres_GCM}%
\end{figure*}

\section{Exploration of the parameter space}
\label{sec:exploration_space}

We now examine the evolution of the tidal torque with the planet semi-major axis ($\smaxis$) and surface pressure ($\psurf$). 

\subsection{Frequency spectra of the tidal torque}

Considering the planet defined in \sect{sec:physical_setup}, we carry out two studies. In study~1, we set $\psurf = 10 $\units{bar} and we compute frequency spectra of the imaginary part of the $\Yquad$-surface pressure component in the low-frequency range for $\smaxis$ varying from 0.3 to 0.9\units{AU}. In study~2, we set $ \smaxis = \smvenus$, that is $\smaxis = 0.723$\units{AU}, and frequency spectra are computed for $\psurf$ varying from 1 to 30\units{bar}. The reference case characterized in the previous section, and parametrized by $\smaxis = \smvenus$ and $\psurf = 10$\units{bar}, is located at the intersection of the two studies. 
 
Limitations concerning the lower bound of the orbital radius range and the upper bound of the surface pressure range come from the spectra of optical properties used in simulations to compute radiative transfers (see \sect{sec:physical_setup}), which were produced for temperatures below 710\units{K}. Indeed, for $\smaxis < 0.3 $\units{AU} or $\psurf> 30 $\units{bar}, the planet surface temperature exceeds this maximum. As this might lead to erroneous estimations of radiative transfers, we choose not to treat extremal cases, although there is no formal limitation  for the GCM to run normally in these conditions. 

Radiative transfers also determine the convergence timescale necessary for the atmosphere to reach a steady state, $\Pconv$. For study~1, we use the timescale obtained in the reference case, that is 5800 Earth Solar days, considering that the steady state is reached more rapidly in mosts cases, where the planet is closer to the star (see Eq.~\ref{taurad}). Similarly, to take into account the dependence of $\Pconv$ on the planet surface pressure in study~2, we set $\Pconv$ to 1100, 2300, 5800 and 14000 Earth Solar days for $\psurf= 1,3,10,30$\units{bar}, respectively.

The obtained frequency spectra are plotted in \fig{fig:spectres_GCM} in linear (left) and logarithmic scales (right) for study~1 (top) and study~2 (bottom). In all plots, points designate the results of GCM simulations with the method described in \sect{sec:method}, while solid lines correspond to the associated cubic splines interpolations. The reference case ($\smaxis = \smvenus$ and $\psurf = 10$\units{bar}) is designated by the solid grey line. Numerical values used to produce these plots are given by \tab{tableS1} for study~1 and \tab{tableS2} for study~2.

We retrieve here the features identified in \sect{sec:frequency_behaviour}. The tidal torque exhibits maxima located at the transition between the low-frequency and high-frequency asymptotic regimes. The corresponding peaks are slightly higher in the negative-frequency range than in the positive-frequency range owing to Coriolis effects and the impact of zonal jets on the angular lag of the tidal bulge. As expected, the amplitude of peaks increases with both the incoming stellar flux and the planet surface pressure. Interestingly, the evolution of $\maxpeak$ and $\fpeak$ with $\smaxis$ and $\psurf$ looks very regular. This suggests that the dependences of the peak maximum and characteristic timescale on the planet surface pressure and distance to star are well approximated by simple power scaling laws, and it is the case indeed, as shown in \sect{ssec:evolution_peak}.


As previously noticed in the study of the reference case, the asymptotic behaviour of the tidal torque in the zero-frequency limit differs from that described by the Maxwell model. Particularly, the logarithmic plot of study~2 (bottom right panel) shows that the torque follows the scaling law $\fgcm \left( \ftide \right) \scale \ftide^{1/2}$ in cases characterized by low surface pressures, that is 1 and 3~bar. These cases correspond to the thin-atmosphere asymptotic limit, where thermal tides are driven by diffusion in the ground in the vicinity of the surface. We note that the simplified linear model of the surface thermal response detailed in \sect{ssec:coupling} and \append{app:surface_response} leads to a surface-generated radiative heating scaling as $\imag{\deltaTsurf} \scale \ftide^{1/2} $ in the zero-frequency limit, which is precisely the dependence observed in \fig{fig:spectres_GCM}.

\subsection{Evolution of the thermal peak with the planet semi-major axis and atmospheric surface pressure}
\label{ssec:evolution_peak}

Let us now quantify the regular dependence of the peak of tidal torque on the planet orbital radius and surface pressure observed in the preceding section. We have thereby to determine how the two parameters defining the peak -- namely its maximum value $ \maxpeak $ and associated timescale $\taupeak$ -- vary with $\smaxis$ and $\psurf$. 


\begin{figure*}
   \centering
      \includegraphics[width=0.25\textwidth,trim = 1.7cm 2.5cm 8.cm 2.5cm, clip]{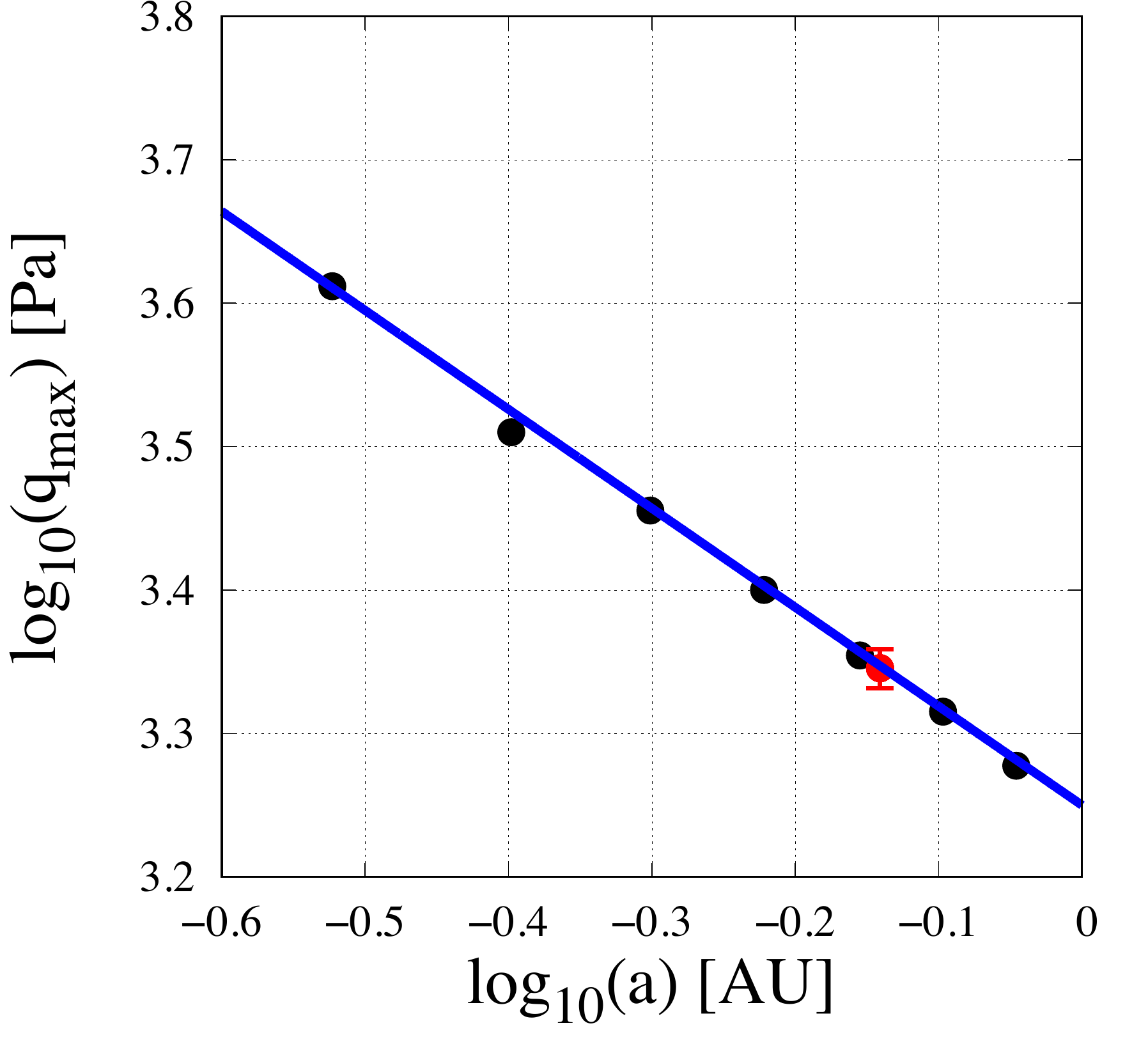} \hspace{0.5cm}
   \includegraphics[width=0.25\textwidth,trim = 1.7cm 2.5cm 8.cm 2.5cm, clip]{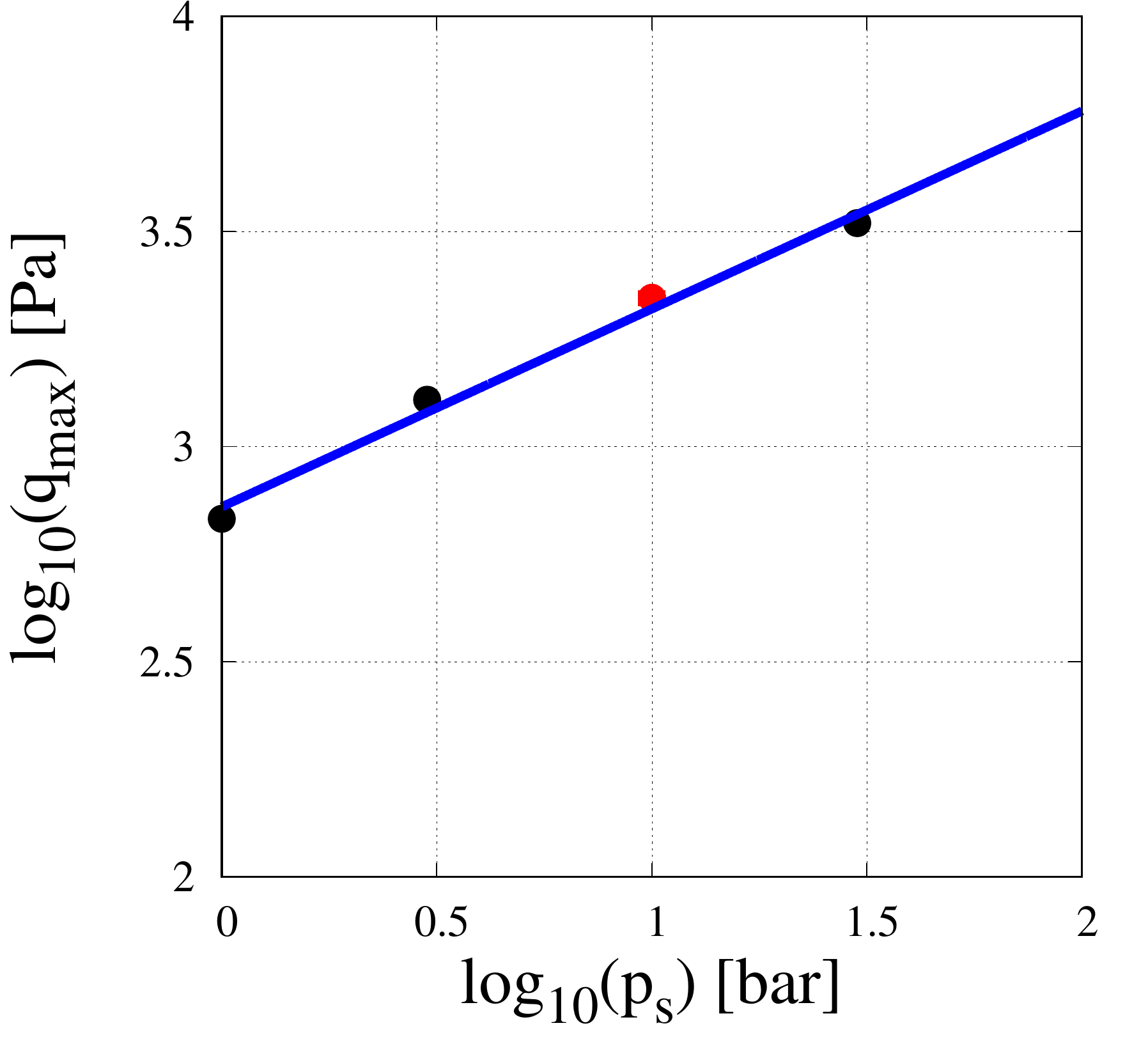} 
  \hspace{0.5cm}
   \includegraphics[width=0.11\textwidth,trim = 21.cm 8cm 1.5cm 2.2cm, clip]{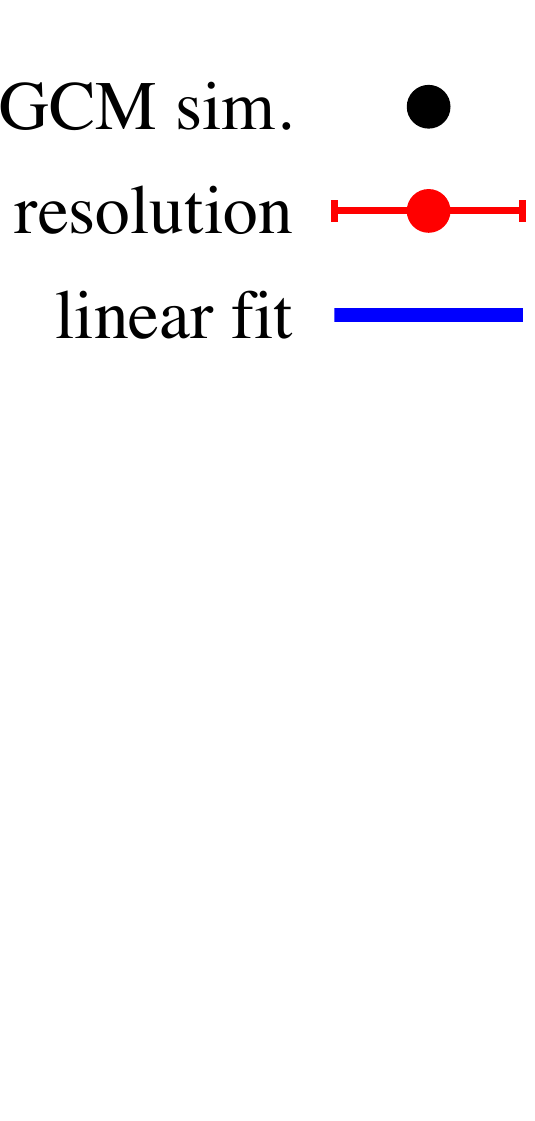}  \\
    \includegraphics[width=0.25\textwidth,trim = 1.7cm 2.5cm 8.cm 2.5cm, clip]{auclair-desrotour_fig7c} \hspace{0.5cm}
   \includegraphics[width=0.25\textwidth,trim = 1.7cm 2.5cm 8.cm 2.5cm, clip]{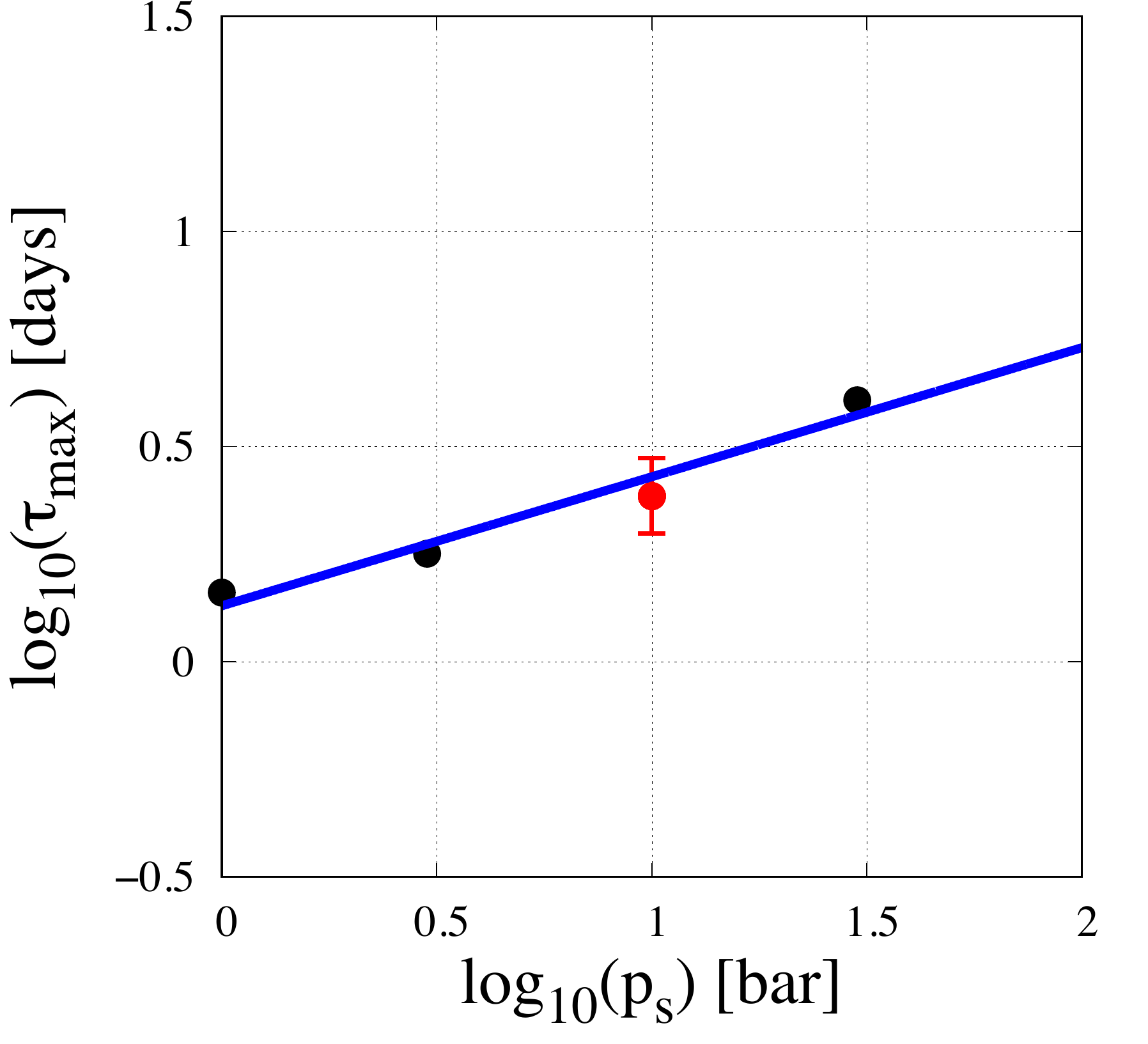} 
\hspace{0.5cm}
   \hspace{0.10\textwidth}
   \caption{Parameters of the Maxwell model given by \eq{maxwell} computed from GCM simulations as functions of the star-planet distance $\smaxis$ (left panels, in AU) and $\psurf$ (right panels, in bar) in logarithmic scales. {\it Top:} Amplitude $\maxpeak$ of the maximum (Pa). {\it Bottom:} Associated characteristic thermal time $\taupeak$ (days). Numerical results obtained with GCM simulations are designated by black points, the corresponding linear regressions (see \tab{scaling_laws}) by solid blue lines. For the reference case, error bars are given to indicate the resolution of the sampling (for details about how they are constructed, see \sect{ssec:evolution_peak}). }
       \label{fig:reglin}%
\end{figure*}


Thus, for each study, we fit numerical values of $\maxpeak$ and $\taupeak$ using a linear regression, formulated as

\begin{equation}
\Yreg = \alphareg \Xreg + \betareg,
\end{equation}

\noindent where $\Yreg$ designates the logarithm of $\maxpeak$ (Pa) or $\taupeak$ (days), $\Xreg$ the logarithm of $\smaxis$ (AU) or $\psurf$ (bar), and $\alphareg$ and $\betareg$ the dimensionless parameters of the fit. The values of these parameters are given by Table~\ref{scaling_laws}, as well as those of the corresponding coefficients of determination $\Rdet^2$. We also compute $\log \left( \maxpeak / \taupeak \right) $ for comparison with the theoretical scaling law given by \eq{SLana} in the case where the dependence of the tidal torque on the forcing frequency is approximated by a Maxwell function. 

Linear regressions are plotted in \fig{fig:reglin} (blue solid line). In order to provide an estimation of the variability of numerical results, error bars are given for the reference case. These error bars do not literally correspond to a margin of error, but indicate the resolution of the sampling for the frequency and maximum of the tidal torque. For $\maxpeak$, the amplitude of the error bar is the departure between the maxima of the interpolating function and data. For $\taupeak$, the two bounds of the error bar are the values associated with the nearest points of the sampling, designated by the subscripts $\iinf$ and $\isup$, such that $\tauinf \leq \taupeak \leq \tausup $. \rec{These error bars depend on the ratio between the size of a frequency interval and the width of the thermal peak. For example, the thermal peak is undersampled for $\smaxis = 0.3$~\units{UA}, which makes the fit less reliable in this case.}

\begin{table}[htb]
 \textsf{\caption{\label{scaling_laws} Scaling laws of $ \taupeak $ and $ \maxpeak $ obtained using the LMDZ GCM for a dry terrestrial planet with a homogeneous $\nitrogen$ atmosphere. The scaling law of $\maxpeak/\taupeak$ is computed using the formers and should be compared to \eq{SLnum}. Units: $\smaxis$ is given in AU, $ \psurf $ in bar, $\taupeak$ in days, $\maxpeak$ in Pa, the parameters of the linear fit $ \alphareg $, $ \betareg $ and $\Rdet^2$ are dimensionless. }}
\centering
{\setlength{\tabcolsep}{1.5pt}
    \begin{tabular}{c c r l c}
      \hline
      \hline 
      \textsc{Case} & \multicolumn{3}{c}{\textsc{Scaling laws of $\maxpeak$ and $\taupeak$}} & $\Rdet^2$ \\
      \hline \\
       & $\log \left( \maxpeak \right) = $ & $ -0.69  \log \left( \smaxis \right) $ & $ + \, 3.25$ & \hspace{3pt} 0.998 \\
        $\psurf = 10$\units{bar} & $\log \left( \taupeak \right) = $ & $ 0.86 \log \left( \smaxis \right) $ & $ + \, 0.48 $ & \hspace{3pt} 0.901\\
       & $\log \left( \dfrac{\maxpeak}{\taupeak} \right) = $ & $ -1.55 \log \left( \smaxis \right) $ & $ + \, 2.77 $ & -- \\
       
      \vspace{0.1mm}\\
       \hline \\
       & $\log \left( \maxpeak \right) = $ & $ 0.46 \log \left( \psurf \right) $ & $ + \, 2.86 $ & \hspace{3pt} 0.990 \\
              $\smaxis = \smvenus$ & $\log \left( \taupeak \right) = $ & $ 0.30 \log \left( \psurf \right) $ & $ + \, 0.13 $ & \hspace{3pt} 0.959 \\
        & $\log \left( \dfrac{\maxpeak}{\taupeak} \right) = $ & $ 0.16 \log \left( \psurf \right) $ & $ + \, 2.73 $ & -- \\
       \vspace{0.1mm}\\
       \hline
\end{tabular}}
\end{table}

%

Comparing coefficients of determination in \tab{scaling_laws}, we observe that a better fit is systematically obtained for $\maxpeak$ than for $\taupeak$. This difference may be explained by the aspect of spectra displayed in \figs{fig:spectres_GCM}{fig:spectres_model}. Since the peak of the tidal torque computed with the GCM is both flatter and larger than that of the Maxwell function, the position of the maximum is more sensitive to small fluctuations than the maximum itself. As a consequence, the variability of $\maxpeak$ is less than the variability of $\taupeak$.

Hence, the linear regression fits particularly well the dependence of $\maxpeak$ on $\smaxis$, while the plot of $\taupeak$ exhibits a relatively important variability with respect to the linear tendency. Note however that differences with the fit are not significative since they remain small compared to the width of the peak. Concerning the dependence of $\taupeak$ on $\smaxis$, one may also observe that the slope, given by $\alphareg = 0.86$, is almost twice smaller than that predicted by the scaling law of the radiative timescale given by \eq{taurad}, that is $\taupeak \scale \norb^{-1} \scale \smaxis^{3/2}$.

As regards the ratio $\maxpeak / \taupeak$ however, we recover numerically the scaling law predicted by the theoretical model (\eq{SLana}) with a good approximation. This scaling law is numerically expressed in the units of \tab{scaling_laws} as 

\begin{equation}
\log \left( \frac{\maxM}{\tauM} \right) = -\frac{3}{2} \log  \left( \smaxis \right) + 2.49,
\label{SLnum}
\end{equation}

\noindent if we assume that $\effheat = 1-\Asurf$ (i.e. the flux reemitted by the ground is entirely absorbed by the atmosphere). 

As may be seen, the dependence of $\maxpeak / \taupeak$ on the surface pressure is small ($\alphareg = 0.13$) for want of being zero, as predicted by the model. Regarding the dependence on $\smaxis$, the relative difference between the numerical and theoretical values of $\alphareg$ (i.e. 1.55 and $3/2$, respectively) is around 3\%. However, the value of $\betareg$ computed from GCM simulations (2.77) is higher than that predicted by the model (2.49), despite the fact that this latter is an upper estimation. This difference illustrates the limitations of the Maxwell model, which fails to describe the sharp variations of the tidal torque with the tidal frequency when $ \abs{\ftide}  \sim \fpeak$. 
 

\begin{figure*}[htb]
   \centering
   \includegraphics[height=0.25\textheight,trim = 1.5cm 2.1cm 8.5cm 2.5cm,clip]{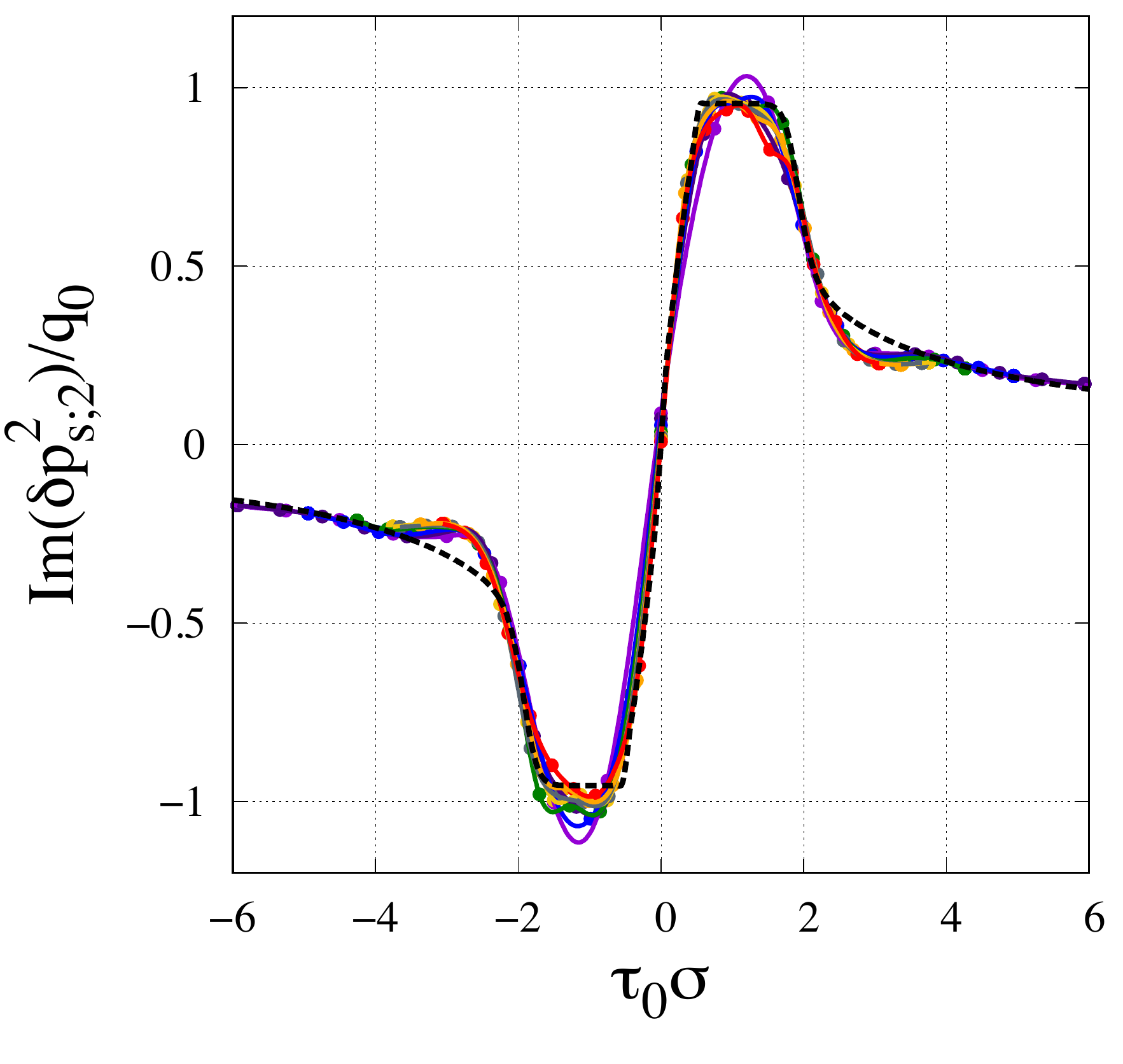} \hspace{0.5cm}
   \includegraphics[height=0.25\textheight,trim = 1.5cm 2.1cm 1.5cm 2.5cm,clip]{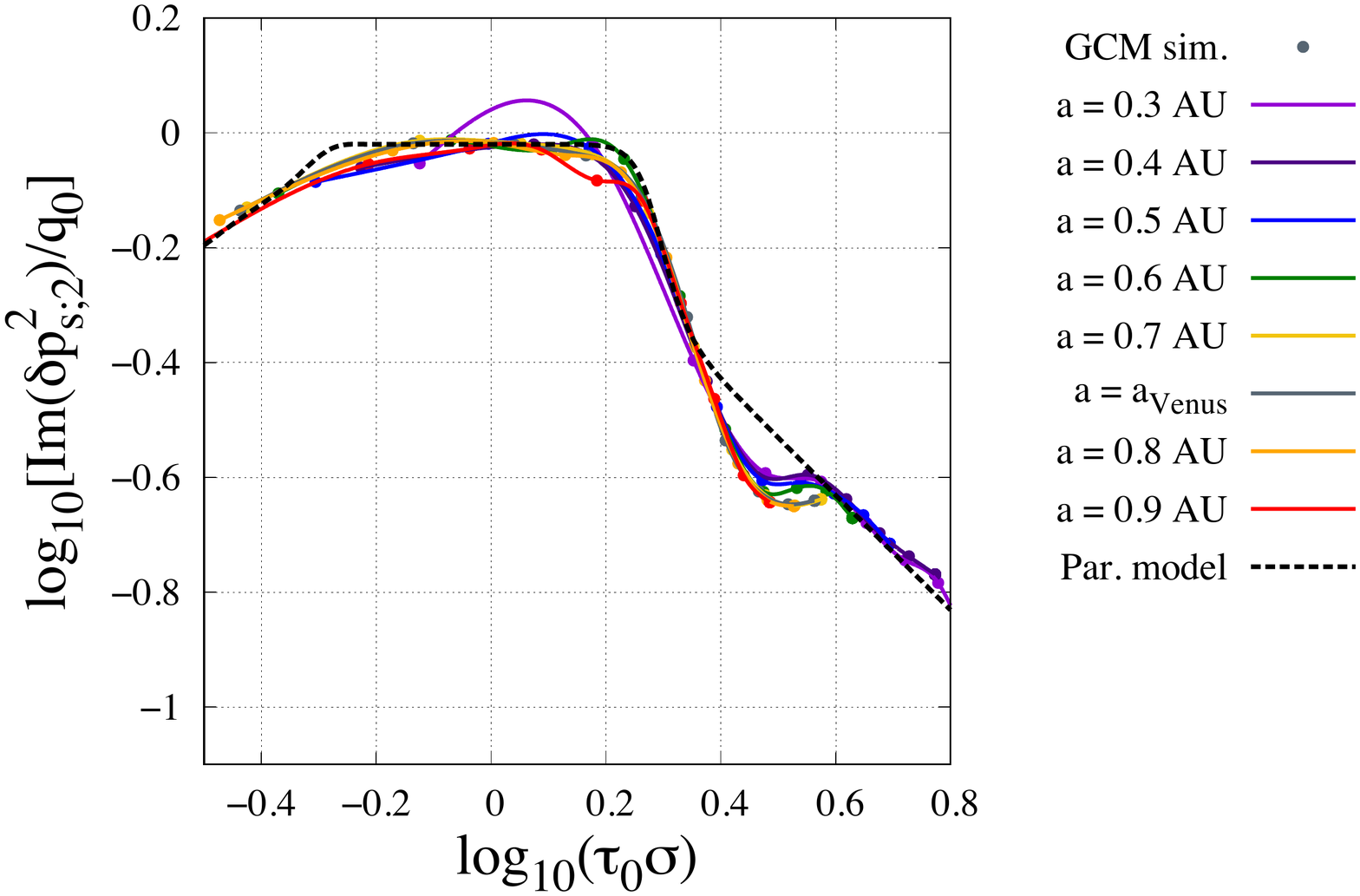} \\
   \includegraphics[height=0.25\textheight,trim = 1.5cm 2.1cm 8.5cm 2.5cm,clip]{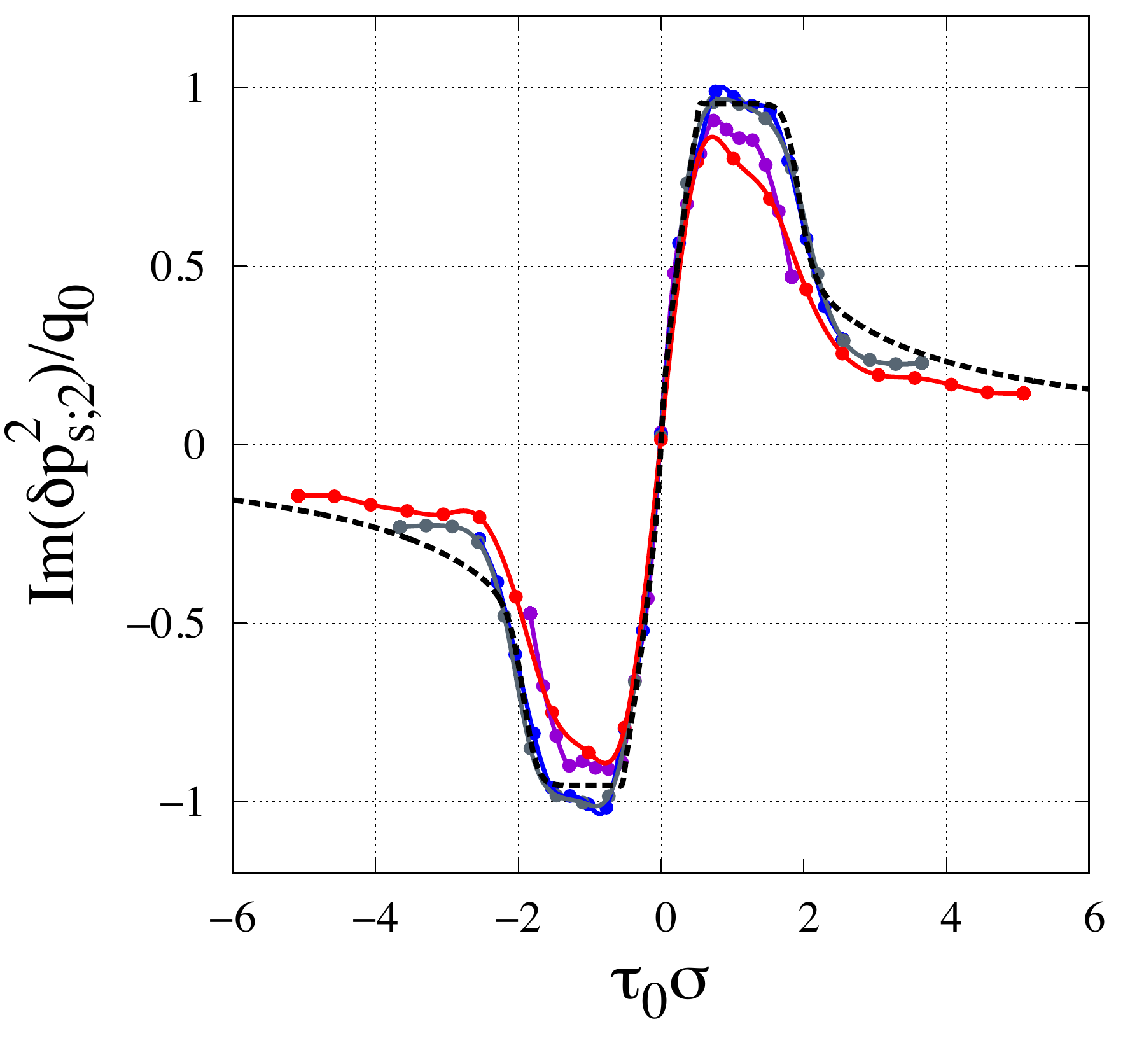} \hspace{0.5cm}
   \includegraphics[height=0.25\textheight,trim = 1.5cm 2.1cm 1.5cm 2.5cm,clip]{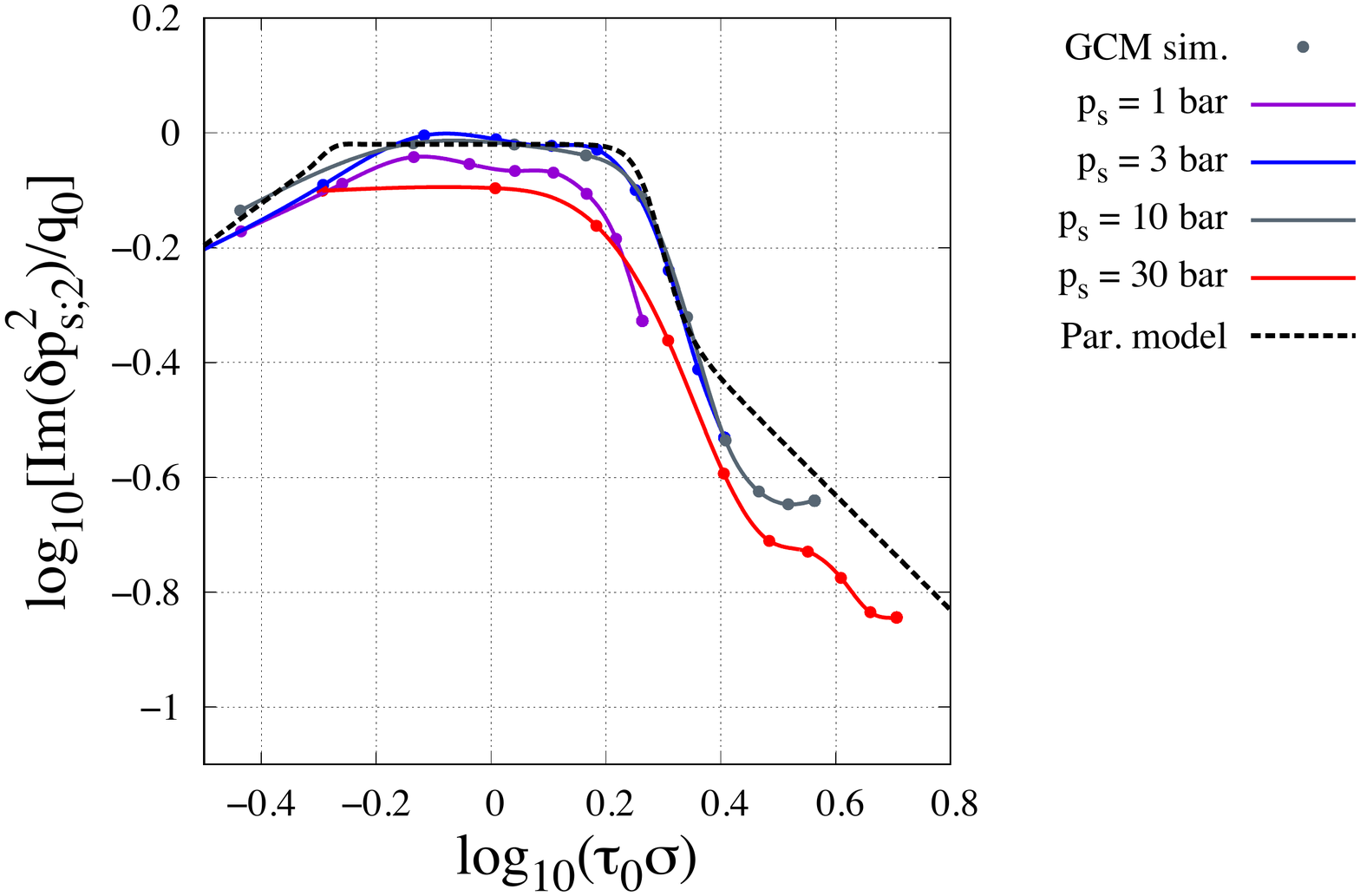}
   \caption{Normalized imaginary part of the $\Yquad$ harmonic of surface pressure oscillations as a function of the normalized tidal frequency in linear (left panels) and logarithmic scales (for $\omeganorm >0$). {\it Top:} Spectra obtained with GCM numerical simulations for a fixed surface pressure, $ \psurf = 10 $\units{bar}, and various values of the star-planet distance : $\smaxis = 0.3-0.9$\units{AU} with a step $\Delta \smaxis = 0.1$\units{AU}. {\it Bottom:} Spectra obtained for a fixed star planet-planet distance, $a = \smvenus$ and various values of the surface pressure : $\psurf = 1,3,10,30$\units{bar}. Results are designated by points and interpolated with cubic splines. The reference case of the study ($\smaxis = \smvenus$ and $ \psurf = 10$\units{bar}) corresponds to the grey line in all plots. The generic formula of the tidal torque derived in this study and given by \eq{model_generic} is designated by the black dashed line. The characteristic torque $\maxcar$ and timescale $\taucar$ used for the rescaling are given by \eqsto{psl1}{psl4} as functions of $\smaxis$ and $\psurf$.}
       \label{fig:spectres_GCM_renorm}%
\end{figure*}

\subsection{Scaling laws and generic formula for the tidal torque}

\comments{Renormalisation et formule générique pour le couple de marée.}

By proceeding to a quantitative study of the evolution of the tidal torque maximum with the planet orbital radius and surface pressure, we demonstrated in the preceding section the regularity observed in \fig{fig:spectres_model}. The scaling laws given by \tab{scaling_laws} and plotted in \fig{fig:reglin} show that frequency-spectra have the same aspect from the moment that the horizontal and vertical axes are rescaled following the obtained dependences on $\smaxis$ and $\psurf$. In this section, our purpose is to compute this rescaling in a robust way, by taking into account the whole set of data at our disposal rather than the maximal value of the torque and the associated timescale only. Combining this rescaling with the parametrized model given by \eq{fonction_para}, we will obtain a novel generic formula for the frequency-behaviour of the thermally generated atmospheric tidal torque.

The parameter with respect to which axes are rescaled, $\smaxis$ or $\psurf$, is denoted by $\qpar$, and the considered case is subscripted $\jcas$. A given family is thus composed of $\Ncas$ couples of numerical vectors $\left( \xvj , \yvj \right)$, with $1 \leq \jcas \leq \Ncas$ (see \tab{tableS1} and \tab{tableS2}), associated with the value $\aj$ of the varying parameter $\qpar$. For a given couple of vector, one may introduce the associated interpolating function 

\begin{equation}
\finterpj \left( \xv \right) \define \finterp \left( \xvj , \yvj , \xv \right) .
\end{equation}

\noindent We also introduce the renormalized vectors $ \xvnormj $ and $ \yvnormj $, defined by 

\beqtwo{\xvnormj \define \aj^{-\parx} \xvj }{\yvnormj \define \aj^{-\pary} \yvj .}

\noindent where $\parx$ and $\pary$ are the exponents characterizing the renormalization. 

The size of frequency domains covered by the $\xvnormj$ vectors varies with $\qpar$ in the general case. As a consequence, rescaling axes requires to define the bounds of the largest common interval,  

\begin{align}
& \xvnorminf \left( \parx \right) = \max \left\{ \aj^{-\parx} \xvjfirst  \right\}_{\jcas}, \\
& \xvnormsup \left( \parx \right) = \min \left\{  \aj^{-\parx} \xvjlast \right\}_{\jcas},
\end{align}

\noindent the notation $\Nxv$ referring to the size of the frequency sampling (typically $\Nxv = 21$, see \tab{tableS1} and \tab{tableS2}), and $\xvjfirst$ and $\xvjlast$ to the lower and upper bounds of the interval sampled by $\xvj$, respectively. 

The values of $\parx$ and $\pary$ are obtained by minimizing the squared difference function 

\begin{equation}
\Ffit \left( \parx , \pary \right) = \frac{ \displaystyle \sum_{\jcas < \kcas} \integ{ \! \! \left[ \aj^{-\pary} \finterpj \left( \aj^{\parx} \xvnorm \right) - \ak^{-\pary} \finterpk \left( \ak^{\parx} \xvnorm \right)  \right]^2 \! \!}{\xvnorm}{\xvnorminf \left( \parx \right)}{\xvnormsup \left( \parx \right)}  }{\xvnormsup \left( \parx \right) - \xvnorminf \left( \parx \right)} .
\end{equation}

Note that the parameters derived from these calculations, $\maxcar$ and $\taucar$, slightly differ from $\maxpeak$ and $\taupeak$. They stand for the characteristic amplitude and timescale of the peak and not for its maximum value and corresponding forcing period, as $\maxpeak$ and $\taupeak$. These parameters are defined by functions of $\qpar$ (that is $\smaxis$ or $\psurf$), by

\beqtwo{\taucar \define \taupeakref \left( \frac{\qpar}{\qparref} \right)^{-\parx}}{\maxcar \define \maxpeakref \left( \frac{\qpar}{\qparref} \right)^{\pary},}

\noindent where $\qparref$ designates the value of $\qpar$ in the reference case (typically $\smvenus$ or 10~bar), $\maxpeakref = 2.24$\units{kPa} the corresponding maximum of the peak and $\taupeakref = 2.18$\units{days} the associated timescale. Hence, we end up for the first family ($\psurf = 10$\units{bar} and variable $\smaxis$) with the scaling laws 

\begin{align}
\label{psl1}
& \log \left( \maxcar \right) = - 0.61 \log \left( \smaxis \right) + 3.26, \\
& \log \left( \taucar \right) = 0.68 \log \left( \smaxis \right) + 0.43, 
\end{align}

\noindent and for the second family ($\smaxis = \smvenus$ and variable $\psurf$), with

\begin{align}
& \log \left( \maxcar \right) = 0.48 \log \left( \psurf \right) + 2.87, \\
\label{psl4}
& \log \left( \taucar \right) = 0.30 \log \left( \psurf \right) + 0.038. 
\end{align}

\noindent Let us remind here the used units: $\smaxis$ is given in AU, $\psurf$ in bar, $\maxcar$ in Pa, and $\taucar$ in days. 

The last step consists in combining these scaling laws with the parametrized model derived in the reference case (\eq{fonction_para}). Proceeding to the change of variables associated with the renormalization, we obtain the generic parametrized model function 

\begin{equation}
\Im \left\{ \deltapsquad \right\}_{\ipara} \define \maxcar 10^{\Fpara \left( \log \abs{\taucar \ftide} \right) } \sign \left( \ftide \right).
\label{model_generic}
\end{equation}

\noindent We \rec{remind} here that $\Fpara$ is the function

\begin{align}
\label{fonction_para2}
\Fpara \left( \xmod \right) \define & \left( \alf \xmod + \blf \right) \Flf \left( \xmod \right) + \left( \ahf \xmod + \bhf \right) \Fhf \left( \xmod \right) \\
  & + \btrans \left[ 1 - \Flf \left( \xmod \right) - \Fhf \left( \xmod \right)  \right],  \nonumber
\end{align}

\noindent where $\Flf$ and $\Fhf$ are the activation functions defined by

\beqtwo{\Flf \left( \xmod \right) \define \frac{1}{1 + \expo{\left( \xmod - \xlf \right)/\dlf}} }{\Fhf \left( \xmod \right) \define \frac{1}{1 + \expo{- \left( \xmod - \xhf \right) / \dhf}}.}

\noindent The parameters characterizing the generic formula given by \eq{model_generic} take the values

\begin{equation}
\begin{array}{llll}
\alf = 0.734, & \blf = 0.171, & \dlf = 0.010 , & \xlf = -0.277 ,   \\ 
\ahf = -1 , & \bhf = -0.031, & \dhf = 0.023, & \xhf = 0.290. \\
 \btrans = -0.020. & & &
\end{array}
\end{equation}



The spectra of \fig{fig:spectres_GCM} are replotted in \fig{fig:spectres_GCM_renorm} using the normalized variables derived from the axes rescaling. In addition to numerical results and their interpolating functions, the tidal torque described by the generic parametrized model (\eq{model_generic}) is plotted as a function of the normalized tidal frequency $\taucar \ftide$ (dashed black line). \fig{fig:spectres_GCM_renorm} clearly shows the relevance of the rescaling as regards the first family, where the dependence of the torque on the star-planet distance is investigated. After rescaling, spectra look similar and the model matches them fairly well. As regards the second family, we observe a greater variability of $\maxcar$ and $\taucar$ with a net separation between the reference and 30~bar cases. However, the frequency behaviour of the torque does not change much from one case to another and the parametrized function given by \eq{model_generic} remains a reasonable approximation of its main features.

\section{Conclusions}
\label{sec:conclusions}

In order to better understand the behaviour of the atmospheric torque created by the thermal tide, we computed the tidal response of the atmosphere hosted by a terrestrial planet using the LMDZ general circulation model. This work builds on both the early study by \cite{Leconte2015}, which was a first attempt to characterize the atmospheric tidal response with this approach, and the early analytical works based upon the linear theory of atmospheric tides \citep[e.g.][]{ADLM2017a,ADLM2017b}. It is motivated by the need to merge these two different approaches together in a self-consistent picture. Our aim was to proceed to a methodic comparison of their predictions while exploring the parameter space. 

Hence, we considered the simplified case of a dry Venus-sized terrestrial planet orbiting a Sun-like star circularly and hosting a nitrogen-dominated atmosphere. Following the method by \cite{Leconte2015}, we computed the atmospheric torque created by the semidiurnal thermal tide as a function of the tidal frequency by extracting the $Y_2^2$ component of the surface pressure anomaly in simulations. 

As a first step, we characterized the variation of the torque with the forcing frequency for a reference case ($\psurf = 10$\units{bar} and $\smaxis = \smvenus$), and explained its various features with an independent analytical model. 
As a second step, we explored the parameters space by focusing on the dependence of the tidal torque on the planet orbital radius and atmospheric surface pressure. The obtained results were then used to derive scaling laws characterizing the torque, renormalize the pressure anomaly and forcing period, and finally propose a novel generic parametrized function to model the frequency-behaviour of the torque in a realistic way in the case of a nitrogen-dominated atmosphere. 

The first investigation confirmed and extended the results obtained by \cite{Leconte2015}. We showed that the torque follows two different asymptotic regimes. In the high-frequency range, the torque decays inversely proportionally to the tidal frequency until it exhibits a resonance peak. These two features are both explained by the analytical solution derived using the ab initio linear theory of atmospheric tides. Particularly, the peak corresponds to a resonance associated with the Lamb mode, an acoustic type wave of wavelength comparable with the planet radius. 

In the low-frequency range the torque, which is zero at synchronization, increases following a power law of index ranging from $0.5$ to $0.7$ until it reaches a maximum. While the increase and presence of a maximum are predicted by the analytical solution, the exponent and the value of amplitude of the peak differ significantly. These discrepancies result from the complex interactions between mechanisms laying beyond the scope of standard analytical treatments but resolved in 3D GCM simulations, such as the non-linear effects inherent to the atmospheric dynamics in the vicinity of synchronization, and the strong radiative coupling between the atmosphere and the planet surface. Typically, the low-frequency asymptotic regime of the tidal response is characterized by diurnal oscillations of large amplitude. The resulting differences in the day- and night-side temperature profiles significantly affect the stratification of the atmosphere. This clearly violates the small perturbation approximation upon which the analytic approach is based, and induces a non-linear coupling between the diurnal and semidiurnal oscillations that is important enough to modify the dependence of the tidal torque on the forcing frequency. 

The parametrized function that we propose in the present work (given by \eq{fonction_para2}) appears as a good compromise as it matches numerical results in a more satisfactory way than the Maxwell model while being defined by a reasonably small number of parameters. It is thus perfectly suited to be implemented in evolutionary models of the rotational dynamics of a planet. Nevertheless, the Maxwell-like analytic solution derived by early studies \citep[e.g.][]{ID1978,ADLM2017a} provides a first order of magnitude approximation of the torque. It also predicts a relationship between the maximum of the thermal peak and the associated characteristic timescale. By establishing scaling laws governing the evolution of these features with the planet orbital radius and surface pressure, we retrieved numerically this relationship, which is $\maxpeak / \taupeak \scale \smaxis^{-3/2}$. 

The fact that scaling laws match well numerical results reveals that the torque and the tidal frequency can be normalized by the characteristic amplitude and frequency associated with the low-frequency regime. This was confirmed by the rescaling of spectra, which shows that numerical results obtained in all of the treated cases actually describe the same frequency dependence, whatever the star-planet distance and surface pressure. The combination of the parametrized function and scaling laws derived in this work thus leads to a generic empirical formula for the atmospheric tidal torque in the vicinity of synchronous rotation.

In spite of its limitations in the low-frequency regime, the analytic approach remains complementary with GCM calculations owing to the high computational cost of this method (several days of parallel computation on 80 processors are necessary to produce a spectrum with a sampling of 21 points in frequency). Results obtained from simulations can be used to improve the linear analysis, which provides in return a diagnosis of the physical and dynamical mechanisms involved in the tidal response. 

As the study showed evidence of the interest of the numerical method using GCMs in characterizing the atmospheric tidal torque of terrestrial planets, several prospects can be considered for future works. First, the effects of clouds and optical thickness should be investigated owing to their strong impact on the tidal response. The case of an exo-Earth hosting a cloudy atmosphere may be treated in a similar way as the idealized planet of the present study. Second, it would be interesting to better characterize the dependence of the tidal torque on the atmospheric structure using ab initio analytic models. Third, numerical results and the derived generic parametrized function may be coupled to evolutionary models in order to quantify in a realistic way the contribution of the atmosphere to the evolution of the planet rotation over long timescales.

\begin{acknowledgements}
     The authors acknowledge funding by the European Research Council through the ERC grant WHIPLASH 679030. \rec{They also thank the anonymous reviewer for helpful comments that improved the manuscript.}
\end{acknowledgements}

\bibliographystyle{aa} 
\bibliography{auclair-desrotour} 

\appendix 

\section{Normalized spherical harmonics}
\label{app:normSH}

With the notations introduced in \sect{sec:basic_principle}, the normalized spherical harmonics associated with the \rec{degrees} $\llat$ and $\mm$ ($\llat \in \Nset$ and $- \llat \leq \mm \leq \llat$) are defined by \citep[e.g.][]{Arfken2005}

\begin{equation}
\Ylm \left( \col, \lon \right) \define \left( - 1 \right)^\mm \sqrt{\frac{2 \llat +1}{4 \pi} \frac{\left( \llat - \mm \right) !}{\left( \llat + \mm \right) !}}  \Plm \left( \cos \col \right) \expo{\inumber \mm \lon},
\end{equation}

\noindent where the $\Plm$ designate the associated Legendre functions, expressed, for $\xpoly \in \left[ -1,1 \right]$, as

\begin{equation}
\Plm \left( \xpoly \right) \define \left( - 1 \right)^\mm  \left( 1 - \xpoly^2 \right)^{\mm/2}  \DDn{}{\xpoly}{\mm} \Pl \left( \xpoly \right), 
\end{equation}
 
 \noindent the $\Pl$ being the Legendre polynomials, defined by 
 
 \begin{equation}
\Pl \left( \xpoly \right) \define \frac{1}{2^\llat \llat !}  \DDn{}{\xpoly}{\llat} \left[ \left( \xpoly^2 - 1 \right)^\llat \right]. 
 \end{equation}

\section{Linear analytical model for the high-frequency regime}
\label{app:analytical_model}

\comments{Nouvel appendice pour détailler le calcul de la solution analytique utilisée pour la discussion sur le régime des hautes fréquences.}

In \sect{ssec:analytical_model}, we give an expression of the surface pressure anomaly as a function of the tidal frequency (see \eqs{dps1}{dps2}). This expression is a solution of the thermally generated tidal response derived in the framework of the linear theory of atmospheric tides, as described for instance in \cite[]{CL70}. We detail here the calculations that allowed us to obtain it.

In the linear analysis, the wind velocity $\Vvect$, pressure $\pressure$, density $\density$ and temperature $\temp$ are written as the sum of a spherically symmetrical constant component, subscripted $\ibgd$, and a time-dependent small perturbation, identified by the symbol $\delta$. Hence, background equilibrium quantities depend on the radial coordinate only, this later being here the altitude with respect to the planet surface $\zz$. Perturbed quantities are functions of the time~$\time$, altitude $\zz$, colatitude $\col$, and longitude $\lon$. The constant component of $\Vvect$ is ignored, which enforces a solid rotation condition and allows us to simply denote by $\Vvect$ the velocity vector of tidal winds.

The tidal response of the atmosphere in the accelerated frame rotating with the planet is governed by the perturbed momentum equation \citep[e.g.][p.~128]{Siebert1961} 

\begin{equation}
\dd{\Vvect}{\time} + 2 \spinvect \vprod \Vvect  = - \frac{1}{\density} \grad \deltap + \frac{\deltarho}{\density} \gvect,
\end{equation}

\noindent the equation of mass conservation,

\begin{equation}
\Dpart{\density}{\time} + \rhobgd \div \Vvect = 0,
\end{equation}

\noindent the conservation of energy,

\begin{equation}
\frac{\Rspec}{\gad - 1} \Dpart{\temp}{\time} = \frac{\ggravi \Hatm}{\rhobgd} \Dpart{\density}{\time} + \Jtide,
\end{equation}

\noindent and the perfect gas law,

\begin{equation}
\pressure = \density \Rspec \temp ,
\end{equation}

\noindent the symbol $\dd{}{\time}$ designating the partial time derivative, $\grad$ the gradient operator, $\div $ the divergence and $\Dpart{}{\time}  \define \dd{}{\time} + \Vvect \sprod \grad$ the material derivative. 

Note that we have neglected the force resulting from the tidal gravitational potential in the momentum equation as we focus on the thermally generated component of the tidal response. The only source term is thus the net absorbed heat per unit mass $\Jtide$ in the right-hand side of the conservation of energy. Moreover, dissipative processes are not taken into account since they play a role in the low-frequency regime mainly, where their characteristic associated timescales are comparable or greater than the tidal period. The considered regime is thus adiabatic, which is an appropriate approximation for studying the atmospheric tidal response in the high-frequency range, as demonstrated by early studies of the Earth's case \citep[see][for a review]{LC1969}.

Other approximations are convenient to simplify analytic calculations. First, considering that $\Hatm \ll \Rpla$, we assume the hydrostatic approximation, as discussed in \sect{sec:basic_principle}. Second, we neglect Coriolis components associated with a vertical displacement. This approximation, known as the traditional approximation \citep[e.g.][]{Unno1989}, is appropriate provided that the fluid layer is stably-stratified or thin with respect to the planet radius, which is the case here. Thus the preceding equations reduce to

\begin{align}
\label{tide_eq1}
\dd{\Vtheta}{\time} - 2 \spinrate \cos \col \Vphi = & - \frac{1}{\Rpla} \dd{}{\col} \left( \frac{\deltap}{\rhobgd} \right), \\
\label{tide_eq2}
\dd{\Vphi}{\time} + 2 \spinrate \cos \col \Vtheta = &  - \frac{1}{\Rpla \sin \theta} \dd{}{\lon} \left( \frac{\deltap}{\rhobgd} \right), \\
\label{tide_eq3}
\dd{\deltap}{\zz} = & - \ggravi \deltarho, \\
\label{tide_eq4}
\dd{\deltarho}{\time} + \DD{\rhobgd}{\zz} \Vr + \rhobgd \dd{\Vr}{\zz} = & - \rhobgd \divh \Vvect \\
\label{tide_eq5}
\frac{\Rspec}{\gad - 1} \left( \dd{\deltaT}{\time} + \DD{\Tbgd}{\zz} \Vr \right) = & \frac{\ggravi \Hatm}{\rhobgd} \left( \dd{\deltarho}{\time} + \DD{\rhobgd}{\zz} \Vr \right) + \Jtide, \\
\label{tide_eq6}
\frac{\deltap}{\pbgd} = & \frac{\deltarho}{\rhobgd} + \frac{\deltaT}{\Tbgd},
\end{align}

\noindent where the horizontal part of the velocity divergence $\divh \Vvect$ is defined by 

\begin{equation}
\divh \Vvect \define \frac{1}{\Rpla \sin \col} \left[  \dd{}{\col} \left( \Vtheta \sin \col \right) + \dd{\Vphi}{\lon} \right]. 
\label{divh_velocity}
\end{equation}

Beside, we introduce the atmospheric pressure heigh scale,

\begin{equation}
\Hatm \define \frac{\Rspec \Tbgd}{\ggravi } ,
\label{Hatm}
\end{equation}

\noindent and the pressure altitude,

\begin{equation}
\xx \define \int_0^z \frac{\ddroit z}{H},
\label{xpress}
\end{equation}

\noindent which will be used instead of $\zz$ for convenience in the following.

Since it is periodical in time and longitude, the perturbation can be written as a combination of Fourier modes, each one being associated with a forcing frequency $\ftide$ and a longitudinal \rec{degree} $\mm$. Fourier coefficients are functions of the colatitude ($\col$) and altitude ($\zz$). Any perturbed quantity $\quanti$ is thus expressed as

\begin{equation}
\quanti \left( \xx , \col , \lon \right) = \sum_{\ftide,\mm} \quanti^{\mm , \ftide} \left( \zz , \col \right)  \expo{\inumber \left( \ftide \time + \mm \lon \right)}
\end{equation}

\noindent where the $\quanti^{\mm , \ftide}$ are the Fourier coefficients of the expansion. Under the assumed approximations, \eqs{tide_eq1}{tide_eq2} are decoupled from the radial momentum equation, which allows us to write $\Vthetams$ and $\Vphims$ as functions of $\deltapn/\rhobgd$. By substituting horizontal winds by the obtained expressions in $\divh \Vvect $ (\eq{divh_velocity}), and introducing the variable 

\begin{equation}
\Gpress = - \frac{1}{\gad \pbgd} \Dpart{\pressure}{\time},
\label{Gpress}
\end{equation}

\noindent the whole system of governing equations given by \eqsto{tide_eq1}{tide_eq6} can be put after some manipulations into the form \citep[see][]{LC1969}

\begin{equation}
\Fms \Gms = \Laplace \left[ \left(  \DD{\ln \Hatm}{\xx} + \kad \right) \Gms - \frac{\kad \Jms}{\gad \ggravi \Hatm}  \right].
\end{equation}

\noindent Here, $\Fms$ is an operator depending on the $\xx$ coordinate only and $\Laplace$ the Laplace's tidal operator, which depends on the $\col$ coordinate only and is formulated as \citep[e.g.][]{LS1997}

\begin{align}
 \label{hough_expansion}
\Laplace \define & \frac{1}{\sin \col} \dd{}{\col} \left( \frac{\sin \col}{1 - \spinpar^2 \cos^2 \col} \dd{}{\col} \right) \\
 & - \frac{1}{1 - \spinpar^2 \cos^2 \col} \left( \mm \spinpar \frac{1 + \spinpar^2 \cos^2 \col}{1 - \spinpar^2 \cos^2 \col} + \frac{\mm^2}{\sin^2 \col} \right), \nonumber
\end{align}

\noindent the quantity $\spinpar \define 2 \spinrate / \ftide$ designating the so-called spin parameter. 

The above separation of coordinates allows us to expand the Fourier coefficients of $\Gpress$ as

\begin{equation}
\Gms \left( \xx , \col \right) = \sum_n \Gmsn \left( \xx \right) \Thetan \left( \col \right).
\label{Gms}
\end{equation}

\noindent The Fourier coefficients of $\deltap$, $\deltarho$ and $\deltaT$ are written likewise. In \eq{Gms}, the integer $\nn$ corresponds to the latitudinal wavenumber \rec{of a spherical mode modified by the planet spin rotation} (\rec{in the static case, where $\spinpar = 0$, $\nn = \llat - \left| \mm \right|$}). Similarly, the Hough functions $\Thetan$ correspond to the associated Legendre functions (see \append{app:normSH}) modified by rotation. Hough functions are the eigenvectors of the Laplace operator \citep[e.g.][]{LS1997}. They are associated with the eigenvalues $\Lambdan$ through the relationship 

\begin{equation}
\Laplace \Thetan = \Lambdan \Thetan,
\end{equation}

\noindent and determine the equivalent depth of the mode associated with the triplet $\left( \nn , \mm , \ftide \right)$ \citep[e.g.][]{Taylor1936},

\begin{equation}
\heqmsn \define \frac{\Rpla^2 \ftide^2}{\ggravi \Lambdan}. 
\label{heqmsn}
\end{equation}

In the absence of resonances, the semidiurnal tidal response is generally dominated by the fundamental gravity mode, indicated by $\nn = 0\,$\footnote{We follow here the indexing notation by \cite{LS1997}, which associates g-modes with positive $\nn$ and r-modes to strictly negative $\nn$.}, which corresponds to the associated Legendre function $\LegF{2}{2}$ in the static case \citep[e.g.][]{ADL2018a}. In the high-frequency regime, $\Houghval{0}{2}{\spinpar} \approx \Houghval{0}{2}{1} \approx 11.1$, from the moment that $ \norb \ll \abs{\spinrate} $.

 Note that this value, denoted by $\Lambdao$ in the following, can be modified by dissipative processes. For instance, by including friction with the planet surface using a Raleigh drag of constant characteristic frequency $\fdrag$, one may show that the eigenvalue of the modes tends to the value of the static case, that is $\Houghval{0}{2}{\spinpar} \approx 6$, if $\ftide / \fdrag \rightarrow 0$ \citep[see e.g.][]{Volland1974a,ADLM2017b}.

As we focus on the $\nn = 0$ mode, we can drop the subscripts and superscripts $\nn$, $\mm$ and $\ftide$ to lighten notations. The function $\Gmsn$ is now simply denoted by $\Gpress$, and so on for the tidal heat source, pressure, density, temperature, wind velocity components, eigenvalues and equivalent depths. The usual change of variable $\Gpress = \expo{\xx/2} \yvar $ leads to the vertical structure equation in its canonical form,

\begin{equation}
\DDn{\yvar}{\xx}{2} + \kvert^2 \, \yvar = \frac{\kad \Jtide}{\gad \ggravi \heq} \expo{-\xx / 2},
\label{vertical_structure}
\end{equation} 

\noindent where we have introduced the dimensionless vertical wavenumber $\kvert$, defined by 

\begin{equation}
\kvert^2 \define \frac{1}{4}  \left[ \frac{4}{\heq} \left( \kad \Hatm + \DD{\Hatm}{\xx} \right) -1 \right].
\end{equation}

The vertical structure equation describe the behaviour of a forced harmonic oscillator, and $\kvert$ thus corresponds to the inverse of a length scale of the variation of perturbed quantities across the vertical coordinate. Since the tidal response is adiabatic, $\kvert^2 \in \Rset $, and its sign directly determines the nature of waves across the vertical axis. The condition $\kvert^2 > 0 $ indicates a propagating mode. Conversely, $\kvert^2 <0$ corresponds to an evanescent mode. 

Computing analytic solutions turns out to be a very challenging problem except for a few simplified configurations. Therefore, we treat here the idealized case of the isothermal atmosphere, which is one of these configurations. Note that we acknowledge the limitations of this academic atmospheric structure regarding real ones, where convective instability leads to a strong temperature gradient near the planet surface. However, this approach appears to be sufficient for the purpose of this appendix. 

In the isothermal approximation, the temperature profile is supposed to be invariant with the radial coordinate. In light of \eq{Hatm}, it immediately follows that $\Hatm $ is a constant, $\ddroit \Hatm / \ddroit \xx = 0$, and 

\begin{equation}
\kvert^2 = \frac{1}{4} \left[ \frac{4 \kad \Hatm}{\heq} - 1  \right]. 
\end{equation}

\noindent The above expression shows the existence of a turning point for $\heq = 4 \kad \Hatm$, where the sign of $\kvert^2$ changes. This turning point occurs at the frequency

\begin{equation}
\fturning = \sqrt{\frac{4 \kad \Hatm \Lambdao \ggravi}{\Rpla^2}}.
\end{equation}

In the reference case of the study, GCM simulations provide $\Tsurf \approx 316 $\units{K}, which, combined with $\Rspec \approx 297$\units{J.kg^{-1}.K^{-1}}, gives $H \approx 10.6$\units{km} in the isothermal approximation. An estimation of the normalized frequency $\fturningnorm = \fturning / \left( 2 \norb \right)$ using \eq{Hatm} thus gives $\fturningnorm \approx 270 $, showing that the turning point occurs in the high-frequency range and must therefore be taken into account in the calculation of an analytical solution. The condition $\kvert^2 > 0$ ($\abs{\ftide} < \fturning$) corresponds to an oscillatory regime, while $\kvert^2 < 0$ ($\abs{\ftide} > \fturning$) corresponds to an evanescent one. 

To solve the vertical structure equation, we have to choose a vertical profile for the tidal heat power per unit mass $\Jtide$. Following \cite{Lindzen1968}, we opt for a profile of the form

\begin{equation}
\Jtide = \Jsurf \expo{- \tauJ \xx},
\label{Jtide}
\end{equation}

\noindent where $\Jsurf$ stands for the heat absorbed at the planet surface and $\tauJ$ is a dimensionless optical depth characterizing the decay of heating across the vertical coordinate. This profile is derived from the Beer's law \citep[e.g.][]{Heng2017} applied to an isothermal atmosphere of constant extinction coefficient, where the variation of flux is supposed to be proportional to the local pressure,

\begin{equation}
 \frac{\ddroit \flux}{\flux} = - \tauJ \expo{-\xx} \ddroit \xx. 
\end{equation}

Integrating this expression from the surface to $\xx$ and introducing the surface flux $\fluxsurf$, we obtain 

\begin{equation}
\flux = \fluxsurf \expo{\tauJ \left( \expo{- \xx} - 1 \right)},
\end{equation}

\noindent which, linearized in the vicinity of the surface ($\xx \ll 1$) and assuming $\tauJ \gg 1$, can be approximated by 

\begin{equation}
\flux \approx \fluxsurf \expo{- \tauJ \xx}. 
\end{equation}

\noindent Since the absorbed heat per unit mass is given by 

\begin{equation}
\Jtide = -  \frac{1}{\Hatm \rhobgd} \DD{\flux}{\xx},
\end{equation}

 \noindent and $\rhobgd \left( \xx \right) \scale \expo{- \xx}$, we retrieve \eq{Jtide}, with 
 
 \begin{equation}
 \Jsurf = \frac{\ggravi \tauJ}{\psurf} \fluxsurf. 
 \end{equation}

With the preceding choices and approximations, the general solution of \eq{vertical_structure} can be written as

\begin{equation}
\yvar = \Aint \expo{\inumber \kvert \xx} + \Bint \expo{ - \inumber \kvert \xx} + \yspec \expo{- \left( \tauJ + \frac{1}{2} \right) \xx},
\end{equation}

\noindent the notations $\Aint$ and $\Bint$ designating integrating constants, and $\yspec$ the constant factor of the particular solution. The integrating constant $\Bint$ can be easily eliminated by setting the appropriate upper boundary condition. First, consider the oscillatory case, where $\kvert^2 > 0$. If the convention $\kvert > 0$ is adopted, the first term of the solution is associated with upward propagation of energy and the second term with downward propagation \citep[][]{Wilkes1949,LC1969}. 

Since there is no energy source at $\xx = + \infty$, we expect that the energy propagation is only upward at the upper boundary. It immediately follows that $\Bint = 0 $. This boundary condition is known as the radiation condition \citep[][]{LC1969}. When $\kvert^2 < 0 $, the radiation condition is not appropriate any more since $\kvert$ is now a pure imaginary number. In this case the wave is evanescent, and the first and second term of the solution correspond to decaying and diverging components, respectively, from the moment that the convention $\imag{\kvert} > 0$ is adopted. In this case, a non-divergence condition is applied at the upper boundary on the energy flux \citep[e.g.][]{Wilkes1949,Lindzen1968}, which leads to $\Bint = 0$ again. 

At the lower boundary, the fact that fluid particles cannot penetrate through the planet surface is enforced by a rigid wall condition. The vertical velocity is set to $\Vr = 0 $ at $\xx = 0$. As the vertical velocity is given by \citep[][]{LC1969}

\begin{equation}
\Vr = \gad \heq \expo{\xx / 2} \left[ \DD{\yvar}{\xx} + \left( \frac{\Hatm}{\heq} - \frac{1}{2} \right) \yvar \right],
\end{equation}

\noindent the rigid wall condition is expressed as 

\begin{equation}
\DD{\yvar}{\xx} + \left( \frac{\Hatm}{\heq} - \frac{1}{2} \right) \yvar = 0 ,
\end{equation}

\noindent at $\xx = 0$. Thus the solution is

\begin{equation}
\yvar = \frac{\kad \Jsurf}{\gad \ggravi \heq \left( \tauJ \left( \tauJ + 1 \right) + \frac{\kad \Hatm}{\heq} \right) }  \left[  \frac{1 + \tauJ - \frac{\Hatm}{\heq}}{\inumber \kvert + \frac{\Hatm}{\heq} - \frac{1}{2} } \expo{\inumber \kvert \xx} + \expo{- \left( \tauJ + \frac{1}{2} \right) \xx }  \right] ,
\end{equation}

\noindent with the conventions $\kvert > 0$ if $\kvert^2 > 0$ and $\imag{\kvert} > 0$ else. We recover here the solution previously obtained by \cite{Lindzen1968} for different values of $\tauJ$.

The surface pressure variation is readily deduced from $\yvar$ using \eq{Gpress}, 

\begin{equation}
\frac{\deltapsurf}{\psurf} = \inumber  \frac{\gad}{\ftide} \yvar \left( 0 \right).
\end{equation}

\noindent Taking the imaginary part of the preceding expression, we finally obtain, for $\kvert^2 >0 $ (i.e. $\abs{\ftide} < \fturning$),

\begin{equation}
\imag{\deltapsurf} = \ftide^{-1} \psurf \frac{\kad \Jsurf}{g \Hatm} \frac{\frac{\Hatm}{\heq} \left( \tauJ + \frac{1}{2} + \kad \right) - \frac{1}{2} \left( \tauJ + 1 \right) }{\left[ \tauJ \left( \tauJ + 1 \right) + \frac{\kad \Hatm}{\heq}  \right] \left( \frac{\Hatm}{\heq} - \frac{1}{\Gamma_1} \right) } ,
\label{sol1}
\end{equation}

\noindent and, for $\kvert^2 <0 $ (i.e. $\abs{\ftide} > \fturning$), 

\begin{equation}
\imag{\deltapsurf} =  \frac{ \ftide^{-1} \psurf \frac{\kad \Jsurf}{\ggravi \heq} \left[  \tauJ + \frac{1}{2} \left( 1 - \sqrt{1 - \frac{4 \kad \Hatm}{\heq}} \right) \right]}{ \left[ \tauJ \left( \tauJ + 1 \right) + \frac{\kad \Hatm}{\heq}  \right] \left[ \frac{\Hatm}{\heq} - \frac{1}{2} \left( 1 + \sqrt{1 - \frac{4 \kad \Hatm}{\heq}} \right) \right] }. 
\label{sol2}
\end{equation}

Although they have been obtained using an idealized atmospheric structure, these two expressions of the semidiurnal surface pressure anomaly inform us about the frequency-behaviour of the thermally generated atmospheric tidal torque in the high-frequency range. First consider \eq{sol1}. By assuming that the condition $ 1 \ll \Hatm / \heq \ll \kad^{-1} \tauJ^2 $ is satisfied, that is for $\abs{\ftide} \ll \fturning$ and a small thickness of the heated layer, the torque scales as $\torque \scale \sigma^{-1}$. This corresponds to what may be observed in \fig{fig:spectres_model} in the interval $30 < \abs{\omeganorm} < 200$. In the zero-frequency limit, $\heq \rightarrow 0$ by scaling as $\heq \scale \sigma^2$ if the dependence of the eigenvalue $\Lambdao$ on the forcing frequency is ignored (see \eq{heqmsn}). It follows that $\torque \scale \sigma$ in this asymptotic regime. The transition between the two regimes occurs for $\abs{\ftide} \approx \fthermal $ with 

\begin{equation}
\fthermal = \frac{\fturning}{2 \sqrt{\tauJ \left( \tauJ + 1 \right)}} \ll \fturning.
\end{equation}

Since $\Hatm / \heq \gg 1 $ when $\abs{\ftide} \ll \fLamb $ (let us remind that $\fLamb$ designates the Lamb frequency given by \eq{fLamb}), \eq{sol1} can be put into the form of the Maxwell function, 

\begin{equation}
\imag{\deltapsurf}  \approx \frac{2 \maxthermal \tauthermal \ftide }{1 + \left( \tauthermal \ftide \right)^2},
\end{equation}

\noindent where the associated characteristic timescale $\tauthermal$ and maximal amplitude $\maxthermal$ are

\beqtwo{\tauthermal = \fthermal^{-1}}{\maxthermal = \frac{\kad \Jsurf \left( \tauJ + \frac{1}{2} + \kad \right)}{ \ggravi \Hatm \fturning \sqrt{\tauJ \left( \tauJ + 1 \right)} } \psurf .}

Let us now consider the case where $\kvert^2 < 0$, described by \eq{sol2}. We notice that the surface pressure anomaly diverges for $\heq = \gad \Hatm$. This feature was discussed by early studies \citep[e.g.][]{LB1972}. It corresponds to the resonance associated with an horizontally propagating acoustic mode of large wavelength known as the Lamb mode \citep[e.g.][]{Lindzen1968,Bretherton1969,LB1972,Platzman1988,Unno1989}. In the reference case, with the values previously used to estimate $\fturningnorm$, we obtain $\omegaLamb \approx 308$. The resonance can be observed in \fig{fig:spectres_model} for a smaller frequency ($\omegaLamb \approx 260 $) because of the departure between the isothermal atmospheric structure and the realistic one computed in GCM simulations, as discussed in \sect{ssec:resonance_lamb}.

\section{Dependence of the tidal torque on the atmospheric composition}
\label{app:dependence_composition}
The atmospheric torque generated by the thermal tide depends on the atmospheric composition, which has a strong impact on the vertical distribution of the tidal heating through clouds formation and the optical thickness of the gas mixture. In the study, we treat the case of a terrestrial planet hosting a cloudless $\nitrogen$-dominated atmosphere with a small amount of $\carbondiox$. Hence, we ignore the effects of clouds and compute the thermal tide of an optically thin atmosphere in the visible frequency range, where the major part of the stellar flux reaches the planet surface without being absorbed. 

 In this appendix, we consider the case of a planet hosting a Venus-like $\carbondiox$-dominated atmosphere with a mixture of water and sulfuric acid ($\sulfuricacid$) in the same reference configuration ($\smaxis = \smvenus$ and $\psurf = 10$\units{bar}). We do not attempt to reproduce exactly the composition and dynamics of the Venus atmosphere, which is a complex problem beyond the scope of this study \citep[see e.g.][]{Lebonnois2010,Lebonnois2016}, but to simply retrieve its main features (optical opacity, clouds absorption, etc.). As a consequence, we opt for a generic approach excluding a fine tuning of the atmospheric properties. We set the thermal capacity per unit mass of the gas ($\Cpgaz$) to 1000\units{J.kg^{-1}.K^{-1}}, which is the typical value of $\Cpgaz$ in the case of Venus \citep[e.g.][]{Seiff1985}, where the parameter decreases from 1181\units{J.kg^{-1}.K^{-1}} near the surface to 904\units{J.kg^{-1}.K^{-1}} at an altitude of 50~km. 
 
 
\begin{figure}[htb]
   \centering
   \includegraphics[width=0.49\textwidth,trim = 1.5cm 2.2cm 1.5cm 2.1cm,clip]{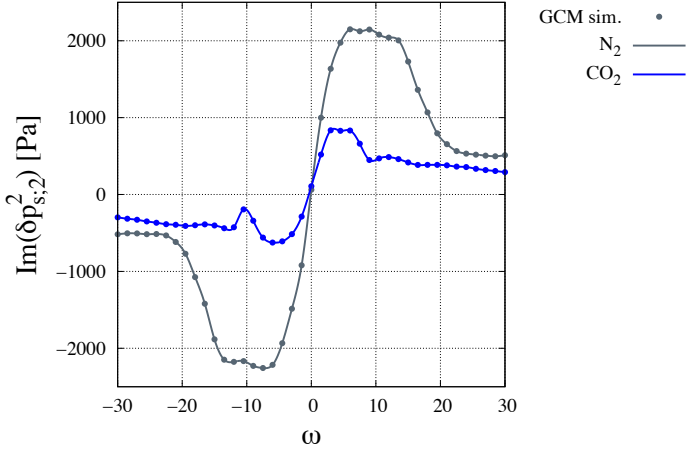} 
   \caption{Imaginary part of the surface pressure variation created by the semidiurnal thermal tide as a function of the normalized tidal frequency $\omeganorm = \left( \spinrate - \norb \right) / \norb$ for a Venus-sized planet of surface pressure $\psurf = 10$\units{bar} and orbital radius $\smaxis = \smvenus$. Results obtained using GCM simulations are designated by points. Spectra are plotted for two cases: the reference case of the study, where the atmosphere is mainly composed of $\nitrogen$ (black line), and the case treated in \append{app:dependence_composition}, where the atmosphere is composed of $\carbondiox$ with a mixture of water vapour and sulfuric acid.  }
       \label{fig:composition}%
\end{figure}

 Similarly, the mean molecular mass is set to $ 43.45$\units{g.mol^{-1}}, and the volume mixing ratio of water vapor to 20\units{ppm} \citep[][]{Moroz1979}. We set the diameter of water particles to 3\units{\mu m}, which is a typical value in the lower cloud \citep[e.g.][]{KH1980}. To take into account the impact of sulfuric acid on the saturation pressure of water vapor $\psatwater$, we use the prescription given by \cite{GV1964} for aqueous sulfuric acid (see Eq.~(24) of their article). This prescription is written as a function of the local temperature $\temp$, as
 
\begin{equation}
\ln \left( \psatwater \right) = A \ln \left( \frac{298}{\temp} \right) + \frac{B}{\temp} + C + D \temp + E \temp^2,
\end{equation}

\noindent where $A$, $B$, $C$, $D$, and $E$ are empirical constants. The optical properties of the atmosphere used to compute radiative transfers are pre-computed using the HITRAN 2008 database \citep[][]{Rothman2009} for the Venus atmospheric mixture (instead of the Earth mixture used in the study). The spectrum of the atmospheric tidal torque due to the semidiurnal tide is plotted in \fig{fig:composition} with the spectrum of the $\nitrogen$ reference case for comparison. 

Because of the opacity of the atmosphere in the visible range, the fraction of the incoming stellar flux reaching the planet surface is less in the case of the $\carbondiox$ atmosphere than in the case of the $\nitrogen$ atmosphere. Particularly, the resonance peak is strongly attenuated. We also observe a greater impact of Coriolis effects, the asymmetry of the tidal torque between negative and positive frequency ranges being more significant. 

This difference can be explained by the vertical distribution of tidal heating. As mentioned above, the major part of the stellar flux reaches the planet surface if the atmosphere is composed of $\nitrogen$, which means that the tidal torque is mainly due to density variations occurring in the vicinity of the ground, where friction predominate over Coriolis forces. In the case of the $\carbondiox$ atmosphere, an important fraction of the incoming energy flux is absorbed at the cloud level. The contribution of this fraction is thus strongly affected by Coriolis effects through zonal mean flows characterizing the equilibrium dynamical state. 


\section{Simplified ab initio analytical model for the ground thermal response}
\label{app:surface_response}

As mentioned in \sect{ssec:coupling}, we follow along the line by \cite{Bernard1962} to study the thermal response of the planet surface. In this approach, we consider the surface-atmosphere interface, located at the altitude $\zz = 0$, and write the power flux budget for a small perturbation in the framework of a frequency linear analysis. Hence, any quantity $\quanti$ can be expressed as $\quanti = \quantisig \expo{\inumber \ftide \time}$, where $\ftide$ is the forcing frequency introduced in \sect{sec:basic_principle}. In the following, we omit the superscript $\ftide$ and use $\quanti$ in place of $\quantisig$, given that we work in the frequency domain. 

A variation of the effective incoming stellar flux (i.e. where the reflected component has been removed), denoted $\deltaFinc$, is absorbed by the planet surface. A fraction $\deltaQsol$ of this power is transmitted to the ground by thermal conduction, and an other fraction, $\deltaQatm$, is transmitted to the atmosphere through turbulent thermal diffusion. Finally, the increase of surface temperature $\deltaTsurf$ generated by $\deltaFinc$ induces a radiative emission, $\deltaFrad$, which is expressed as $\deltaFrad = 4 \SBconstant \Tsurf^3 \deltaTsurf$ in the black body approximation (we recall that $\SBconstant$ and $\Tsurf$ are the Stefan-Boltzmann constant and mean surface temperature introduced in \sect{ssec:extraction}, respectively). Since the atmosphere is heated by both the incoming stellar flux and the surface thermal forcing, it undergoes a radiative cooling, similarly as the surface. The flux emitted downward to the surface is denoted $\deltaFatm$. Thus, the power budget of the thermal perturbation at the interface is expressed as

\begin{equation}
 \deltaFinc - 4 \SBconstant \Tsurf^3 \deltaTsurf + \deltaFatm -\deltaQsol - \deltaQatm = 0.
 \label{budget_flux}
\end{equation}

To study the surface thermal response without having to consider the full atmospheric tidal response in its whole complexity, it is necessary to ignore the coupling induced by $\deltaFatm$. This amounts to assuming either that the emission of the atmosphere toward the planet surface is negligible, or that it is proportional to $\deltaTsurf$ \citep[see][]{Bernard1962}. Thus, introducing the surface effective emissivity $\emisurf$, radiative terms can be reduced to 

\begin{equation}
4 \SBconstant \Tsurf^3 \deltaTsurf - \deltaFatm =  4 \SBconstant \Tsurf^3 \emisurf \deltaTsurf.
\end{equation}

The next step consists in defining the thermal exchanges resulting from diffusive processes, $\deltaQsol$ and $\deltaQatm$. These flux are directly proportional to the gradient of the temperature profile anomaly in the vicinity of the interface, and are expressed as 

\begin{equation}
\begin{array}{rll}
 \deltaQsol = \ksol \left( \dd{\deltaT}{\zz}  \right)_{\zz = 0^-}, & \mbox{and} & \deltaQatm = - \katm \left( \dd{\deltaT}{\zz} \right)_{\zz = 0^+},
\end{array}
\label{deltaQ}
\end{equation}

\noindent where $\dd{}{\zz}$ designates the partial derivative in altitude, $\deltaT$ the profile of temperature variations, and $\ksol$ and $\katm$ the thermal conductivities of the ground and of the atmosphere at $\zz = 0$, respectively. By introducing the mean density profile of the planet $\rhobgd$ and the thermal capacity per unit mass of the ground $\Csol$ (the analogous parameter for the atmosphere being $\Cpgaz$; see \sect{sec:physical_setup}), the corresponding diffusivities can be defined by

\begin{equation}
\begin{array}{rll}
 \Ksol \define \dfrac{\ksol}{\rhobgd \left( 0^- \right) \Csol} & \mbox{and} &  \Katm \define \dfrac{\katm}{\rhobgd \left( 0^+ \right) \Cpgaz}.
\end{array}
\end{equation}

Temperature variations in the vicinity of the interface are described by the heat transport equation. We assume that diffusive processes predominates in the $\zz \rightarrow 0$ limit. Moreover, since the typical horizontal length scale is far greater than the vertical one in the thin layer approximation, the horizontal component of the Laplacian describing diffusive processes can be neglected with respect to the vertical component $\dd{}{\zz}$ in both the solid and atmospheric regions. It follows that 

\begin{align}
& \inumber \ftide \deltaT = \Ksol \ddd{\deltaT}{\zz}{\zz}  &  {\rm for} \ \zz \leq 0, \\
& \inumber \ftide \deltaT = \Katm \ddd{\deltaT}{\zz}{\zz} & {\rm for} \ \zz >0.
\end{align}

Solving these two equations with constant $\Ksol$ and $\Katm$, and ignoring the diverging term in solutions, we end up with 

\begin{align}
& \deltaT \left( \zz \right) = \deltaTsurf \expo{\left[ 1 + \sign \left( \ftide \right) \inumber \right] \zz / \skinhsol}, & {\rm for} \ \zz \leq 0, \\
& \deltaT \left( \zz \right) = \deltaTsurf \expo{- \left[ 1 + \sign \left( \ftide \right) \inumber \right] / \skinhatm } & {\rm for} \  \zz > 0,
\label{deltaTdiff}
\end{align}

\noindent where we have introduced the frequency-dependent skin thicknesses of heat transport by thermal diffusion in the ground ($\skinhsol$) and atmosphere ($\skinhatm$), expressed as 

\begin{equation}
\begin{array}{rcl}
 \skinhsol = \sqrt{ \dfrac{2 \Ksol}{\abs{\ftide}}} & \mbox{and} & \skinhatm = \sqrt{ \dfrac{2 \Katm}{\abs{\ftide}}}. 
 \end{array}
\end{equation}

The transfer function $\Bsolf$ defined by \eq{Bs_model} comes straightforwardly when the profiles of $\deltaT$ are substituted by \eq{deltaTdiff} in \eqs{deltaQ}{budget_flux} successively. Introducing $\Bsolstat = \left( 4 \SBconstant \Tsurf^3 \emisurf  \right)^{-1}$, we obtain $\deltaTsurf = \Bsolf \deltaflux$, with 

\begin{equation}
\Bsolf = \frac{\Bsolstat}{1 + \left[ 1 + \sign \left( \ftide \right) \inumber \right] \sqrt{\tausurf \abs{\ftide}}},
\label{Bs_model2}
\end{equation}

\noindent where $\tausurf$ can be interpreted as the characteristic timescale of the surface thermal response. The parameter $\tausurf$ is a function of the thermal inertia of the ground $\Isol \define \rhobgd \left( 0^- \right) \Csol \sqrt{\Ksol}$ and of the atmosphere $\Iatm \define \rhobgd \left( 0^+ \right) \Cpgaz \sqrt{\Katm} $,

\begin{equation}
\tausurf \define \frac{1}{2} \left( \frac{\Isol + \Iatm}{4 \SBconstant \Tsurf^3 \emisurf} \right)^2. 
\end{equation}

The above expression shows that $\tausurf$ compares the efficiency of diffusive processes to that of the radiative cooling of the surface. The thermal time increases with the interface thermal inertia and decays when the surface temperature increases, scaling as $\tausurf \scale \Tsurf^{-6}$. 

The expression of $\Bsolf $ given by \eq{Bs_model2} highlights two asymptotic regimes. In the low-frequency regime, where $\abs{\ftide} \ll \tausurf^{-1}$, the surface responds instantaneously to the forcing $\delta F$, leading to a surface temperature oscillation in phase with the incoming stellar flux. At $\ftide = 0$, the incoming flux is equal to the radiative flux ($\deltaFinc = 4 \SBconstant \Tsurf^3 \emisurf$), and $\Bsolf = \Bsolstat$. In the hight-frequency regime, where $\abs{\ftide} \gg \tausurf^{-1}$, the amplitude of the surface temperature variations decays and tends to zero in the limit $\abs{\ftide \tausurf} \rightarrow + \infty $. At the transition, that is $\abs{\ftide}= \tausurf^{-1} $, $\real{ \Bsolf}= \left( 2/5 \right) \Bsolstat $ and $\imag{\Bsol} = - \left( 1/5 \right) \Bsolstat $.

As discussed in \sect{ssec:coupling}, the transfer function obtained using GCM simulations is well approximated by \eq{Bs_model2} in the low-frequency regime (see \fig{fig:Bgr_Nyquist}). But it diverges from the model when the forcing frequency increases, typically for $\abs{\ftide} \gtrsim \tausurf^{-1}$. This divergences seems to result mainly from the fact that we ignored the radiative coupling between the atmosphere and the surface associated with $\deltaFatm$, although it may be very strong. Particularly, this is the case for resonances of the atmospheric tidal response, where $\deltaFatm$ is increased similarly as the amplitude of pressure and temperature oscillations.


\section{\rec{Tables of values obtained with GCM simulations for the exploration of the parameter space}}

\begin{table*}[htb]
 \textsf{\caption{\label{tableS1} Values of the imaginary part of the semidiurnal surface pressure anomaly $\deltapsquad$ (Pa) obtained from GCM simulations in study~1, where $\psurf = 10$\units{bar}, and used to plot the spectra of \fig{fig:spectres_GCM} (top panels). The first column corresponds to the normalized tidal frequency $\omeganorm = \left( \spinrate - \norb \right) / \norb$.  }}
\centering
\small{
    \begin{tabular}{c r r r r r r r r }
      \hline
      \hline 
      $\omega$ & 0.3~AU & 0.4~AU & 0.5~AU & 0.6~AU & 0.7~AU & $\smvenus$ & 0.8~AU & 0.9~AU  \\
      \hline \\
-30 & -473.07 & -545.09 & -539.98 & -533.95 & -525.30 & -517.29 & -471.78 & -434.51 \\ 
 -27 & -511.43 & -585.93 & -607.25 & -595.19 & -512.35 & -507.07 & -465.63 & -482.23 \\ 
 -24 & -642.57 & -644.54 & -685.95 & -580.62 & -520.41 & -513.39 & -525.40 & -651.49 \\ 
 -21 & -704.17 & -744.95 & -703.56 & -582.58 & -592.16 & -611.03 & -772.29 & -1035.44 \\ 
 -18 & -804.42 & -824.14 & -666.45 & -696.53 & -1020.13 & -1074.53 & -1293.65 & -1489.11 \\ 
 -15 & -952.30 & -778.00 & -855.21 & -1309.89 & -1773.11 & -1903.43 & -1924.55 & -1760.51 \\ 
 -12 & -977.99 & -1063.36 & -1732.51 & -2454.48 & -2273.65 & -2199.45 & -2027.54 & -1890.20 \\ 
 -9 & -1472.86 & -2614.20 & -2777.14 & -2534.91 & -2237.35 & -2243.82 & -2102.46 & -1926.23 \\ 
 -6 & -3822.28 & -3250.69 & -2932.75 & -2575.46 & -2274.10 & -2203.40 & -2007.00 & -1750.41 \\ 
 -3 & -3588.82 & -2847.24 & -2059.37 & -1753.74 & -1530.60 & -1486.34 & -1389.68 & -1213.63 \\ 
 0 & 334.42 & 234.58 & 151.64 & 89.75 & 55.15 & 49.50 & 32.02 & 15.13 \\ 
 3 & 3374.24 & 2785.68 & 2297.94 & 1964.36 & 1691.86 & 1638.27 & 1482.97 & 1241.96 \\ 
 6 & 3654.44 & 3058.04 & 2678.35 & 2433.68 & 2210.35 & 2145.61 & 1960.61 & 1728.26 \\ 
 9 & 1529.77 & 2386.06 & 2634.64 & 2356.54 & 2185.29 & 2135.41 & 2020.46 & 1839.10 \\ 
 12 & 973.74 & 1182.24 & 1720.70 & 2255.56 & 2092.64 & 2043.52 & 1925.20 & 1832.17 \\ 
 15 & 941.31 & 808.44 & 932.12 & 1302.36 & 1654.98 & 1732.38 & 1796.60 & 1618.30 \\ 
 18 & 797.69 & 812.74 & 694.15 & 762.96 & 969.00 & 1068.94 & 1276.22 & 1492.46 \\ 
 21 & 686.87 & 737.32 & 688.13 & 593.61 & 640.17 & 651.72 & 779.55 & 989.06 \\ 
 24 & 627.03 & 643.14 & 658.42 & 603.18 & 533.36 & 531.05 & 558.75 & 674.29 \\ 
 27 & 506.85 & 585.93 & 603.91 & 596.34 & 512.01 & 504.31 & 482.52 & 496.22 \\ 
 30 & 486.10 & 545.09 & 538.44 & 534.47 & 524.68 & 511.96 & 470.95 & 445.30 \\ 
       \vspace{0.1mm}\\
       \hline
\end{tabular}
}
\end{table*}

\begin{table}[htb]
 \textsf{\caption{\label{tableS2} Values of the imaginary part of the semidiurnal surface pressure anomaly $\deltapsquad$ (Pa) obtained from GCM simulations in study~2, where $\smaxis = \smvenus$, and used to plot the spectra of \fig{fig:spectres_GCM} (bottom panels). The first column corresponds to the normalized tidal frequency $\omeganorm = \left( \spinrate - \norb \right) / \norb$.  }}
\centering
\small{
    \begin{tabular}{c r r r r}
      \hline
      \hline 
      $\omega$ & 1~bar & 3~bar & 10~bar & 30~bar   \\
      \hline \\
-30 & -354.77 & -335.07 & -517.29 & -541.19 \\ 
 -27 & -506.33 & -486.33 & -507.07 & -547.23 \\ 
 -24 & -611.39 & -741.93 & -513.39 & -636.13 \\ 
 -21 & -673.92 & -1022.06 & -611.03 & -700.94 \\ 
 -18 & -664.71 & -1213.72 & -1074.53 & -737.02 \\ 
 -15 & -678.58 & -1242.91 & -1903.43 & -767.81 \\ 
 -12 & -680.65 & -1272.90 & -2199.45 & -1607.23 \\ 
 -9 & -665.48 & -1283.57 & -2243.82 & -2830.42 \\ 
 -6 & -495.17 & -1004.48 & -2203.40 & -3252.87 \\ 
 -3 & -322.70 & -657.66 & -1486.34 & -2989.54 \\ 
 0 & 25.37 & 37.06 & 49.50 & 51.20 \\ 
 3 & 358.98 & 712.00 & 1638.27 & 2989.06 \\ 
 6 & 504.55 & 1024.26 & 2145.61 & 3020.61 \\ 
 9 & 610.53 & 1249.16 & 2135.41 & 2598.07 \\ 
 12 & 679.89 & 1229.86 & 2043.52 & 1640.01 \\ 
 15 & 660.94 & 1198.29 & 1732.38 & 961.75 \\ 
 18 & 642.94 & 1179.86 & 1068.94 & 734.32 \\ 
 21 & 638.42 & 1003.31 & 651.72 & 703.06 \\ 
 24 & 586.54 & 726.95 & 531.05 & 632.88 \\ 
 27 & 489.53 & 488.95 & 504.31 & 551.61 \\ 
 30 & 352.31 & 371.83 & 511.96 & 539.69 \\ 
        \vspace{0.1mm}\\
       \hline
\end{tabular}
}
\end{table}

\rec{The values used to plot the frequency-spectra of \fig{fig:spectres_GCM} are given by \tab{tableS1} for study~1 (dependence on the star-planet distance) and \tab{tableS2} for study~2 (dependence on the planet surface pressure). In both cases, the first column corresponds to the normalized tidal frequency $\omeganorm = \left( \spinrate - \norb \right) / \norb$.}

\end{document}